\documentclass[a4paper,fleqn,usenatbib]{mnras}

\usepackage[T1]{fontenc}
\usepackage{ae,aecompl}

\usepackage{graphicx}	
\usepackage{subfig}
\usepackage{rotating,pdflscape,multirow}

\title[Nitrogen Abundances for a Sample of Be Stars]{Rotational Mixing in Be Stars: Nitrogen Abundances for a Sample of Be Stars from the MiMeS Survey}

\author[A. Ahmed and T. A. A. Sigut]{A. Ahmed$^{1,2}$\thanks{E-mail:
ahamza@sci.cu.edu.eg (AA); asigut@uwo.ca (TAAS)} and T. A. A. Sigut$^{1,3}$ \\
$^{1}$Department of Physics and Astronomy, The University of Western Ontario \\ 
London, Ontario, N6A 3K7, Canada\\ 
$^{2}$Department of Astronomy, Cairo University, Giza, 12613, Egypt \\
$^{3}$Centre for Planetary Science and Exploration, The University of Western
Ontario \\ London, Ontario, N6A 3K7, Canada}

\date{Accepted XXX. Received YYY; in original form ZZZ}

\pubyear{2017}

\begin{document}
\label{firstpage}
\pagerange{\pageref{firstpage}--\pageref{lastpage}}
\maketitle

\begin{abstract}

Photospheric nitrogen abundances for a sample of 26 Be stars and 16 normal B stars were found using high-resolution spectra from the Magnetism in Massive Stars (MiMeS) spectroscopic survey. Nitrogen abundances were obtained using non-LTE equivalent widths and line profiles, and Monte Carlo simulation was used to determine the error bounds of the measured nitrogen abundances due to uncertainties in the adopted stellar parameters, continuum normalization, and atomic data. In addition, the effects of the gravitational darkening and disk contamination on the measured Be star nitrogen abundances were investigated. About one third of the sample Be stars and half of the normal B-type stars have nitrogen enrichments that may be due to rotational mixing in these rapidly rotating objects. Corrections for gravitational darkening and disk contamination do not significantly change the overall results for the Be star sample. The average nitrogen abundance of the B and Be star samples coincide with the solar abundance, although the dispersion in the nitrogen abundances is much larger in the Be star sample. The Be star sample also has a significant fraction of sub-solar nitrogen abundance objects which are not present in significant numbers in the B star sample. This may point to yet unresolved systematic errors in the analysis of the Be stars.

\end{abstract}




\section{Introduction}

Theoretical models of the main sequence evolution of rotating stars predict changes in their surface CNO abundances through rotationally-induced mixing. The amount of mixing depends on the initial mass and rotational velocity of the star and increases with age \citep[e.g.\/][]{MM12,Pal13}. Over time, rotational mixing causes an enhancement of the atmospheric nitrogen abundance and a reduction in the carbon and oxygen abundances as stars age on the main-sequence \citep{mey00,mae09,prz2010,brott11,Eks12,Mae14}. Despite this theoretical picture, observational support for this phenomenon is mixed. Comparison to observed samples of massive O and B stars is complicated by the fact that the mixing is not a function of stellar rotation alone, but also depends on stellar mass, age, metallicity, and binarity, a point emphasized by \citet{mae09}. Abundance studies typically find nitrogen enrichment in only a fraction of of main sequence B stars, typically around one-third \citep{gie92,mae09,hun09,lyu12,Mae14, NP14}.

In providing evidence for or against the predicted amount rotational mixing, accurate estimates of the photospheric nitrogen abundances of Be stars may play an important role.  Be stars are the fastest rotating main-sequence stars, with apparent rotational velocities in excess of $80\,\%$ of their critical velocities \citep{Str31,por96,tow04,kel04,cra05}. It is well established that a Be star is a main sequence B-type star surrounded by a thin, equatorial, circumstellar disk in Keplerian rotation, and this disk and its variability are the origin of the observational characteristics of the Be phenomena, namely optical and infrared emission lines, an infrared excess, and net continuum polarization \citep{Riv13}. Be stars may represent an evolved stage on the main-sequence \citep{mcs05,mar06,wis06,tar12}. Consequently, if rotational mixing is as an efficient mechanism of matter transport as predicted by theoretical models, Be stars should should be prime targets to show nitrogen enrichment in their atmospheres. Estimates of the nitrogen abundances for a sample of Be stars in the Magellanic Clouds were presented by \citet{dun11}, but, surprisingly, the distribution of nitrogen abundances in the Be star sample was indistinguishable from the nitrogen abundance distribution in a similarly-selected sample of normal B stars. However, an ad-hoc treatment of the contamination of the Be spectra from the circumstellar disk is an important limiting factor in the work of \cite{dun11}.

Performing an abundance analysis for a hot, rapidly rotating Be star is not an easy task due to several difficulties: (1) There is the potential of departure from local thermodynamic equilibrium (LTE hereafter) in the excitation and ionization state of the gas in its atmosphere \citep{mil78}; (2) Rapid rotation results in shallow and broad profiles with strong line blending; (3) Rapid rotation causes gravitational darkening where the stellar effective temperature and surface gravity become dependent on latitude and the stellar rotational velocity \citep{von24}; and (4) The photospheric spectrum is potentially contaminated by emission (both line and continuum) from the circumstellar disk. As a consequence of all of these aforementioned difficulties, Be stars are typically either excluded \citep{gie92,lyu12} or treated in an approximate way \citep{len05,dun11} in studies of abundances in early-type stars.

This current work aims to address many of these issues and presents the results of an abundance analysis for a sample of Be stars using high-resolution spectra from the MiMeS (Magnetism in Massive Stars) survey conducted with the ESPaDoNs spectropolarimeter on the CFHT \citep{wade16}. To account for non-LTE effects, the recent non-LTE N\,{\sc ii} line profiles of \citet{AS15} are employed. In addition, the effect of gravitational darkening on the line profiles and the effect of circmustellar disk contamination on the spectra are estimated using the {\sc bedisk} \citep{sig07} and {\sc beray} \citep{sig11} codes which are designed specifically to model combined star+disk spectra.

The structure of the paper as follows: a short review of previous works is presented in Section~\ref{prev_works}.  Section~\ref{Sample_data} discusses the MiMeS sample and the stellar parameters adopted for the analysis. The continuum normalization of the observed spectra and the measurement of the N\,{\sc ii} equivalent widths are also discussed. Section~\ref{Nabund_res} presents the measured nitrogen abundances of the Be stars in the MiMeS sample. Corrections to the abundances due to gravitational darkening and disk contamination are discussed in subsections \ref{gd_sec} and \ref{Disk_paras}. As a comparison and control group, measured nitrogen abundances for a sample of normal B stars also included in the MiMeS survey are presented in Section~\ref{Nabund_res_B_stars}. Conclusions are presented in Section~\ref{disc}, and future directions are discussed in Section~\ref{future_plan}.

\section{Previous Works}\label{prev_works}

\cite{lyu12} studied a sample of 22~Galactic B-type stars with masses between 5 and 11$\;M_{\odot}$. Results did not yield CNO abundances consistent with the predicted changes via rotational mixing from theoretical models \citep{brott11,Eks12}. \citet{nie11} studied 13~narrow-lined, early B-type stars with $v\sin\,i\leq$ 60 $\rm km\,s^{-1}$ in the Ori~OB1 association. They did not find abundances consistent with the predicted changes of the surface abundances by rotational mixing. \citet{nie12} measured the elemental abundances of 20~early B-type stars within $500\;$pc of the sun using echelle spectra with high resolution and high signal-to-noise. About one third of the sample was found to have nitrogen abundances consistent with surface enrichment through rotational mixing. 

All of these mentioned studies used samples of stars with low projected rotational velocities, $v\,\sin i\leq\approx\,60\;\rm km\,s^{-1}$, and therefore may be slow rotators without efficient rotational mixing. Although $v\sin i$ is the projected rotational velocity, only a few percent of each sample could be rapid rotators seen nearly pole-on, assuming a random distribution of inclination angles. 

The VLT-FLAMES survey \citep{evans05} is a large survey of massive stars which spectroscopically studied about 50~O-type stars and 500~B-type stars in seven clusters in the Milky Way and Magellanic clouds. The stellar parameters effective temperature ($T_{\rm eff}$), surface gravity ($\log\,g$), projected rotational velocity ($v\sin i$), and elemental abundances of the B-type stars are given in \citet{hun07}, \citet{hun09}, and \citet{tru07}. Nitrogen abundance analysis for a sample of early B-type stars from the VLT-FLAMES survey was performed by \citet{hun08b} and \citet{hun09}. \citet{hun08b} found that 20$\%$ of a sample of 135 early B-type stars in the Large Magellanic Cloud (LMC) are rapid rotators with nitrogen enrichments in agreement with the predictions of rotational mixing (see Fig.~1 of \citealt{hun08b}).  A further 40$\%$ of the sample are slow rotators with $v\sin i \leq 50\;\rm km\,s^{-1}$ and nitrogen enrichments that cannot be explained by rotational mixing. Instead, different mechanisms, such as mass transfer in close binaries, are suggested. \citet{hun09} extended this investigation to samples of early B-type stars in the Milky Way Galaxy and the Small Magellanic Cloud (SMC). About 20$\%$ of the SMC sample were rapidly rotating and nitrogen enriched, and another 20$\%$ of the sample were slowly rotating and nitrogen enriched. No nitrogen enrichment was found in the Galactic sample \cite[see Figs.~6 and 7 of][]{hun09}.

\citet{dun11} showed that there is no difference in the observed nitrogen abundance distributions between Magellanic Cloud B and Be stars (see Fig.~5 of \citealt{dun11}). This is surprising because Be stars, as noted above, are the fastest rotating main-sequence stars \citep{Fuk82,yud01}. Rapid rotation is thought to be the key driver of the Be phenomena; therefore, low $v\,\sin i$ Be stars are still expected to be rapidly rotating stars but seen at low inclination angles.

Despite all of these results, \citet{mae09} emphasize that nitrogen enrichment is a function of stellar mass, rotation, metallicity, age and multiplicity, and not just a function of stellar rotation alone as assumed by \citet{hun08b}. As a consequence, \citet{mae09} argued that the inclusion of field stars, the broad mass range of the stars ($10-30\,M_\odot$), the low number of B stars with projected rotational velocities higher than $250\;\rm km\,s^{-1}$, and the expected binary error, significantly bias the results of \citet{hun08b}. \citet{mae09} reanalyzed the data and showed that the atmospheric nitrogen abundance increases with increasing stellar rotation for B stars with masses in a narrow range and with the same metallicity and age (see Fig.~3 of \citealt{mae09}). This work was complemented by \citet{Mae14} for a larger sample of B stars from \citet{hun09}. Again, \citet{Mae14} argued that observed nitrogen abundances provides evidence for rotational mixing, and the observed scatter in the abundances is attributed to the low quality of data. \citet{Mae14} re-investigated the accuracy of the estimated stellar parameters and abundances for a sub-sample of B stars from \citet{hun09}, about $10\%$ of the sample, by comparing the observed spectra with synthetic, non-LTE spectra computed with the stellar parameters and abundances of \citet{hun09}. \citet{Mae14} did not obtain a good match in many cases and found that many of the stars showed evidence of binarity. Consequently, \citet{Mae14} suggested a careful reassessment of the results of \citet{hun09}.

\citet{koh12} introduced a new method of testing the predicted nitrogen enrichment due to rotational mixing in early B-type stars by comparing their ages determined from evolutionary tracks in the HR diagram with the predicted ages assuming that the stars were enriched due to rotational mixing on the main-sequence. The latter ages were calculated based on measured masses, nitrogen surface abundances, and rotational velocities. The rotational velocities were measured by fitting line profiles using non-LTE {\sc tlusty} models from \citet{hun08a}. The stars were selected such that their nitrogen abundances are not larger than the maximum predicted nitrogen abundance enhancements by rotational mixing at velocities equal to their measured rotational velocities. It was found that the ages calculated for seven out of 17 stars agree with those measured using the evolutionary tracks, i.e.\ an observed nitrogen enrichment of less than a half of these stars can be explained by rotational mixing.


\section{The MiMeS B and Be Star Samples}\label{Sample_data}

The MiMeS survey represents an extensive spectroscopic survey of local, Galactic, B-type stars for the purpose of detecting and classifying magnetic fields in these objects \citep{wade16}. From the MiMeS survey, two samples of stars were constructed: The first consists of 26 early-type Be stars with spectral types between B0 and B3 and luminosity classes between V and II. The second consists of 16 normal, early-type B-type stars, also with spectral types between B0 and B3 and luminosity classes between V and II. While covering a similar range of spectral types and luminosity classs, the Be star sample is somewhat cooler, with a mean $T_{\rm eff}$ of $19,700\;$K compared to $22,800\;$K for the normal B star sample. In addition, the distribution in $T_{\rm eff}$ is somewhat different with 18 of the 26 Be stars cooler than $20,000\;$K, while only three of the 16 B stars are cooler than $20,000\;$K. There is also a difference in the $\log\,g$ distribution, with the Be sample consisting mostly of $\log g < 4.0$ objects, while the B star sample is mostly $\log g\approx\,4.0$.

As part of the MiMeS survey, high resolution spectra for these stars were taken with the Echelle spectropolarimeter ESPaDOnS on the Cassegrain focus of the 3.6m Canada-France-Hawaii Telescope (CFHT) and the NARVAL spectropolarimeter on the T\'{e}lescope Bernard Lyot (TBL).  ESPaDOnS spectra have a resolving power of 80,000 in the spectroscopic star-only option, used for observing bright stars, and a resolving power of 68,000 in the spectroscopic star + sky mode used for observations of faint stars and in polarimetric mode in which all Stokes parameters are measured. The latter mode was adopted in producing the spectra for the MiMeS survey.
ESPaDOnS has a spectral coverage from 3700 to 10,500\AA\ in a single exposure, divided into 40 overlapped orders, with three gaps in the far red region of the spectrum: $9224-9234\,$\AA, $9608-9636\,$\AA, and $10026-10074\,$\AA.

\subsection{Continuum Normalization}

The initial processing of spectra was done by the CFHT observatory staff using the Libre-Esprit software\footnote{http://www.cfht.hawaii.edu/Instruments/Spectroscopy/Espadons/} which performs bias subtraction, wavelength calibration, and flat field division. The wavelength calibration was done via exposures of a thorium comparison lamp. Also, this software provides tools for normalizing the spectra, subtracting telluric lines, and calculating the polarization Stokes parameters.  More details are provided by \citet{wade16}.

As many of the stars in our sample have large $v\sin i$ and the observed spectral lines are shallow and broad, a careful renormalization of the spectra was performed. This refinement of the normalization was carried out using the {\sc continuum} package of IRAF, the image reduction and analysis facility \citep{tod93}. Fitting of the spectra was done both for the entire spectral orders that contained the lines of interest and also over short ranges of wavelength in each spectral order around the lines of interest. This renormalization was done mainly using a first-order linear spline function or a third-order Legendre polynomial. 
%
%

\subsection{Stellar Parameters of Be Stars Sample}\label{stellar_paras}

%
%
\begin{table*}
\centering{
\caption{Adopted Stellar Parameters for the MiMeS Be Star Sample. \label{parameters_table}}}
\begin{center}
{\small
\begin{tabular}{crrrrrrrrrrrrrrrrrrrrrrrrrrr}
\hline\hline
\multicolumn{1}{l}{HD}	&	\multicolumn{2}{c}{Literature Parameters} &	\multicolumn{1}{c}{Revised Parameters} &	\multicolumn{1}{c}{Spect. Type}  & \multicolumn{4}{c}{\hrulefill $v\,\sin i \; (\rm km\,s^{-1}) $ }\hrulefill & Source \\
& \multicolumn{1}{c}{$T_{\rm eff}\;\rm(K)$} & \multicolumn{1}{c}{$\log g$} & \multicolumn{1}{c}{($T_{\rm eff}\,(\rm K)$, $\log g$)} &	\multicolumn{1}{c}{\footnotesize{SIMBAD}} & \multicolumn{1}{c}{\footnotesize{F05}} & \footnotesize{SIMBAD} & \multicolumn{2}{c}{\footnotesize{Current Estimates}} & \\
& & & & & & & \footnotesize{TLUSTY} & \footnotesize{N\,{\sc ii} $\lambda\,$3995} & \\ 
\hline 				
\multicolumn{1}{l}{11415}  & \multicolumn{1}{l}{15147$\pm$455}	&	\multicolumn{1}{l}{3.50$\pm$0.20}	& \multicolumn{1}{l}{(15600, 3.5)}	&	\multicolumn{1}{l}{B3 III}		&\multicolumn{1}{c}{$-$}	&	30.0	 & 45.0 & 35.0 & \multicolumn{1}{l}{Tak10}	\\	
\multicolumn{1}{l}{20336}	&	\multicolumn{1}{l}{18684$\pm$517}	&	\multicolumn{1}{l}{3.87$\pm$0.07} &	\multicolumn{1}{c}{$-$}&	\multicolumn{1}{l}{B2.5 Vne}	&	328.0$\pm$21.0	&	328.0	& 285.0 & 285.0 & \multicolumn{1}{l}{F05}\\
\multicolumn{1}{l}{33328}	& \multicolumn{1}{l}{21137$\pm$514}	&	\multicolumn{1}{l}{3.45$\pm$0.08}	& \multicolumn{1}{c}{$-$}	&	\multicolumn{1}{l}{B2 IV ne}	&	318.0$\pm$22.0	&	150.0	& 290.0 & :265.0 & \multicolumn{1}{l}{F05}\\
\multicolumn{1}{l}{45725}	&	\multicolumn{1}{l}{17810$\pm$455}	&	\multicolumn{1}{l}{3.90$\pm$0.07} &  \multicolumn{1}{c}{$-$}&	\multicolumn{1}{l}{B3 Ve}		&	330.0$\pm$20.0	&	260.0	&	290.0 & 280.0	& \multicolumn{1}{l}{F05}\\
\multicolumn{1}{l}{49567} 	&	\multicolumn{1}{l}{17270$\pm$1010}	&	\multicolumn{1}{c}{3.80$\pm$0.20} & \multicolumn{1}{l}{(17270.0, 3.3)} & \multicolumn{1}{l}{B3 II-III}	&\multicolumn{1}{c}{$-$}&	85.0	&	80.0 & 80.0 	& \multicolumn{1}{l}{Zor09}\\
\multicolumn{1}{l}{54309}	&	\multicolumn{1}{l}{20859$\pm$397}	&  \multicolumn{1}{l}{3.59$\pm$0.05}	& \multicolumn{1}{l}{(20859, 3.4)}&	\multicolumn{1}{l}{B2 V:nn} 	&	195.0$\pm$10.0	&	195.0	&	195.0 & 220.0 & \multicolumn{1}{l}{F05}\\
\multicolumn{1}{l}{56139}	&	\multicolumn{1}{l}{19537$\pm$331}	&	\multicolumn{1}{l}{3.62$\pm$0.04}	& \multicolumn{1}{l}{(19537, 3.4)} &	\multicolumn{1}{l}{B2IV-Ve} 	&	 85.0$\pm$4.0	& 100.0	& 50.0 & 75.0 & \multicolumn{1}{l}{F05}\\
\multicolumn{1}{l}{58050} 	&	\multicolumn{1}{l}{19961$\pm$465} 	&  \multicolumn{1}{l}{3.93$\pm$0.06}	& \multicolumn{1}{l}{(19961, 3.8)} &	\multicolumn{1}{l}{B2 Ve}		&	130.0$\pm$8.0	&	130.0 & 130.0 & 110.0	& \multicolumn{1}{l}{F05}\\
\multicolumn{1}{l}{58343} 	&	\multicolumn{1}{l}{16531$\pm$409}	&	\multicolumn{1}{l}{3.62$\pm$0.06}	&\multicolumn{1}{c}{$-$} &	\multicolumn{1}{l}{B2 Vne} 	&	 43.0$\pm$2.0	&	43.0	 & 50.0 & 40.0 & \multicolumn{1}{l}{F05}\\
\multicolumn{1}{l}{58978}	&	\multicolumn{1}{l}{24445$\pm$476}	&	\multicolumn{1}{l}{4.15$\pm$0.06}	&  \multicolumn{1}{l}{(24445, 3.4)}&	\multicolumn{1}{l}{	B1 II}	&	370.0$\pm$21.0	&	155.0	& 310.0 & 300.0	& \multicolumn{1}{l}{F05}\\
\multicolumn{1}{l}{65875}	&	\multicolumn{1}{l}{20205$\pm$532}	&	\multicolumn{1}{l}{3.84$\pm$0.07}	& \multicolumn{1}{l}{(19900, 3.30)} &	\multicolumn{1}{l}{B2.5 Ve}	&	153.0$\pm$10.0	&	140.0	& 190.0 & 140.0	& \multicolumn{1}{l}{F05}\\	
\multicolumn{1}{l}{67698}	& 	\multicolumn{1}{l}{17400$\pm$500}	&	\multicolumn{1}{l}{3.60$\pm$0.01}	& \multicolumn{1}{l}{(15500, 3.70)}&	\multicolumn{1}{l}{B3 III/IV}	&\multicolumn{1}{c}{$-$}&	150.0	& 90.0 & 75.0 & \multicolumn{1}{l}{LL06} \\
\multicolumn{1}{l}{120324} & \multicolumn{1}{l}{20000$\pm$500}		&	\multicolumn{1}{c}{$-$}	& \multicolumn{1}{l}{(20000, 4.0)}&	\multicolumn{1}{l}{B2V}		& \multicolumn{1}{c}{$-$}&	155.0	 &	140.0 & 120.0 & \multicolumn{1}{l}{H00}\\
\multicolumn{1}{l}{143275}	& \multicolumn{1}{l}{31478$\pm$500}	&	\multicolumn{1}{c}{$-$}	 &	\multicolumn{1}{l}{(31478, 3.5)} & \multicolumn{1}{l}{B0.2IV}		&	 \multicolumn{1}{c}{$-$}&	175.0	 & :190.0 & 180.0  & \multicolumn{1}{l}{H00}\\
\multicolumn{1}{l}{174237} & \multicolumn{1}{l}{17683$\pm$556}		&  \multicolumn{1}{l}{3.76$\pm$0.08} & \multicolumn{1}{l}{(17683, 3.65)}&	\multicolumn{1}{l}{B2.5 Ve}	&	163.0$\pm$11.0	&	163.0	& 150.0  & 130.0	& \multicolumn{1}{l}{F05}	\\
\multicolumn{1}{l}{178175} & \multicolumn{1}{l}{18939$\pm$286}		&	\multicolumn{1}{l}{3.49$\pm$0.04}	&  \multicolumn{1}{l}{(22000, 3.5)}& \multicolumn{1}{l}{B2 V}		&	105.0$\pm$5.0	&	105.0	&	150.0 & 140.0 & \multicolumn{1}{l}{F05}\\
\multicolumn{1}{l}{187567}	& \multicolumn{1}{l}{23110$\pm$500}		&	\multicolumn{1}{l}{4.00} & \multicolumn{1}{l}{(23110, 3.7)} & \multicolumn{1}{l}{B2.5 IV}	&\multicolumn{1}{c}{$-$}&	140.0	& 195.0 & 245.0 & \multicolumn{1}{l}{Cat13}\\
\multicolumn{1}{l}{187811} & \multicolumn{1}{l}{18086$\pm$583}		&	\multicolumn{1}{l}{3.81$\pm$0.08}	&	\multicolumn{1}{l}{(17800, 3.8)} &\multicolumn{1}{l}{B2.5 Ve}	&	245.0$\pm$17.0	&	245.0	&	220.0 & :145.0 & \multicolumn{1}{l}{F05}\\
\multicolumn{1}{l}{189687}	& \multicolumn{1}{l}{18106$\pm$379}		&	\multicolumn{1}{l}{3.46$\pm$0.05}	& \multicolumn{1}{c}{$-$}&	\multicolumn{1}{l}{B3IV}		&  200.0$\pm$11.0 & 	200.0 & 200.0 & 200.0 & \multicolumn{1}{l}{F05}\\
\multicolumn{1}{l}{191610}	& \multicolumn{1}{l}{18353$\pm$516}		&	\multicolumn{1}{l}{3.72$\pm$0.07}	& \multicolumn{1}{c}{$-$}&	\multicolumn{1}{l}{B2.5 V}		&	300.0$\pm$20.0 &	300.0	&	270.0 & 270.0 & \multicolumn{1}{l}{F05}\\
\multicolumn{1}{l}{192685} & \multicolumn{1}{l}{18000$\pm$500}		&	\multicolumn{1}{l}{3.5$\pm$0.20}	&	\multicolumn{1}{l}{(18500, 3.7)}&	\multicolumn{1}{l}{B3 V} 		&\multicolumn{1}{c}{$-$}&	160.0	&  180.0 & 180.0	& \multicolumn{1}{l}{Cote96}\\
%
\multicolumn{1}{l}{203467} &	\multicolumn{1}{l}{17087$\pm$521}	&	\multicolumn{1}{l}{3.38$\pm$0.07} &\multicolumn{1}{c}{$-$} &	\multicolumn{1}{l}{B3 IVe}		&	153.0$\pm$10.0	&	120.0	& 200.0	& 150.0	& \multicolumn{1}{l}{F05}\\
\multicolumn{1}{l}{205637} & 	\multicolumn{1}{l}{17801$\pm$470}	&	\multicolumn{1}{l}{3.44$\pm$0.06} &	\multicolumn{1}{c}{$-$}&   \multicolumn{1}{l}{B3 Vpe}		&	225.0$\pm$14.0	&	225.0 & 220.0	& 300.0	& \multicolumn{1}{l}{F05}\\
\multicolumn{1}{l}{212076} &	\multicolumn{1}{l}{19270$\pm$326}	&	\multicolumn{1}{l}{3.73$\pm$0.04}	&	\multicolumn{1}{l}{(19270, 3.5)} &\multicolumn{1}{l}{B2 IV-Ve}	&	 98.0$\pm$5.0	&	98.0	&	130.0 & 115.0 & \multicolumn{1}{l}{F05}\\
\multicolumn{1}{l}{212571} &	\multicolumn{1}{l}{26061$\pm$736}	& 	\multicolumn{1}{l}{3.92$\pm$0.09}	&	\multicolumn{1}{l}{(26061, 3.7)} & \multicolumn{1}{l}{B1Ve}		&	230.0$\pm$17.0	&	215.0	& 230.0 & 300.0	& \multicolumn{1}{l}{F05}\\
\multicolumn{1}{l}{217050} &	\multicolumn{1}{l}{17893$\pm$509}  &	\multicolumn{1}{l}{3.57$\pm$0.07}	&	\multicolumn{1}{l}{(17893, 3.3)} & \multicolumn{1}{l}{B3IV} &	340.0$\pm$22.0	&	250.0	&	270.0 & 265.0 & \multicolumn{1}{l}{F05}\\
\hline
\end{tabular}}
\end{center}
\noindent{ {\small Sources: Tak10 \citep{tak10}, F05 \citep{fre05}, H00 \citep{H00}, Zor09 \citep{Zor09}, LL06 \citep{LL06}, Cat13 \citep{Cat13}, Cote96 \citep{cote96}.\\ Notes: SIMBAD refers to spectral type classifications and $v\sin i$ estimates adopted from the SIMBAD database \citep{W00}.}}
\end{table*}

The stellar parameters $T_{\rm eff}$ and $\log\,g$ of the 26 Be~stars in the MiMeS sample are given in Table~\ref{parameters_table}. These were adopted from a literature search and are mainly from \citet{fre05}. In \citet{fre05}, stellar effective temperatures and gravities were obtained by comparing observed spectra with non-LTE, synthetic ones over four wavelength ranges between 4300 and $4490\;$\AA, which included lines sensitive to $T_{\rm eff}$ and $\log g$. A hybrid, non-LTE approach was adopted in which line-blanketed, LTE model atmospheres \citep{kur79,kur93} were used as input into detailed, non-LTE line formation calculations which held the model atmosphere fixed. The lines used were He\,{\sc i} $\lambda\,4471$ and Mg\,{\sc ii} $\lambda\,4481$, which are sensitive for variations of stellar effective temperature, and the Balmer H$\gamma$ line which is sensitive to stellar gravity. \citet{fre05} provides two sets of stellar parameters: the first set is directly estimated from the observed spectral lines without corrections for gravitational darkening and these are called the apparent stellar parameters. The other set was obtained including the effects of gravitational darkening following the standard treatment of \citet{von24} \citep[see][]{Collins65}. In the current work, the apparent parameters of \citet{fre05} were adopted and not the ones corrected for gravitational darkening. The main reason for this decision is the potential for overestimation of the effects of gravitational darkening using the standard von Zeipel treatment as compared with the recent reformulation of \citet{lar11}. The potential impact of gravitational darkening on the derived abundances will be discussed in Section~\ref{gd_sec}. 

Some notes on the parameters for specific stars are as follows, mostly for stars for which \citet{fre05} parameters were unavailable. Of these other works, only \citet{LL06} accounted for the effects of gravitational darkening, using the standard von Zeipel treatment. 

The effective temperature and the gravity for the Be star HD~11415 (B3III) was taken from \citet{tak10} with the temperature obtained from a calibration of the Str\"{o}mgren $ubvy$ colour indexes. The effective temperature and gravity of the Be star HD~49567 (B3II/III) was taken from \citet{Zor09} who fit the available spectral energy distribution. The effective temperature and gravity of the Be star HD~67698 (B3 III/IV) were obtained from \citet{LL06} as determined by the ionization balances of He, CNO, and Si, and the equivalent widths of hydrogen Balmer lines. The effective temperature of the Be star HD~187567 (B2.5IV) was taken from \citet{Cat13} who used measured Str\"{o}mgren colour indexes and the \citet{MD85} algorithm; a value of $\log\,g=4.0$ was adopted for stars with III/IV luminosity types. For the Be stars HD~120324 (B2V) and HD~143275 (B0.2IV), the measurements of \citet{H00} were adopted. In this work, the temperatures were estimated photometrically following the procedure of \citet{MD85}. Also, \citet{Cat13} provides a $T_{\rm eff}$ estimate for the Be star HD~143275 (26700 K) which is much lower than the value (31478 K) of \citet{H00}.  However, the observed spectra was best fit with {\sc tlusty} models interpolated for stellar parameters close to those of \citet{H00}, as will be discussed in the next section. Finally, the effective temperature and gravity of the Be star HD~192685 (B3V) were obtained from \citet{cote96} who compared de-reddened, observed spectra for $\lambda< 8000\,$\AA\ with the LTE, line-blanketed model atmospheres of \citet{kur79,kur93}.

\subsection{Testing the Adopted Stellar Parameters}

A further check on the adopted $T_{\rm eff}$ and $\log g$ parameters for the Be star sample was performed by comparing the MiMeS spectra with the {\sc tlusty} non-LTE, stellar atmosphere models of \citet{Lanz07} over spectral regions that included the helium and hydrogen lines usually used for estimating stellar parameters, e.g. He\,{\sc i} $\lambda\,4471$, He\,{\sc i} $\lambda\,4009$, Mg\,{\sc ii} $\lambda\,4481$, and the H$\epsilon$ line. We note that the \citet{Lanz07} grids were computed for a fixed solar abundance table and a single microturbulence value, $\xi_t=2\;\rm km\,s^{-1}$, which limits their ability to match the stellar spectra in detail. However, the {\sc tlusty} spectra do provide a homogeneous set of theoretical spectra to check the heterogeneous collection of $T_{\rm eff}$ and $\log\,g$ collected from the literature. In addition, there is always the potential for circumstellar contamination of the photospheric spectrum for the Be stars; however, the expected continuum contamination in our sample is small based on direct modeling of the H$\alpha$ emission line, as discussed in Section~\ref{Nabund_corr}.

The {\sc tlusty} grid of models was interpolated for the stellar parameters and rotationally broadened. This comparison also provided an additional estimate of $v\sin i$. The observed spectra were slightly re-normalized in some cases to best match the synthetic spectra. Overall, this comparison shows that the effective temperatures and/or gravities of most of the Be stars in the sample could be slightly corrected to improve the fits. Corrections of $T_{\rm eff}$ usually lay within $\pm\,500\;$K, while the corrections of $\log g$ were within $\pm$ 0.5 dex. 

\begin{figure*}
\centering
\subfloat[]{
\hspace*{-3.0cm}\includegraphics[scale=0.37]{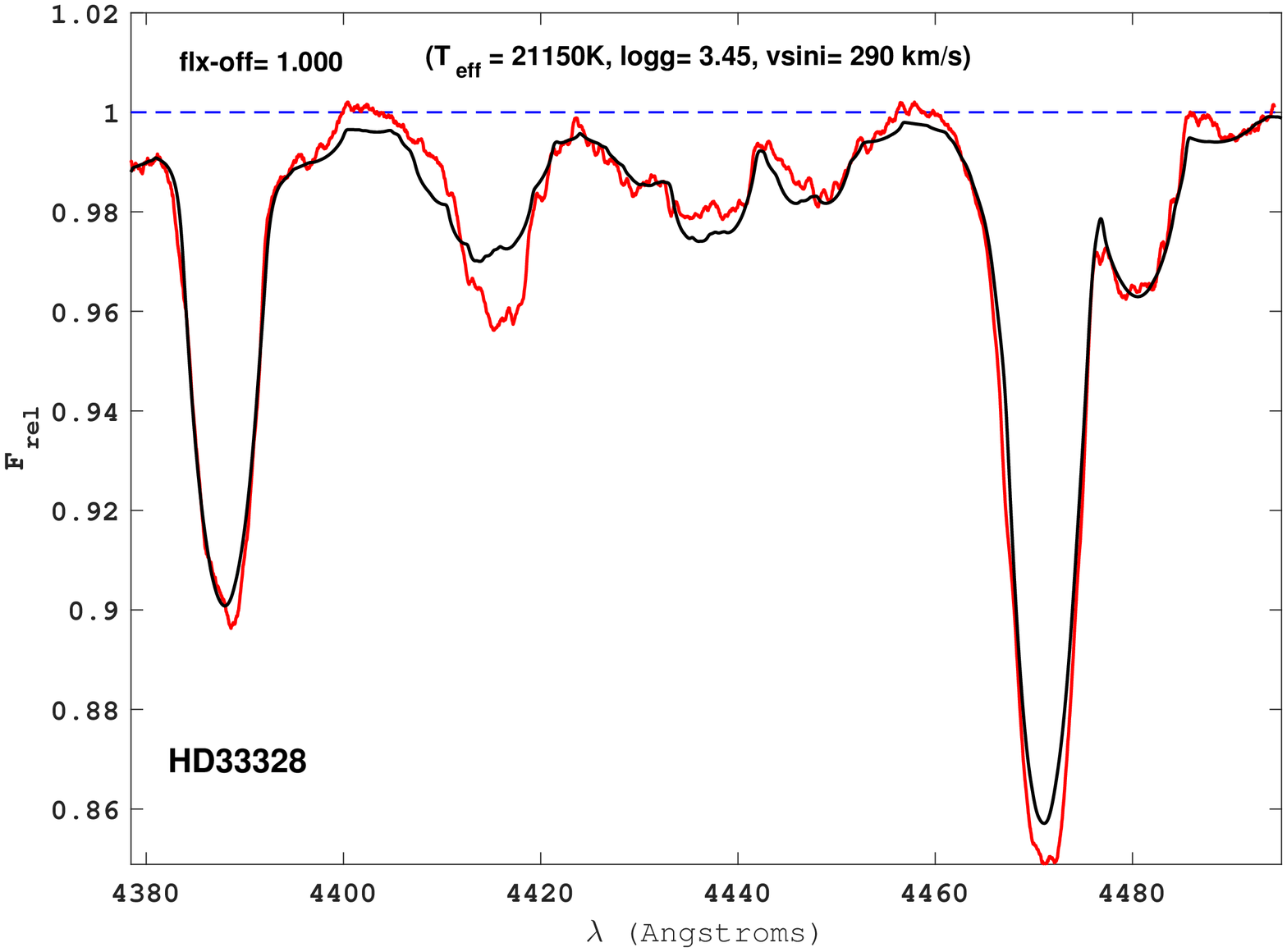}
\label{Obs_vs_TLUSTy_HD33328_ord11}
}
\hspace{0.5cm}
\subfloat[]{
\includegraphics[scale= 0.40]{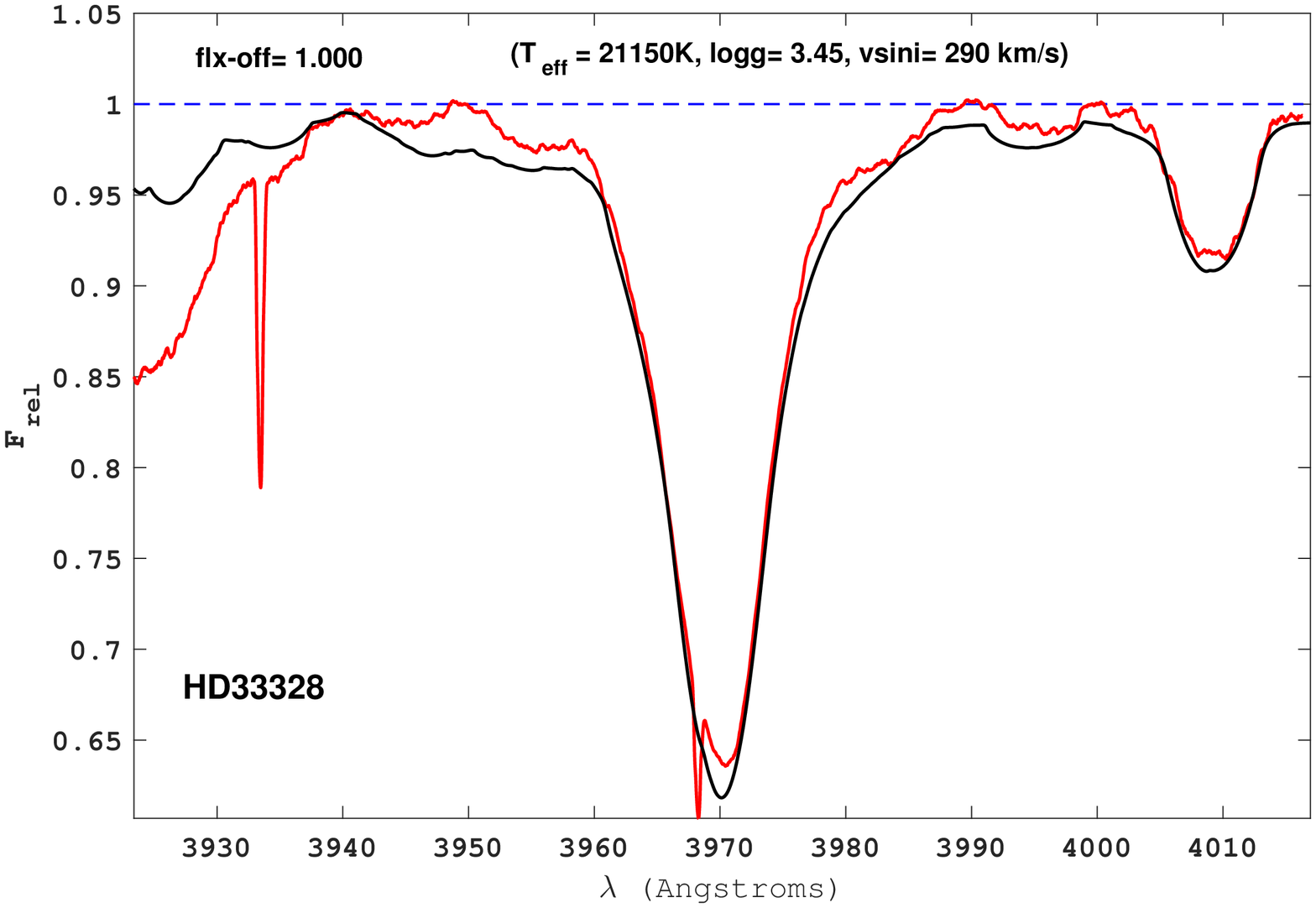}
\hspace*{-3.0cm}
\label{Obs_vs_TLUSTy_HD33328_ord5}
}
\caption{Comparison of the observed spectra of the Be star HD33328 (red line) with the non-LTE fluxes of \citet{Lanz07} (black line), interpolated and rotationally broadened to the literature stellar parameters. The dashed blue lines represents the continuum. The left panel includes He\,{\sc i} $\lambda\,4388$, He\,{\sc i} $\lambda\,4471$ and Mg\,{\sc ii} $\lambda\,4481.2$, while the right panel includes H$\epsilon$ $\lambda\,3970.1$ and He\,{\sc i} $\lambda\,4009$.}
\label{Obs_vs_TLUSTy_HD33328}
\end{figure*}

\begin{figure*}
\centering
\subfloat[]{
\hspace*{-3.0cm}\includegraphics[scale= 0.40]{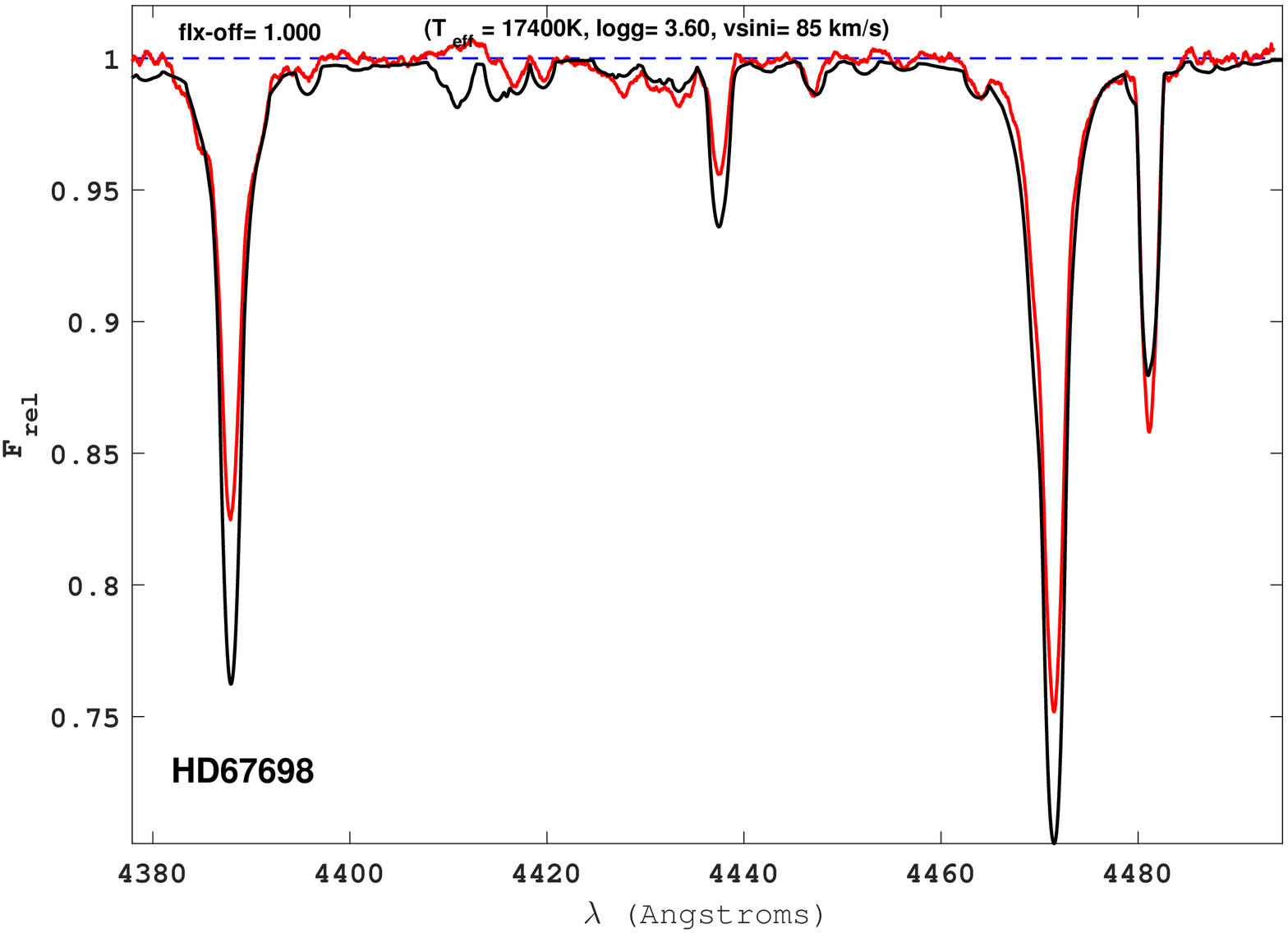}
\label{Obs_vs_TLUSTy_HD67698_ord11}
}
\hspace{0.5cm}
\subfloat[]{
\includegraphics[scale= 0.40]{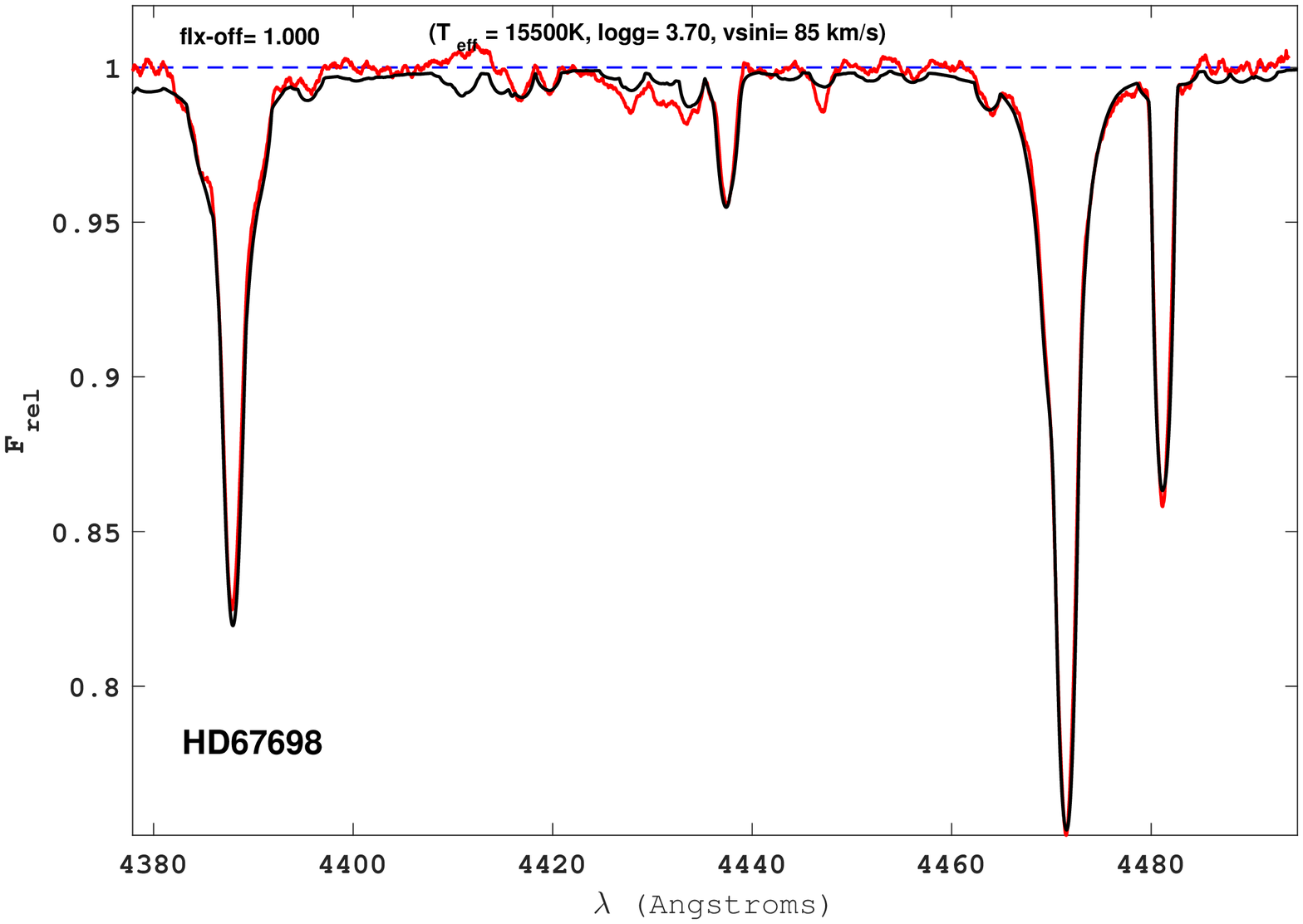}
\hspace*{-3.0cm}
\label{Obs_vs_TLUSTy_HD67698_ord11_2}
}
\caption{The same as Figure~\ref{Obs_vs_TLUSTy_HD33328} but for Be star HD\,67698. In the left panel, the non-LTE synthetic fluxes were interpolated to the stellar parameters of \citet{fre05}, while the right panel shows the stellar parameters at which the non-LTE spectra best match the observed spectra.}
\label{Obs_vs_TLUSTy_HD67698}
\end{figure*}

Figure~\ref{Obs_vs_TLUSTy_HD33328} shows the Be star HD~33328 for which a good match of the observed spectra was obtained using the literature $T_{\rm eff}$ and $\log g$ stellar parameters. On the other hand, Figure~\ref{Obs_vs_TLUSTy_HD67698} shows that the $T_{\rm eff}$ for HD~67698 needs to be corrected significantly (by about 2000~K) to attain a good match. This star and HD~178175, which required a 3,000~K correction in $T_{\rm eff}$ (not shown), represent the largest adjustments to the literature parameters. All revised stellar parameters for the Be star sample are listed in Table~\ref{parameters_table}.

\subsection{The $v\sin i$ Distribution of the Be Stars}

As a by-product of testing the stellar parameters, an estimate of each star's $v\sin i$ was obtained. Table~\ref{parameters_table} lists these $v\sin i$ values based both on the TLUSTY fitting of this section and the profile fitting of the N\,{\sc ii} $\lambda\,3995$ line. Gravitational darkening was not accounted for in these estimates, and this issue is further discussed in Section~\ref{gd_sec} for the N\,{\sc ii}~$\lambda\,3995$ line. Also given is the $v\sin i$ value available from the SIMBAD database \citep{W00} and the estimates of \citet{fre05} and \citet{yud01}. Note that 18 Be stars of the MiMeS sample can be found in the larger Be sample of \citet{fre05}.

\begin{figure}
\centering
\hspace{-0.5cm}
\includegraphics[scale= 0.45]{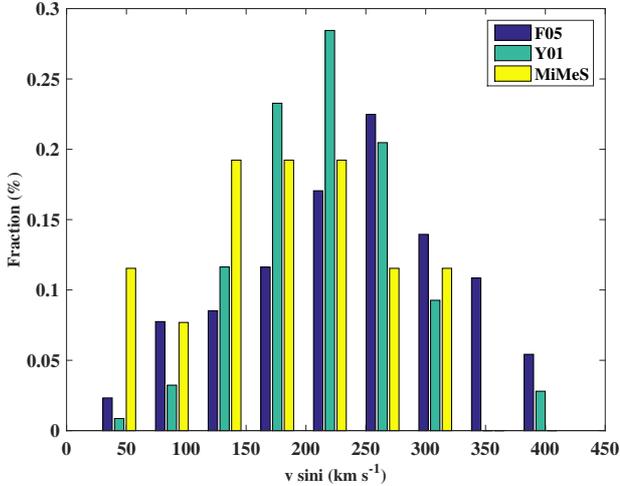}
\caption{The Be star $v\sin i$ distributions from the MiMeS sample (yellow, 26 stars), \citet{fre05} (blue, 192 stars), and \citet{yud01} (green, 463 stars).}
\label{vsini_hist}
\end{figure}

Figure~\ref{vsini_hist} shows the distribution of $v\sin i$ for all three samples. The MiMeS sample of Be stars represents a slightly more slowly rotating sample; the mean $v\sin i$ values are $241\,\rm km\,s^{-1}$ for the \cite{fre05} sample (129 stars), $219\,\rm km\,s^{-1}$ for the \cite{yud01} sample (463 stars), and $183\,\rm km\,s^{-1}$ for the current MiMeS sample. This difference is due mainly to the lack of Be stars with $v\sin i> 310\,\rm km\,s^{-1}$ in the MiMeS sample. However, it is important to keep in mind that (1) the Be stars as a population are expected to be rapidly rotating so the actual difference between these samples may be the inclination distribution, and (2) we have not included a macroturbulent broadening in the calculated line profiles which may be non-negligible for some of the lower surface gravity, more slowly rotating stars in our sample \citep{SD17}.

\subsection{Stellar Parameters for the B-type Stars in the MiMeS Sample}
\label{stellar_paras_Bstars}

The adopted stellar parameters $T_{\rm eff}$ and $\log g$ for the normal B star sample from the MiMeS survey are listed in Table~\ref{parameters_table_Bstars}. Most values were adopted from the literature. Apparent from the table is that the sample of normal B-type stars consists of low $v\sin i$ objects, as sharp lines allow much easier detection of magnetic fields. Sharp lines also allow an estimate of the atmospheric microturbulence to be made, done by forcing strong and weak lines to yield the same abundance. Often this is not the case for more rapidly rotating stars, i.e.\ the Be stars, because of the difficulty in accurately measuring the strengths of weak lines at high rotational broadening.

\begin{table*}
\centering{
\caption{Adopted Stellar Parameters for the MiMeS normal B star sample. \label{parameters_table_Bstars}}}
\begin{center}
{\small
\begin{tabular}{c rrrrrrrrrrrrrrrrrrrrrrrrrrr}
\hline\hline
\multicolumn{1}{l}{HD}	&	\multicolumn{1}{c}{$T_{\rm eff}\,(\rm K)$}	&	\multicolumn{1}{c}{$\log g$}	& \multicolumn{1}{c}{$v\,\sin i$ } & \multicolumn{1}{c}{$\xi_t$} & \multicolumn{1}{c}{$\zeta$}&	\multicolumn{1}{c}{Spectral}  & \multicolumn{2}{c}{$v\,\sin i \; (\rm km\,s^{-1}) $ } & \multicolumn{1}{c}{$\xi_t$($\rm km\,s^{-1}$)} & \multicolumn{1}{l}{Source} \\
& & &	\multicolumn{3}{c}{\footnotesize{\hrulefill ($\;\rm km\,s^{-1}\;$)\hrulefill}} & \multicolumn{1}{c}{\footnotesize{Type}} & \multicolumn{1}{c}{\footnotesize{MiMeS}} & \footnotesize{Current} & \multicolumn{1}{c}{\footnotesize{Current}} &\\ 
&&&&&&&& \footnotesize{N\,{\sc ii} $\lambda\,$3995} & \\
\hline 				
\multicolumn{1}{l}{3360}  & \multicolumn{1}{l}{20750$\pm$200}	&	\multicolumn{1}{l}{3.80$\pm$0.05}&\multicolumn{1}{l}{$20\pm$2} &	\multicolumn{1}{l}{$2\pm1$}&\multicolumn{1}{l}{12$\pm$5}	& \multicolumn{1}{l}{B2IV}			& \multicolumn{1}{c}{17} & \multicolumn{1}{c}{23}& \multicolumn{1}{c}{0.1$\pm$0.5} & \multicolumn{1}{l}{Nieva12}\\
\multicolumn{1}{l}{30836}  & \multicolumn{1}{l}{21900$\pm$450}	&	\multicolumn{1}{l}{3.50$\pm$0.00} &\multicolumn{1}{c}{$-$} &\multicolumn{1}{c}{$-$}&\multicolumn{1}{c}{$-$}		&	\multicolumn{1}{l}{B2III+}	& \multicolumn{1}{c}{35} & \multicolumn{1}{c}{39}& \multicolumn{1}{c}{5.4$\pm$2.7}  & \multicolumn{1}{l}{Underhill79}\\
\multicolumn{1}{l}{35708}  & \multicolumn{1}{l}{20700$\pm$200}	&	\multicolumn{1}{l}{4.15$\pm$0.07}& \multicolumn{1}{l}{25$\pm$2}&\multicolumn{1}{l}{2$\pm$1} &\multicolumn{1}{l}{17$\pm$5}		&	\multicolumn{1}{l}{B2V}			& \multicolumn{1}{c}{25} & \multicolumn{1}{c}{30}& \multicolumn{1}{c}{0.6$\pm$1.5} & \multicolumn{1}{l}{Nieva12}\\
\multicolumn{1}{l}{35468}  & \multicolumn{1}{l}{22000$\pm$1000}	&	\multicolumn{1}{l}{3.60$\pm$0.10} & \multicolumn{1}{l}{46$\pm$8} &\multicolumn{1}{l}{10} & \multicolumn{1}{l}{37}		&	\multicolumn{1}{l}{B2II-III}	& \multicolumn{1}{c}{55} & \multicolumn{1}{c}{52}& \multicolumn{1}{c}{1.5$\pm$2.1} & \multicolumn{1}{l}{Lefever10}\\
\multicolumn{1}{l}{36629}  & \multicolumn{1}{l}{20300$\pm$400}	&	\multicolumn{1}{l}{4.15$\pm$0.10}&\multicolumn{1}{l}{10$\pm$2}&\multicolumn{1}{l}{2$\pm$1}&\multicolumn{1}{l}{5$\pm$1}		&	\multicolumn{1}{l}{B2V}			& \multicolumn{1}{c}{5} & \multicolumn{1}{c}{17}& \multicolumn{1}{c}{2.2$\pm$2.8} & \multicolumn{1}{l}{Nieva11}\\
\multicolumn{1}{l}{36822}  & \multicolumn{1}{l}{30000$\pm$300}	&	\multicolumn{1}{l}{4.05$\pm$0.10}&\multicolumn{1}{l}{28$\pm$2}&\multicolumn{1}{l}{8$\pm$1}&\multicolumn{1}{l}{18$\pm$5}		&	\multicolumn{1}{l}{B0.5 III}		 & \multicolumn{1}{c}{20} & \multicolumn{1}{c}{33}& \multicolumn{1}{c}{ 6.2$\pm$4.0} & \multicolumn{1}{l}{Nieva12}\\
\multicolumn{1}{l}{36959}  & \multicolumn{1}{l}{26100$\pm$200}	&	\multicolumn{1}{l}{4.25$\pm$0.07}&\multicolumn{1}{l}{12$\pm$2}&\multicolumn{1}{l}{0$\pm$1}&\multicolumn{1}{l}{5$\pm$1}	&	\multicolumn{1}{l}{B1.5V}			& \multicolumn{1}{c}{5} & \multicolumn{1}{c}{17}& \multicolumn{1}{c}{0.0$\pm$0.1} & \multicolumn{1}{l}{Nieva11}\\
\multicolumn{1}{l}{36960}  & \multicolumn{1}{l}{29000$\pm$300}	&	\multicolumn{1}{l}{4.10$\pm$0.07}&\multicolumn{1}{l}{28$\pm$3}&\multicolumn{1}{l}{4$\pm$1}&\multicolumn{1}{l}{20$\pm$7}		&	\multicolumn{1}{l}{B0.7V}			& \multicolumn{1}{c}{20} & \multicolumn{1}{c}{33}& \multicolumn{1}{c}{4.2$\pm$.3} & \multicolumn{1}{l}{Nieva12}\\
\multicolumn{1}{l}{46328}  & \multicolumn{1}{l}{27000$\pm$1000}	&	\multicolumn{1}{l}{3.80$\pm$0.10}&\multicolumn{1}{l}{9$\pm$2} &\multicolumn{1}{c}{6}&\multicolumn{1}{c}{11}		&	\multicolumn{1}{l}{B1III}			& \multicolumn{1}{c}{14} & \multicolumn{1}{c}{19}& \multicolumn{1}{c}{0.1$\pm$0.5} & \multicolumn{1}{l}{Lefever10}\\
%
\multicolumn{1}{l}{48977}  & \multicolumn{1}{l}{20000$\pm$1000}	&	\multicolumn{1}{l}{4.20$\pm$0.10}&\multicolumn{1}{l}{29$\pm$1}&\multicolumn{1}{c}{$-$}&\multicolumn{1}{c}{$-$}		&	\multicolumn{1}{l}{B2.5V}			& \multicolumn{1}{c}{20} & \multicolumn{1}{c}{25}& \multicolumn{1}{c}{4.8$\pm$4.6} & \multicolumn{1}{l}{Thoul13}\\
\multicolumn{1}{l}{61068}  & \multicolumn{1}{l}{23800$\pm$1100}	&	\multicolumn{1}{l}{4.01$\pm$0.20}&\multicolumn{1}{l}{10$\pm$9}&\multicolumn{1}{c}{$-$}&\multicolumn{1}{c}{$-$}		&	\multicolumn{1}{l}{B2II}	& \multicolumn{1}{c}{18} & \multicolumn{1}{c}{21}& \multicolumn{1}{c}{0} & \multicolumn{1}{l}{Hubrig09}\\
\multicolumn{1}{l}{66665}  & \multicolumn{1}{l}{28500$\pm$1000}	&	\multicolumn{1}{l}{3.90$\pm$0.10} &\multicolumn{1}{c}{$\leq$10}	&\multicolumn{1}{c}{$-$}&\multicolumn{1}{c}{$-$}	&	\multicolumn{1}{l}{B1V}	& \multicolumn{1}{c}{70} & \multicolumn{1}{c}{16}& \multicolumn{1}{c}{ 0.2$\pm$0.6} & \multicolumn{1}{l}{Petit11}\\
\multicolumn{1}{l}{74560}  & \multicolumn{1}{l}{17000$\pm$500$^*$}	&	\multicolumn{1}{l}{4.00$\pm$0.20$^*$}	&\multicolumn{1}{c}{$-$}&\multicolumn{1}{c}{$-$}&\multicolumn{1}{c}{$-$}	&	\multicolumn{1}{l}{B3 III}			& \multicolumn{1}{c}{60} & \multicolumn{1}{c}{21}& \multicolumn{1}{c}{8.6$\pm$2.8} & \multicolumn{1}{l}{$-$}\\
%
\multicolumn{1}{l}{74575 }  & \multicolumn{1}{l}{22900$\pm$300}	&	\multicolumn{1}{l}{3.60$\pm$0.05} &\multicolumn{1}{c}{11$\pm$2} &\multicolumn{1}{l}{5$\pm$1}&\multicolumn{1}{l}{20$\pm$1}	&	\multicolumn{1}{l}{B1.5III}	& \multicolumn{1}{c}{20} & \multicolumn{1}{c}{28}& \multicolumn{1}{c}{5.5$\pm$1.8} & \multicolumn{1}{l}{Nieva12}\\
\multicolumn{1}{l}{85953}  & \multicolumn{1}{l}{18621}	&	\multicolumn{1}{l}{3.89}&\multicolumn{1}{c}{$\leq$45}&\multicolumn{1}{c}{$-$}&\multicolumn{1}{c}{$-$}		&	\multicolumn{1}{l}{B2III}			& \multicolumn{1}{c}{30} & \multicolumn{1}{c}{31}& \multicolumn{1}{c}{3.2$\pm$4.0} & \multicolumn{1}{l}{Aerts99}\\
\multicolumn{1}{l}{169467}  & \multicolumn{1}{l}{16700$\pm$800}	&	\multicolumn{1}{l}{4.12$\pm$0.20}&\multicolumn{1}{l}{14$\pm$8}&\multicolumn{1}{c}{$-$}&\multicolumn{1}{c}{$-$}	&	\multicolumn{1}{l}{B3IV}	& \multicolumn{1}{c}{20} & \multicolumn{1}{c}{29}& \multicolumn{1}{c}{6.3$\pm$4.4} & \multicolumn{1}{l}{Hubrig09}\\
\hline
\end{tabular}}
\end{center}
\noindent{ {\small Sources: Nieva12 \citep{nie12}, Nieva11 \citep{nie11}, Underhill79 \citep{und79}, Lefever10 \citep{lef10}, Thoul13 \citep{thou13}, Hubrig09 \citep{hub09}, Petit11 \citep{pet11}, Aerts99 \citep{aer99}. \\ Notes: The values marked with asterisks are current estimations. The v$\sin i$ estimates listed in the fourth column are obtained from the same sources as the effective temperatures and gravities, while those listed in the eighth column are obtained from the MiMeS project. The v$\sin i$ values in the ninth column represent current estimates of the present study using the N\,{\sc ii} 3995 \AA\, line.}}
\end{table*}

%
\begin{figure*}
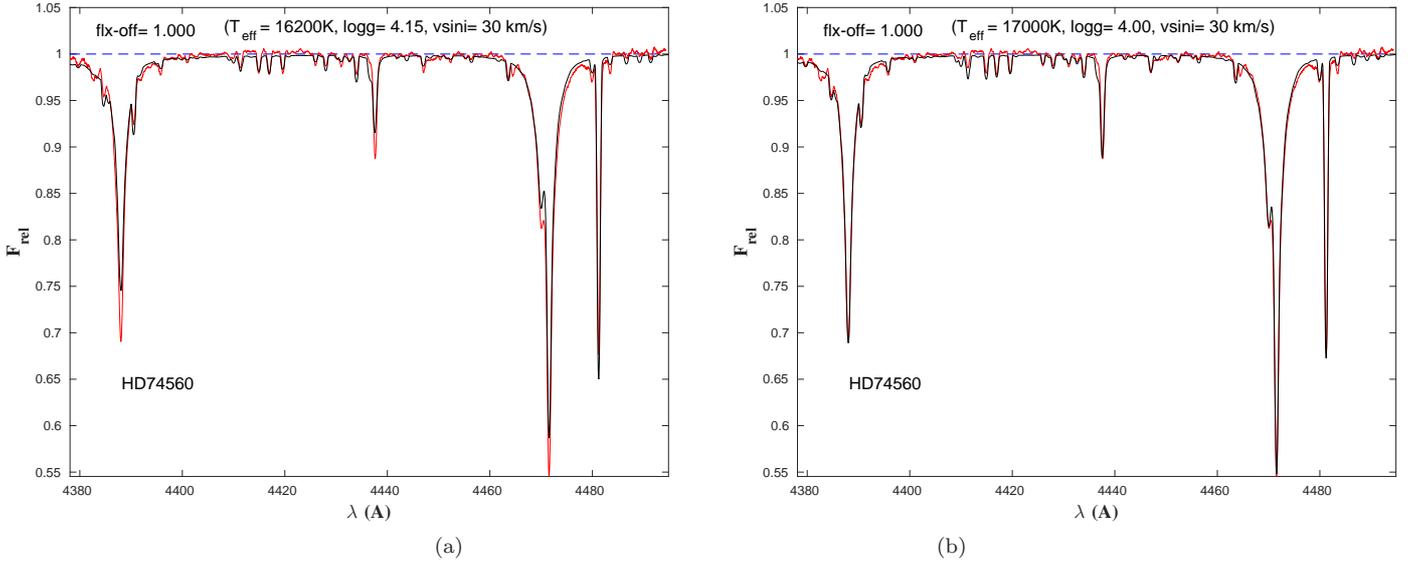

\centering
\subfloat[]{
\hspace*{-3.0cm}\includegraphics[scale= 0.50]{Fig4a.eps}
\label{Obs_vs_TLUSTy_HD74560_ord11_Hubrig_etal09}
}
\hspace{0.5cm}
\subfloat[]{
\includegraphics[scale= 0.50]{Fig4b.eps}
\hspace*{-3.0cm}
\label{Obs_vs_TLUSTy_HD74560_ord11_current}
}
\caption{The same as Figure~\ref{Obs_vs_TLUSTy_HD33328}, but for the B star HD~74560. In the left panel, the non-LTE synthetic spectra was interpolated to the stellar parameters of \citet{hub09}, while the right panel shows the stellar parameters at which the non-LTE synthetic spectra best match the observed spectra.}
\label{Obs_vs_TLUSTy_HD74560_2}
\end{figure*}

The observed MiMeS spectra of the normal B stars were compared with the non-LTE stellar atmosphere models of \citet{Lanz07} to test the adopted parameters of Table~\ref{parameters_table_Bstars}. Good matches were obtained for all stars, with the possible exception of HD~74560. Figure~\ref{Obs_vs_TLUSTy_HD74560_2} shows comparison of the observed spectra of this star with the non-LTE stellar atmosphere models of \citet{Lanz07} at both the stellar parameters of \citet{hub09} and newly estimated parameters (right panel). While new estimates agree within uncertainties with those of \citet{hub09}, the new ones provide a better fit. 

%
%
\subsection{Line Blending in Rapidly Rotating Stars}
\begin{figure}
\begin{center}
\includegraphics[scale=0.40]{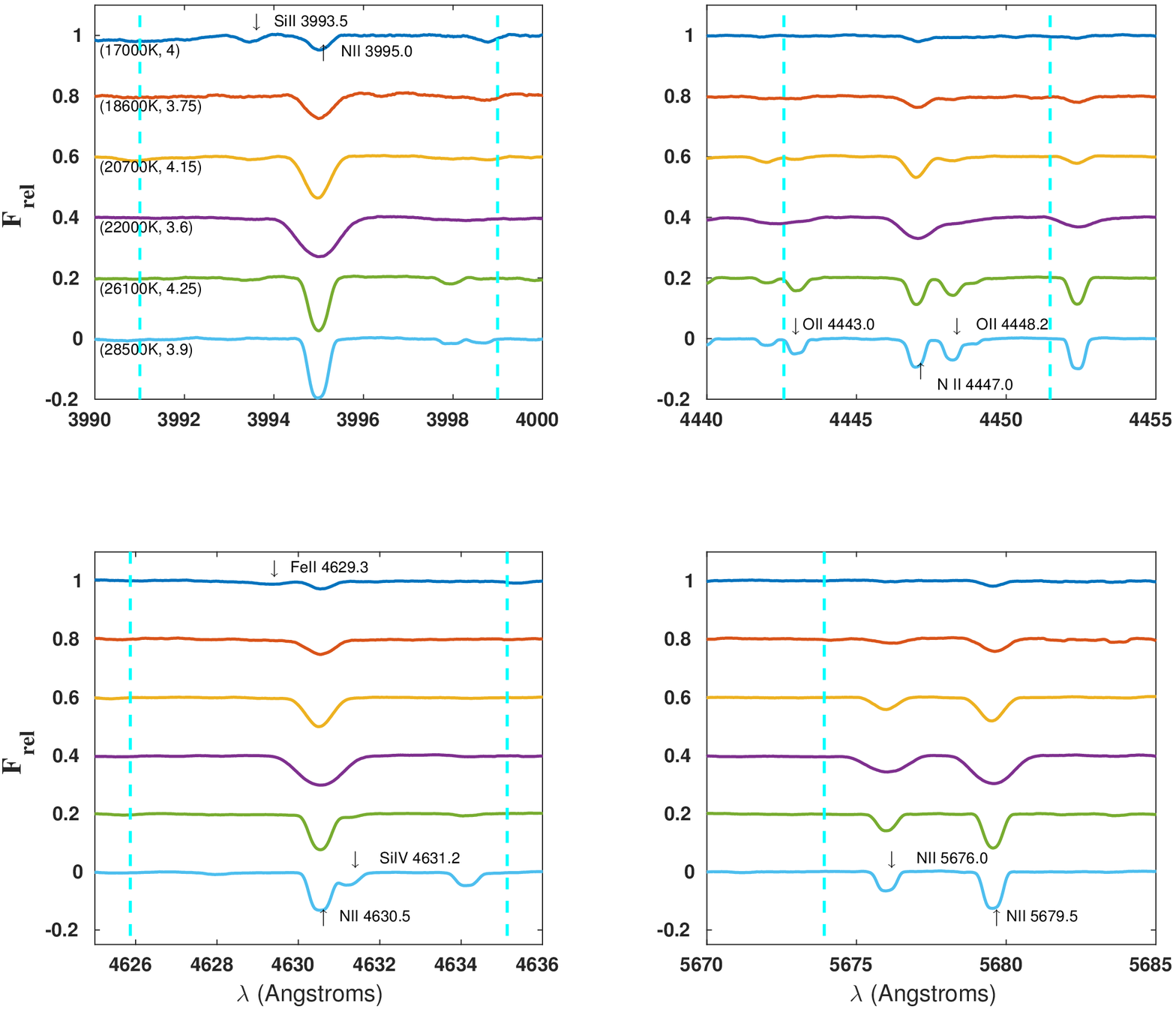}
\caption{Potential blends within $\pm 300\,\rm km\,s^{-1}$ (vertical dashed lines) of the indicated N\,{\sc ii} lines. The upper left panel identifies the stellar parameters of the comparison, normal B, low $v\sin i$ stellar spectra, and the coloured lines refers to these same models in all panels.}
\label{NII_blends}
\end{center}
\end{figure}

Measuring the equivalent widths of the N\,{\sc ii} lines in the Be star sample was potentially complicated by line blending in stars with high rotational velocities. In order to determine the extent of this problem, we first examined the spectra of normal B stars of similar $T_{\rm eff}$ in the B star sample with low projected rotational velocities within a specified wavelength range around each N\,{\sc ii} line of interest. This range was taken to be equal to the total width of the line assuming a projected rotational velocity of $300\;\rm km\,s^{-1}$.  This was done for the four strongest N\,{\sc ii} lines observed in six normal B-type stars with effective temperatures between 17,000~K and 28,500~K, as shown in Figure~\ref{NII_blends}. The top left panel shows the observed lines in the spectral region between 3990 and 4000~\AA\, which includes N\,{\sc ii} $\lambda\;3995$, generally the strongest N\,{\sc ii} line in the spectra of B stars. Only at the lowest effective temperature is $\lambda\;3995$ potentially blended with Si\,{\sc ii} $3993.5$. However, this Si\,{\sc ii} line rapidly decreases in strength for higher $T_{\rm eff}$, and for effective temperatures equal to $\approx$ 18,000~K or higher, the N\,{\sc ii} $\lambda\,3995$ line is not affected by line blending. Consequently, line blending will not affect the equivalent equivalent measurement for N\,{\sc ii} $\lambda\;3995$  for most objects in the Be star sample. 

The upper right panel of the figure shows the observed lines in a spectral region that includes N\,{\sc ii} $\lambda\;4447$ line for the same stars. This nitrogen line is blended with O\,{\sc ii} $\lambda\;4448.2$ at higher $T_{\rm eff}$. This blend does not significantly affect the measured equivalent for stars with $T_{\rm eff}\leq$ 22,000~K, but does become more serious for stars with higher effective temperatures as the strength of O\,{\sc ii} $\lambda\;4448.2$ increases. As this temperature range includes most of the observed Be stars sample, N\,{\sc ii} $\lambda\;4447$ is excluded in the current analysis in many cases. The lower left panel of Figure~\ref{NII_blends} shows lines that could possibly be blended with N\,{\sc ii} $\lambda\;4630.5$. This line is affected by blending only at the lowest and hottest effective temperatures. At $T_{\rm eff}=$ 17,000 K the line is blended with Fe\,{\sc ii} $\lambda\;4629.3$, while at $T_{\rm eff} \approx$ 28,000~K, the line is significantly blended with Si\,{\sc iv} $\lambda\;4631.2$. The lower right panel shows the possible line blends with N\,{\sc ii} $\lambda\;5679.5$ for the same stars. This figure shows that this line is blended with another N\,{\sc ii} line, $\lambda\,5676\;$\AA, in most cases. This blend was treated by fitting the two lines simultaneously and obtaining a single nitrogen estimate.

%
\begin{table*}
\centering{
\caption{Measured Equivalent Widths of the N\,{\sc ii} Lines in the MiMeS Be and B star samples. \label{measured_NII_EWs}}}
\begin{center}
{\small
\begin{tabular}{|c|r|r|r|r|r|r|r|r|r|r|r|}
\hline\hline
\multirow{2}{*}{HD} & \multicolumn{9}{|c|}{\hrulefill Measured N\,{\sc ii} Equivalent Widths (m\AA)\hrulefill}\\
\multicolumn{1}{|c}{}& \multicolumn{1}{|c}{$\lambda$\,3995	\AA}& \multicolumn{1}{|c}{$\lambda$\,4447 \AA}& \multicolumn{1}{|c}{$\lambda$\,4601.5 \AA}&  \multicolumn{1}{|c}{$\lambda$\,4607.2 \AA}& \multicolumn{1}{|c}{$\lambda$\,4613.9 \AA}& \multicolumn{1}{|c}{$\lambda$\,4621.4 \AA}& \multicolumn{1}{|c}{$\lambda$\,4630.5 \AA}& \multicolumn{1}{|c}{$\lambda$\,5676 \AA}& \multicolumn{1}{|c|}{$\lambda$\,5679.6 \AA}\\
\hline
\multicolumn{10}{|c|}{Be stars}\\
%
\multicolumn{1}{r}{11415$\;\;$}& \multicolumn{1}{|c}{ 24.0 $\pm$  7.0 }& \multicolumn{1}{|c}{  8.7 $\pm$  4.4 }& \multicolumn{1}{|c}{  9.4 $\pm$  4.4 }& \multicolumn{1}{|c}{  4.6 $\pm$  2.7 }& \multicolumn{1}{|c}{  3.1 $\pm$  2.0 }& \multicolumn{1}{|c}{ 15.9 $\pm$  4.9 }& \multicolumn{1}{|c}{ 16.5 $\pm$  4.9 }& \multicolumn{1}{|c|}{$-$}& \multicolumn{1}{|c}{  8.0 $\pm$  4.5 }\\
\multicolumn{1}{r}{20336$^*$} & \multicolumn{1}{|c}{ 78.5 $\pm$ 23.2 }%
 & \multicolumn{1}{|c}{ 45.1 $\pm$ 18.0 }& \multicolumn{1}{|c|}{$-$} & \multicolumn{1}{|c|}{$-$} & \multicolumn{1}{|c|}{$-$} & \multicolumn{1}{|c|}{$-$}%
& \multicolumn{1}{|c|}{$-$}%
 & \multicolumn{1}{|c}{ 34.2 $\pm$ 8.4 } & \multicolumn{1}{|c}{ 72.8 $\pm$ 13.6 }\\
\multicolumn{1}{r}{33328$^*$} & \multicolumn{1}{|c}{ 99.9 $\pm$ 24.7 }%
& \multicolumn{1}{|c|}{$-$}& \multicolumn{1}{|c|}{$-$} & \multicolumn{1}{|c|}{$-$} & \multicolumn{1}{|c|}{$-$} & \multicolumn{1}{|c|}{$-$}%
 & \multicolumn{1}{|c}{ 86.5 $\pm$ 14.6 }%
 & \multicolumn{1}{|c}{ 49.8 $\pm$ 6.4 } & \multicolumn{1}{|c}{ 103.4 $\pm$ 9.2 }\\
\multicolumn{1}{r}{45725$^*$} & \multicolumn{1}{|c}{ 72.6 $\pm$ 34.8 }%
& \multicolumn{1}{|c|}{$-$}& \multicolumn{1}{|c|}{$-$} & \multicolumn{1}{|c|}{$-$} & \multicolumn{1}{|c|}{$-$} & \multicolumn{1}{|c|}{$-$}%
& \multicolumn{1}{|c|}{$-$}%
 & \multicolumn{1}{|c}{ 12.2 $\pm$ 7.0 } & \multicolumn{1}{|c}{ 31.7 $\pm$ 13.8 }\\
\multicolumn{1}{r}{49567$\;\;$}& \multicolumn{1}{|c}{ 88.7 $\pm$ 13.3 }& \multicolumn{1}{|c}{ 43.7 $\pm$ 11.7 }& \multicolumn{1}{|c}{ 44.9 $\pm$ 11.1 }& \multicolumn{1}{|c}{ 38.1 $\pm$ 10.2 }& \multicolumn{1}{|c}{ 30.6 $\pm$ 11.1 }& \multicolumn{1}{|c}{ 42.8 $\pm$ 11.7 }& \multicolumn{1}{|c}{ 71.7 $\pm$ 13.0 }& \multicolumn{1}{|c}{ 28.7 $\pm$ 12.8 }& \multicolumn{1}{|c}{ 64.1 $\pm$ 13.8 }\\
\multicolumn{1}{r}{54309$^*$} & \multicolumn{1}{|c}{ 72.6 $\pm$ 15.1 }%
 & \multicolumn{1}{|c}{ 35.5 $\pm$ 9.5 }& \multicolumn{1}{|c|}{$-$} & \multicolumn{1}{|c|}{$-$} & \multicolumn{1}{|c|}{$-$} & \multicolumn{1}{|c|}{$-$}%
& \multicolumn{1}{|c|}{$-$}%
 & \multicolumn{1}{|c}{ 29.2 $\pm$ 5.9 } & \multicolumn{1}{|c}{ 69.3 $\pm$ 9.8 }\\
\multicolumn{1}{r}{56139$^*$} & \multicolumn{1}{|c}{ 40.9 $\pm$ 7.9 }%
& \multicolumn{1}{|c|}{$-$}& \multicolumn{1}{|c|}{$-$} & \multicolumn{1}{|c|}{$-$} & \multicolumn{1}{|c|}{$-$} & \multicolumn{1}{|c|}{$-$}%
& \multicolumn{1}{|c|}{$-$}%
 & \multicolumn{1}{|c}{ 12.1 $\pm$ 3.4 } & \multicolumn{1}{|c}{ 33.7 $\pm$ 7.1 }\\
\multicolumn{1}{r}{58050$^*$} & \multicolumn{1}{|c}{ 54.8 $\pm$ 15.6 }%
& \multicolumn{1}{|c|}{$-$}& \multicolumn{1}{|c|}{$-$} & \multicolumn{1}{|c|}{$-$} & \multicolumn{1}{|c|}{$-$} & \multicolumn{1}{|c|}{$-$}%
 & \multicolumn{1}{|c}{ 43.2 $\pm$ 6.8 }%
 & \multicolumn{1}{|c}{ 16.0 $\pm$ 6.4 } & \multicolumn{1}{|c}{ 41.6 $\pm$ 12.1 }\\
\multicolumn{1}{r}{58343$\;\;$}& \multicolumn{1}{|c}{ 27.0 $\pm$  6.0 }& \multicolumn{1}{|c}{ 14.0 $\pm$  6.7 }& \multicolumn{1}{|c}{ 13.9 $\pm$  8.2 }& \multicolumn{1}{|c}{  7.6 $\pm$  6.0 }& \multicolumn{1}{|c}{  5.7 $\pm$  5.2 }& \multicolumn{1}{|c}{ 16.5 $\pm$  7.5 }& \multicolumn{1}{|c}{ 18.0 $\pm$  4.9 }& \multicolumn{1}{|c|}{$-$}& \multicolumn{1}{|c}{ 10.3 $\pm$  5.6 }\\
\multicolumn{1}{r}{58978$^*$} & \multicolumn{1}{|c}{ 112.6 $\pm$ 27.2 }%
& \multicolumn{1}{|c|}{$-$}& \multicolumn{1}{|c|}{$-$} & \multicolumn{1}{|c|}{$-$} & \multicolumn{1}{|c|}{$-$} & \multicolumn{1}{|c|}{$-$}%
& \multicolumn{1}{|c|}{$-$}%
& \multicolumn{1}{|c|}{$-$}	& \multicolumn{1}{|c|}{$-$}\\
\multicolumn{1}{r}{65875$^*$} & \multicolumn{1}{|c}{ 57.6 $\pm$ 23.4 }%
& \multicolumn{1}{|c|}{$-$}& \multicolumn{1}{|c|}{$-$} & \multicolumn{1}{|c|}{$-$} & \multicolumn{1}{|c|}{$-$} & \multicolumn{1}{|c|}{$-$}%
& \multicolumn{1}{|c|}{$-$}%
 & \multicolumn{1}{|c}{ 19.3 $\pm$ 7.4 } & \multicolumn{1}{|c}{ 48.7 $\pm$ 13.4 }\\
\multicolumn{1}{r}{67698$\;\;$}& \multicolumn{1}{|c}{ 43.4 $\pm$ 11.0 }& \multicolumn{1}{|c}{ 26.4 $\pm$  7.4 }& \multicolumn{1}{|c|}{$-$}& \multicolumn{1}{|c|}{$-$}& \multicolumn{1}{|c|}{$-$}& \multicolumn{1}{|c|}{$-$}& \multicolumn{1}{|c}{ 40.0 $\pm$  8.6 }& \multicolumn{1}{|c|}{$-$}& \multicolumn{1}{|c}{ 20.7 $\pm$ 11.9 }\\
\multicolumn{1}{r}{120324$^*$} & \multicolumn{1}{|c}{ 54.1 $\pm$ 15.4 }%
& \multicolumn{1}{|c|}{$-$}& \multicolumn{1}{|c|}{$-$} & \multicolumn{1}{|c|}{$-$} & \multicolumn{1}{|c|}{$-$} & \multicolumn{1}{|c|}{$-$}%
 & \multicolumn{1}{|c}{ 57.6 $\pm$ 11.7 }%
 & \multicolumn{1}{|c}{ 22.3 $\pm$ 6.2 } & \multicolumn{1}{|c}{ 53.6 $\pm$ 10.7 }\\
\multicolumn{1}{r}{143275$^*$} & \multicolumn{1}{|c}{ 46.6 $\pm$ 20.8 }%
& \multicolumn{1}{|c|}{$-$}& \multicolumn{1}{|c|}{$-$} & \multicolumn{1}{|c|}{$-$} & \multicolumn{1}{|c|}{$-$} & \multicolumn{1}{|c|}{$-$}%
& \multicolumn{1}{|c|}{$-$}%
 & \multicolumn{1}{|c}{ 17.7 $\pm$ 3.7 } & \multicolumn{1}{|c}{ 58.0 $\pm$ 9.3 }\\
\multicolumn{1}{r}{174237$^*$} & \multicolumn{1}{|c}{ 47.7 $\pm$ 14.6 }%
& \multicolumn{1}{|c|}{$-$}& \multicolumn{1}{|c|}{$-$} & \multicolumn{1}{|c|}{$-$} & \multicolumn{1}{|c|}{$-$} & \multicolumn{1}{|c|}{$-$}%
& \multicolumn{1}{|c|}{$-$}%
 & \multicolumn{1}{|c}{ 10.3 $\pm$ 3.3 } & \multicolumn{1}{|c}{ 27.7 $\pm$ 6.9 }\\
\multicolumn{1}{r}{178175$^*$} & \multicolumn{1}{|c}{ 73.2 $\pm$ 12.9 }%
& \multicolumn{1}{|c|}{$-$}& \multicolumn{1}{|c|}{$-$} & \multicolumn{1}{|c|}{$-$} & \multicolumn{1}{|c|}{$-$} & \multicolumn{1}{|c|}{$-$}%
& \multicolumn{1}{|c|}{$-$}%
 & \multicolumn{1}{|c}{ 27.1 $\pm$ 9.5 } & \multicolumn{1}{|c}{ 62.2 $\pm$ 16.1 }\\
\multicolumn{1}{r}{187567$^*$} & \multicolumn{1}{|c}{ 167.5 $\pm$ 16.1 }%
& \multicolumn{1}{|c|}{$-$}& \multicolumn{1}{|c|}{$-$} & \multicolumn{1}{|c|}{$-$} & \multicolumn{1}{|c|}{$-$} & \multicolumn{1}{|c|}{$-$}%
& \multicolumn{1}{|c|}{$-$}%
 & \multicolumn{1}{|c}{ 74.1 $\pm$ 12.6 } & \multicolumn{1}{|c}{ 139.4 $\pm$ 16.0 }\\
\multicolumn{1}{r}{187811$^*$} & \multicolumn{1}{|c}{ 20.3 $\pm$ 17.2 }%
 & \multicolumn{1}{|c}{ 13.7 $\pm$ 8.7 }& \multicolumn{1}{|c|}{$-$} & \multicolumn{1}{|c|}{$-$} & \multicolumn{1}{|c|}{$-$} & \multicolumn{1}{|c|}{$-$}%
 & \multicolumn{1}{|c}{ 16.0 $\pm$ 8.3 }%
 & \multicolumn{1}{|c}{ 4.6 $\pm$ 4.3 } & \multicolumn{1}{|c}{ 14.2 $\pm$ 10.6 }\\
\multicolumn{1}{r}{189687$^*$} & \multicolumn{1}{|c}{ 46.6 $\pm$ 26.4 }%
& \multicolumn{1}{|c|}{$-$}& \multicolumn{1}{|c|}{$-$} & \multicolumn{1}{|c|}{$-$} & \multicolumn{1}{|c|}{$-$} & \multicolumn{1}{|c|}{$-$}%
 & \multicolumn{1}{|c}{ 36.6 $\pm$ 11.7 }%
 & \multicolumn{1}{|c}{ 8.7 $\pm$ 7.9 } & \multicolumn{1}{|c}{ 24.6 $\pm$ 16.8 }\\
\multicolumn{1}{r}{191610$^*$} & \multicolumn{1}{|c}{ 41.6 $\pm$ 44.3 }%
& \multicolumn{1}{|c|}{$-$}& \multicolumn{1}{|c|}{$-$} & \multicolumn{1}{|c|}{$-$} & \multicolumn{1}{|c|}{$-$} & \multicolumn{1}{|c|}{$-$}%
& \multicolumn{1}{|c|}{$-$}%
 & \multicolumn{1}{|c}{ 23.3 $\pm$ 8.3 } & \multicolumn{1}{|c}{ 54.0 $\pm$ 14.4 }\\
\multicolumn{1}{r}{192685$^*$} & \multicolumn{1}{|c}{ 40.5 $\pm$ 17.3 }%
& \multicolumn{1}{|c|}{$-$}& \multicolumn{1}{|c|}{$-$} & \multicolumn{1}{|c|}{$-$} & \multicolumn{1}{|c|}{$-$} & \multicolumn{1}{|c|}{$-$}%
& \multicolumn{1}{|c|}{$-$}%
 & \multicolumn{1}{|c}{ 8.8 $\pm$ 6.3 } & \multicolumn{1}{|c}{ 24.9 $\pm$ 13.6 }\\
\multicolumn{1}{r}{203467$^*$} & \multicolumn{1}{|c}{ 35.3 $\pm$ 18.8 }%
& \multicolumn{1}{|c|}{$-$}& \multicolumn{1}{|c|}{$-$} & \multicolumn{1}{|c|}{$-$} & \multicolumn{1}{|c|}{$-$} & \multicolumn{1}{|c|}{$-$}%
& \multicolumn{1}{|c|}{$-$}%
 & \multicolumn{1}{|c}{ 7.3 $\pm$ 5.7 } & \multicolumn{1}{|c}{ 20.9 $\pm$ 12.5 }\\
\multicolumn{1}{r}{212076$^*$} & \multicolumn{1}{|c}{ 65.4 $\pm$ 9.3 }%
& \multicolumn{1}{|c|}{$-$}& \multicolumn{1}{|c|}{$-$} & \multicolumn{1}{|c|}{$-$} & \multicolumn{1}{|c|}{$-$} & \multicolumn{1}{|c|}{$-$}%
& \multicolumn{1}{|c|}{$-$}%
 & \multicolumn{1}{|c}{ 23.0 $\pm$ 6.1 } & \multicolumn{1}{|c}{ 54.5 $\pm$ 10.8 }\\
\multicolumn{1}{r}{212571$^*$} & \multicolumn{1}{|c}{ 139.6 $\pm$ 27.9 }%
& \multicolumn{1}{|c|}{$-$}& \multicolumn{1}{|c|}{$-$} & \multicolumn{1}{|c|}{$-$} & \multicolumn{1}{|c|}{$-$} & \multicolumn{1}{|c|}{$-$}%
& \multicolumn{1}{|c|}{$-$}%
& \multicolumn{1}{|c|}{$-$}	& \multicolumn{1}{|c|}{$-$}\\
\multicolumn{1}{r}{217050$^*$} & \multicolumn{1}{|c}{ 95.7 $\pm$ 18.4 }%
& \multicolumn{1}{|c|}{$-$}& \multicolumn{1}{|c|}{$-$} & \multicolumn{1}{|c|}{$-$} & \multicolumn{1}{|c|}{$-$} & \multicolumn{1}{|c|}{$-$}%
& \multicolumn{1}{|c|}{$-$}%
 & \multicolumn{1}{|c}{ 22.1 $\pm$ 10.4 } & \multicolumn{1}{|c}{ 51.5 $\pm$ 18.1 }\\
%
%
\multicolumn{10}{|c|}{Normal B-type stars}\\
\multicolumn{1}{r}{   3360$\;\;$} & \multicolumn{1}{|c}{ 103.7 $\pm$ 4.7 } & \multicolumn{1}{|c}{ 62.6 $\pm$ 4.9 } & \multicolumn{1}{|c}{ 61.6 $\pm$ 4.3 } & \multicolumn{1}{|c}{ 55.6 $\pm$ 4.8 } & \multicolumn{1}{|c}{ 47.0 $\pm$ 3.9 } & \multicolumn{1}{|c}{ 51.5 $\pm$ 4.4 } & \multicolumn{1}{|c}{ 93.0 $\pm$ 4.3 } & \multicolumn{1}{|c}{ 51.7 $\pm$ 5.3 } & \multicolumn{1}{|c}{ 98.0 $\pm$ 6.1 }\\ 
\multicolumn{1}{r}{  30836$\;\;$} & \multicolumn{1}{|c}{ 106.3 $\pm$ 8.0 } & \multicolumn{1}{|c}{ 51.4 $\pm$ 5.4 } & \multicolumn{1}{|c}{ 48.7 $\pm$ 5.3 } & \multicolumn{1}{|c}{ 41.8 $\pm$ 5.3 } & \multicolumn{1}{|c}{ 35.5 $\pm$ 5.2 } & \multicolumn{1}{|c}{ 36.6 $\pm$ 6.1 } & \multicolumn{1}{|c}{ 88.3 $\pm$ 7.2 } & \multicolumn{1}{|c}{ 49.5 $\pm$ 7.9 } & \multicolumn{1}{|c}{ 117.2 $\pm$ 8.0 }\\ 
\multicolumn{1}{r}{  35468$\;\;$} & \multicolumn{1}{|c}{ 137.5 $\pm$ 8.8 } & \multicolumn{1}{|c}{ 84.6 $\pm$ 8.0 } & \multicolumn{1}{|c}{ 94.3 $\pm$ 8.8 } & \multicolumn{1}{|c}{ 78.2 $\pm$ 8.9 } & \multicolumn{1}{|c}{ 66.1 $\pm$ 8.0 } & \multicolumn{1}{|c}{ 71.4 $\pm$ 8.0 } & \multicolumn{1}{|c}{ 125.9 $\pm$ 8.9 } & \multicolumn{1}{|c}{ 87.2 $\pm$ 13.2 } & \multicolumn{1}{|c}{ 147.2 $\pm$ 13.3 }\\ 
\multicolumn{1}{r}{  35708$\;\;$} & \multicolumn{1}{|c}{ 89.7 $\pm$ 6.2 } & \multicolumn{1}{|c}{ 49.1 $\pm$ 6.2 } & \multicolumn{1}{|c}{ 53.1 $\pm$ 5.3 } & \multicolumn{1}{|c}{ 44.3 $\pm$ 5.3 } & \multicolumn{1}{|c}{ 37.3 $\pm$ 5.2 } & \multicolumn{1}{|c}{ 43.6 $\pm$ 6.2 } & \multicolumn{1}{|c}{ 77.5 $\pm$ 6.3 } & \multicolumn{1}{|c}{ 37.6 $\pm$ 6.9 } & \multicolumn{1}{|c}{ 74.6 $\pm$ 6.3 }\\ 
\multicolumn{1}{r}{  36629$\;\;$} & \multicolumn{1}{|c}{ 58.6 $\pm$ 3.5 } & \multicolumn{1}{|c}{ 30.7 $\pm$ 3.6 } & \multicolumn{1}{|c}{ 21.7 $\pm$ 2.6 } & \multicolumn{1}{|c}{ 20.2 $\pm$ 2.6 } & \multicolumn{1}{|c}{ 16.7 $\pm$ 2.5 } & \multicolumn{1}{|c}{ 19.3 $\pm$ 3.3 } & \multicolumn{1}{|c}{ 45.9 $\pm$ 3.5 } & \multicolumn{1}{|c}{ 17.4 $\pm$ 4.2 } & \multicolumn{1}{|c}{ 43.7 $\pm$ 4.3 }\\ 
\multicolumn{1}{r}{  36822$\;\;$} & \multicolumn{1}{|c}{ 94.6 $\pm$ 7.1 } & \multicolumn{1}{|c}{ 51.8 $\pm$ 6.2 }& \multicolumn{1}{|c}{$-$} & \multicolumn{1}{|c}{ 27.9 $\pm$ 6.0 }& \multicolumn{1}{|c}{$-$} & \multicolumn{1}{|c}{ 24.4 $\pm$ 6.0 }& \multicolumn{1}{|c}{$-$} & \multicolumn{1}{|c}{ 34.1 $\pm$ 8.4 } & \multicolumn{1}{|c}{ 77.4 $\pm$ 9.7 }\\ 
\multicolumn{1}{r}{  36959$\;\;$} & \multicolumn{1}{|c}{ 85.7 $\pm$ 3.4 } & \multicolumn{1}{|c}{ 49.8 $\pm$ 3.4 } & \multicolumn{1}{|c}{ 38.7 $\pm$ 2.6 } & \multicolumn{1}{|c}{ 35.9 $\pm$ 2.6 } & \multicolumn{1}{|c}{ 34.9 $\pm$ 2.5 } & \multicolumn{1}{|c}{ 32.9 $\pm$ 3.5 } & \multicolumn{1}{|c}{ 74.7 $\pm$ 4.3 } & \multicolumn{1}{|c}{ 42.1 $\pm$ 4.4 } & \multicolumn{1}{|c}{ 83.0 $\pm$ 4.3 }\\ 
\multicolumn{1}{r}{  36960$\;\;$} & \multicolumn{1}{|c}{ 73.2 $\pm$ 9.0 } & \multicolumn{1}{|c}{ 43.0 $\pm$ 5.4 }& \multicolumn{1}{|c}{$-$} & \multicolumn{1}{|c}{ 19.6 $\pm$ 5.4 }& \multicolumn{1}{|c}{$-$} & \multicolumn{1}{|c}{ 22.7 $\pm$ 5.5 }& \multicolumn{1}{|c}{$-$} & \multicolumn{1}{|c}{ 26.7 $\pm$ 8.2 } & \multicolumn{1}{|c}{ 63.0 $\pm$ 9.8 }\\ 
\multicolumn{1}{r}{  46328$\;\;$} & \multicolumn{1}{|c}{ 113.3 $\pm$ 3.3 } & \multicolumn{1}{|c}{ 66.8 $\pm$ 3.4 } & \multicolumn{1}{|c}{ 47.7 $\pm$ 2.6 } & \multicolumn{1}{|c}{ 48.7 $\pm$ 2.5 } & \multicolumn{1}{|c}{ 45.9 $\pm$ 2.6 } & \multicolumn{1}{|c}{ 42.1 $\pm$ 3.5 } & \multicolumn{1}{|c}{ 95.5 $\pm$ 3.8 } & \multicolumn{1}{|c}{ 63.5 $\pm$ 4.5 } & \multicolumn{1}{|c}{ 111.1 $\pm$ 4.1 }\\ 
\multicolumn{1}{r}{  48977$\;\;$} & \multicolumn{1}{|c}{ 31.8 $\pm$ 4.8 } & \multicolumn{1}{|c}{ 11.8 $\pm$ 3.9 } & \multicolumn{1}{|c}{ 21.4 $\pm$ 4.6 } & \multicolumn{1}{|c}{ 12.6 $\pm$ 4.4 } & \multicolumn{1}{|c}{ 12.3 $\pm$ 3.2 }& \multicolumn{1}{|c}{$-$} & \multicolumn{1}{|c}{ 30.6 $\pm$ 4.3 }& \multicolumn{1}{|c}{$-$} & \multicolumn{1}{|c}{ 26.9 $\pm$ 5.2 }\\ 
\multicolumn{1}{r}{  61068$\;\;$} & \multicolumn{1}{|c}{ 106.5 $\pm$ 4.7 } & \multicolumn{1}{|c}{ 66.2 $\pm$ 4.9 } & \multicolumn{1}{|c}{ 55.8 $\pm$ 2.5 } & \multicolumn{1}{|c}{ 59.9 $\pm$ 4.8 } & \multicolumn{1}{|c}{ 55.5 $\pm$ 3.9 } & \multicolumn{1}{|c}{ 52.0 $\pm$ 4.4 } & \multicolumn{1}{|c}{ 96.8 $\pm$ 4.2 } & \multicolumn{1}{|c}{ 68.5 $\pm$ 5.2 } & \multicolumn{1}{|c}{ 111.4 $\pm$ 6.0 }\\ 
\multicolumn{1}{r}{  66665$\;\;$} & \multicolumn{1}{|c}{ 96.0 $\pm$ 3.3 } & \multicolumn{1}{|c}{ 53.2 $\pm$ 3.4 } & \multicolumn{1}{|c}{ 38.1 $\pm$ 2.6 } & \multicolumn{1}{|c}{ 38.2 $\pm$ 2.6 } & \multicolumn{1}{|c}{ 34.4 $\pm$ 2.5 } & \multicolumn{1}{|c}{ 32.3 $\pm$ 3.5 } & \multicolumn{1}{|c}{ 77.4 $\pm$ 3.4 } & \multicolumn{1}{|c}{ 45.6 $\pm$ 4.4 } & \multicolumn{1}{|c}{ 87.3 $\pm$ 4.2 }\\ 
\multicolumn{1}{r}{  74560$\;\;$} & \multicolumn{1}{|c}{ 25.4 $\pm$ 3.5 } & \multicolumn{1}{|c}{ 12.6 $\pm$ 3.3 } & \multicolumn{1}{|c}{ 7.4 $\pm$ 2.4 } & \multicolumn{1}{|c}{ 6.5 $\pm$ 2.3 } & \multicolumn{1}{|c}{ 3.6 $\pm$ 2.1 } & \multicolumn{1}{|c}{ 12.2 $\pm$ 3.3 } & \multicolumn{1}{|c}{ 17.2 $\pm$ 3.8 } & \multicolumn{1}{|c}{ 1.4 $\pm$ 1.8 } & \multicolumn{1}{|c}{ 12.9 $\pm$ 4.0 }\\ 
\multicolumn{1}{r}{  74575$\;\;$} & \multicolumn{1}{|c}{ 158.2 $\pm$ 5.5 } & \multicolumn{1}{|c}{ 90.0 $\pm$ 5.2 } & \multicolumn{1}{|c}{ 87.8 $\pm$ 4.3 } & \multicolumn{1}{|c}{ 83.4 $\pm$ 5.2 } & \multicolumn{1}{|c}{ 69.5 $\pm$ 4.3 } & \multicolumn{1}{|c}{ 67.0 $\pm$ 4.4 } & \multicolumn{1}{|c}{ 149.9 $\pm$ 7.9 } & \multicolumn{1}{|c}{ 102.5 $\pm$ 8.0 } & \multicolumn{1}{|c}{ 190.3 $\pm$ 10.5 }\\ 
\multicolumn{1}{r}{  85953$\;\;$} & \multicolumn{1}{|c}{ 48.3 $\pm$ 5.3 } & \multicolumn{1}{|c}{ 30.8 $\pm$ 5.2 } & \multicolumn{1}{|c}{ 25.4 $\pm$ 5.2 } & \multicolumn{1}{|c}{ 19.8 $\pm$ 5.0 } & \multicolumn{1}{|c}{ 17.5 $\pm$ 5.0 } & \multicolumn{1}{|c}{ 21.2 $\pm$ 5.8 } & \multicolumn{1}{|c}{ 43.8 $\pm$ 7.0 } & \multicolumn{1}{|c}{ 14.4 $\pm$ 6.7 } & \multicolumn{1}{|c}{ 40.7 $\pm$ 7.7 }\\ 
\multicolumn{1}{r}{ 169467$\;\;$} & \multicolumn{1}{|c}{ 28.6 $\pm$ 4.8 } & \multicolumn{1}{|c}{ 10.5 $\pm$ 3.9 } & \multicolumn{1}{|c}{ 11.2 $\pm$ 4.3 } & \multicolumn{1}{|c}{ 9.0 $\pm$ 4.0 } & \multicolumn{1}{|c}{ 5.5 $\pm$ 2.9 }& \multicolumn{1}{|c}{$-$} & \multicolumn{1}{|c}{ 21.9 $\pm$ 5.0 }& \multicolumn{1}{|c}{$-$} & \multicolumn{1}{|c}{ 16.0 $\pm$ 4.8 }\\  
\hline
\end{tabular}}
\end{center}
\noindent{ {\small Notes: A dashed entry means that it was not possible to measure the indicated line. The quoted uncertainties are due to the continuum placement. Equivalent widths marked with asterisk are the equivalent widths of the best-fit, non-LTE synthetic profiles computed for the adopted $T_{\rm eff}$ and $\log g$ but allowing the nitrogen abundance and $v\sin i$ to vary.}}
\end{table*}

\subsection{Equivalent Width Measurements}

N\,{\sc ii} equivalent widths were measured in three ways. The first approach was by direct integration of the observed line over its profile. This approach is appropriate for low $v\sin i$ velocities where the lines are well defined and unblended, and this was the procedure followed for all of the normal, B-type sample stars. However, this method was applicable only to four Be stars: HD~11415, HD~ HD49567, HD~58343 and HD~67698.

\begin{figure}
\centering
\hspace{-0.5cm}
\includegraphics[scale= 0.45]{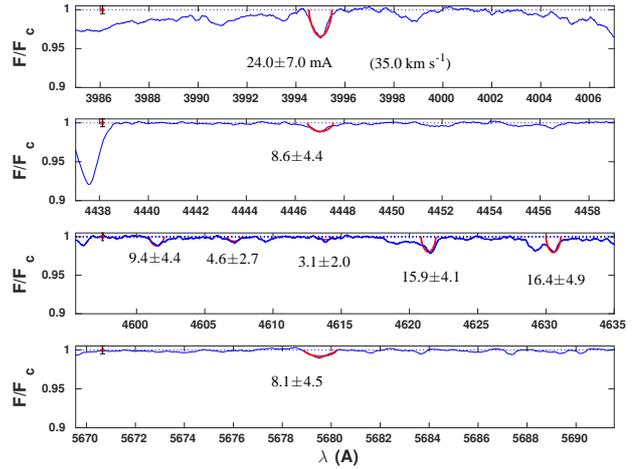}
\caption{Equivalent widths of N\,{\sc ii} lines (from top to bottom panel, N\,{\sc ii} $\lambda\,3995$, $\lambda\,4447$, $\lambda\,4621/4630$, and $\lambda\,5676/5679$), with estimated errors, in the Be star HD 11415 obtained by rotational profile fitting. The blue line is the observed spectrum, and the red lines are pure rotational profiles for $v\sin i=35\;\rm km\,s^{-1}$. The error bar on the left upper corner of each panel represents the adopted error in the continuum level.}
\label{NIIEWs_HD11415}
\end{figure}

For the more rapidly rotating stars in the Be sample, a second approach was to fit the observed lines with pure rotational profiles following \citet{gray05}. This better constrains the expected shape of the line profiles, especially in case of blends, and also provides an estimate of $v \sin i$ for each line considered, acting as a consistency check. Figure~\ref{NIIEWs_HD11415} is an example of measuring equivalent widths of the observed N\,{\sc ii} lines using rotational profile fitting for the Be star HD 11415 (B3III). 

\begin{figure}
\centering
\includegraphics[scale= 0.40]{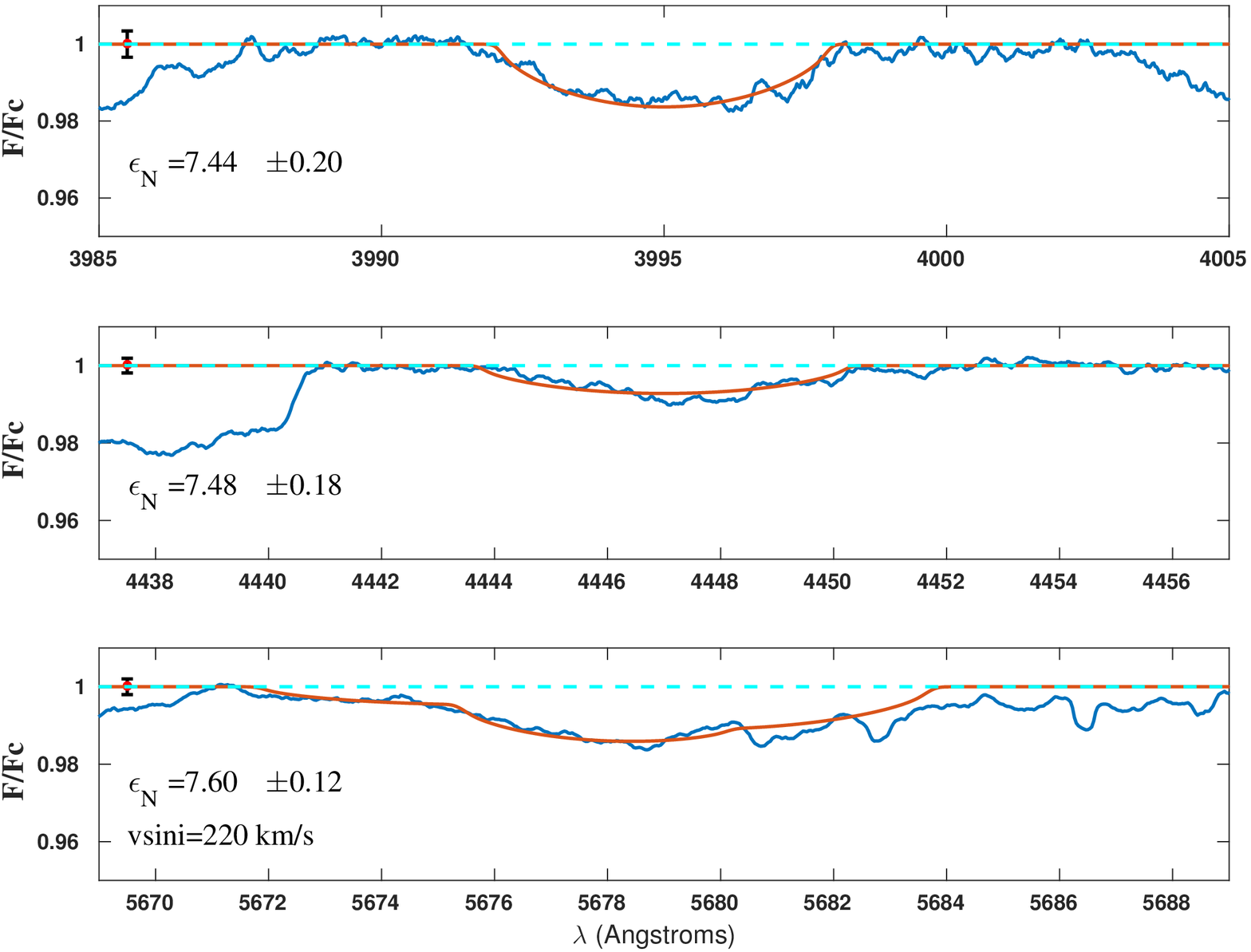}
\caption{The nitrogen abundance of the Be star HD 54309 measured from three N\,{\sc ii} lines, $\lambda\,3995$ (top panel), $\lambda\,4447$ (middle panel), and $\lambda\,5676/5679$ (bottom panel). The blue lines represent the observed spectrum, the blue dotted lines represent the continuum, and the solid red lines are rotationally-broadened, non-LTE line profiles from \citet{AS15} for $v\sin i = 220 \;\rm km\,s^{-1}$. The error bar on the left upper corner of each panel represents the adopted error in the continuum determination.}
\label{NIIEWs_HD54309}
\end{figure}

The third approach was by direct synthetic line profile fitting in which the equivalent widths and abundances were obtained by direct comparison with the rotationally-broadened, non-LTE, synthetic lines computed by \cite{AS15}. The equivalent width and the abundance of the synthetic line profile that best matched the observed one was adopted for each star. This approach was used for stars with high projected rotational velocities where the line profiles are shallow and the shapes of the profiles are less well defined. To select the best synthetic profiles, a figure-of-merit for each model line was computed as the average percentage flux difference, defined as
\begin{equation}
{\cal F}\equiv\frac{1}{N}\sum_i \frac{|F_i^{\rm obs}-F_i^{\rm mod}|}{F_i^{\rm obs}}\,,
\label{eq:fom}
\end{equation}
where $F_{i}^{\rm obs}$ is the observed flux at wavelength $\lambda_i$ in the normalized MiMeS spectrum and $F_{i}^{\rm mod}$ is the synthetic model flux at wavelength $\lambda_i$. The sum is over the $N$ observed wavelengths points in the line. Figure~\ref{NIIEWs_HD54309} gives an example of this method for the Be star HD~54309. The abundances of the synthetic line profile that best match the observed ones are written in the lower left corner of each panel, as is the adopted $v\sin i$. This third procedure was followed for all the Be stars in the sample, with the exception of the four stars mentioned above.

The measured equivalent widths of all N\,{\sc ii} lines in the two samples are listed in Table~\ref{measured_NII_EWs}. The assigned uncertainties represent the change in the equivalent widths caused by changing the continuum level by $\pm 0.5\%$. As discussed above, N\,{\sc ii} $\lambda\,3995$ is least affected by blending in most of the sample.


\section{Nitrogen Abundances in the Be Star Sample}
\label{Nabund_res}

In this section, the nitrogen abundances in the Be star sample are obtained. Section~\ref{Nabund_res_obs} presents abundances based solely on the measured equivalent widths or profiles fits without making any corrections for gravitational darkening or disk emission. The effect of gravitational darkening with be discussed in Section~\ref{gd_sec}, and the effect of circumstellar disk emission, in Section~\ref{Disk_paras}. 

\subsection{Uncorrected Nitrogen Abundances}
\label{Nabund_res_obs}

Initial estimates of the nitrogen abundances of the Be stars, without consideration of gravitational darkening and circumstellar disk emission, were obtained from the N\,{\sc ii} curves-of-growths computed by \citet{AS15} within the hybrid, non-LTE framework (LTE, line-blanketed model atmospheres / non-LTE line transfer for the trace species nitrogen) using a Monte-Multi technique \citep{sig96}. The calculation of \citet{AS15} accounts for uncertainties in the basic N\,{\sc ii} atomic data by constructing 100 atomic models, all randomly realized with the atomic data varied within their assigned uncertainties. A non-LTE solution is converged for each atomic model and a set of curves-of-growth are provided for each N\,{\sc ii} line; therefore, each line is represented by one hundred curves-of-growth, each one corresponding to one of the atomic models. For a given set of stellar parameters $T_{\rm eff}$ and $\log g$, one hundred estimates of the nitrogen abundance are computed using these curves-of-growth, and the dispersion in these estimates represents the uncertainty in the abundance due to the adopted set of N\,{\sc ii} atomic data. This process is referred to as a Monte-Multi abundance analysis \citep{sig96}.

However, the fundamental stellar parameters $T_{\rm eff}$ and $\log g$ and the measured N\,{\sc ii} equivalent widths are themselves uncertain with the respective uncertainties given in Tables~\ref{parameters_table} and \ref{measured_NII_EWs}. To account for these uncertainties, a set of one hundred random sets of stellar parameters and equivalent widths were generated within their respective uncertainties. Each set of $(T_{\rm eff}, \log g, \rm EW)$ was then used as the basis of Monte-Multi abundance analysis as described above. The final nitrogen abundance for each available line is the average of all estimated abundances for all random sets, and the dispersion of abundances is its uncertainty due to errors in the atomic data, the stellar parameters, and the measured equivalent widths. This error estimate assumes these uncertainties are uncorrelated.

A further caveat in any abundance analysis is the atmospheric microturbulence. Ideally, a mix of strong and weak lines allows one to uniquely fix the microturbulent velocity ($\xi_t$ hereafter) by forcing strong and weak lines to yield the same abundance. In the present sample of Be stars, however, this classic procedure is only rarely possible because only stronger lines are measured or, in some cases, only the single N\,{\sc ii} $\lambda\,3995$ line is available. In these cases, the microturbulent velocity must be fixed in advance and represents a further source of uncertainty. In the current analysis, the microturbulent velocity was set to one of two fixed values, 2.0 or 5.0~$\rm km\,s^{-1}$, where a microturbulent velocity of 5.0 $\rm km\,s^{-1}$ represents an average value over main-sequence B-type stars \citep{gie92}. 

\begin{figure}
\begin{center}
\vspace{0.5cm}
\hspace{-0.5cm}
\includegraphics[scale=0.40]{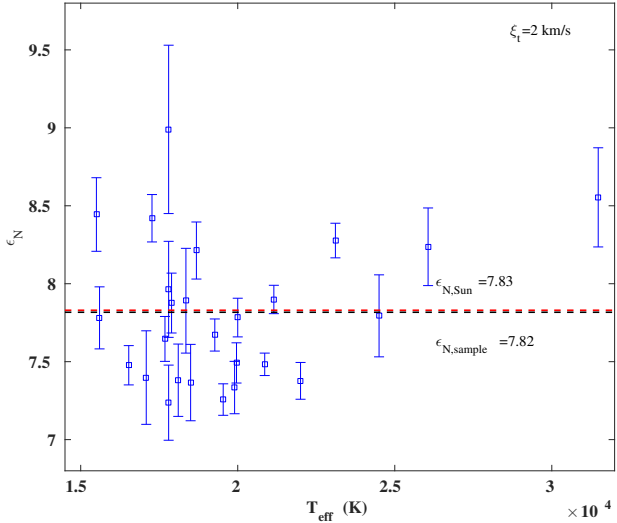}
\caption{Nitrogen abundances, uncorrected for gravitational darkening and disk emission, for the MiMeS Be star sample as a function of stellar $T_{\rm eff}$. The microturbulence was assumed to be $\xi_t=\,2.0\,\rm km\,s^{-1}$. The red dashed line marks the solar nitrogen abundance, while the dashed black line, nearly co-incident with the solar abundance, marks the the mean value of the Be star sample nitrogen abundance.}
\label{N_abund_Teff}
\end{center}
\end{figure}

Figure~\ref{N_abund_Teff} shows the estimated nitrogen abundances for a microturbulence of $\xi_t = 2.0\;\rm \rm km\,s^{-1}$ as a function of adopted stellar $T_{\rm eff}$ for the Be sample. The average nitrogen abundance for the sample is $\epsilon_{\rm N}=7.82$, which coincides with the solar value, $\epsilon^{\odot}_{\rm N}=7.83 \pm 0.11$ \citep{lod03}. However, the dispersion in the nitrogen abundance is large, $\sigma_{\rm N}=0.48$. On the individual level, about a third of the sample stars (7 out of 26 stars) have nitrogen abundances higher than the solar value, with $\epsilon_{\rm N}\sim 8.4$ or about a factor of $3.7$ larger. Slightly less than half the sample (10 out of 26 stars) have a sub-solar nitrogen abundance of $\epsilon_{\rm N}\sim 7.4$ or about a factor of 2.7 lower. The abundances of the remaining stars (9 stars) have the solar nitrogen abundance within their estimated errors. There is no obvious correlation between $T_{\rm eff}$ and the measured nitrogen abundance. The individual abundance for each star is given in Table~\ref{N_abundances_obser}. Adopting the higher value of $\xi = 5.0\;\rm \rm km\,s^{-1}$ reduces the average nitrogen abundance by only 0.1 dex and does not alter the above conclusions. 

\begin{table}
\centering{
\caption{Measured Nitrogen Abundances in the MiMeS Be Star Sample. 
\label{N_abundances_obser}}}
\begin{center}
{\small
\begin{tabular}{cccccc}
\hline\hline
HD &	$T_{\rm eff}\,(\rm K)$ & $\log g$	 & \multicolumn{2}{c}{$\epsilon_N\,\pm\,\Delta\epsilon_N$ (dex)}\\
	 &	& & \multicolumn{1}{c}{{\small $\xi_t=\,2.0\, \rm  km\,s^{-1}$}} & \multicolumn{1}{c}{{\small $\xi_t=\,5.0\,\rm  km\,s^{-1}$}}\\ 
\hline
\multicolumn{5}{c}{Be stars}\\
\multicolumn{1}{l}{  11415 }& 15600.0 & 3.5&7.78 $\pm$ 0.20 & 7.75 $\pm$ 0.15 \\
\multicolumn{1}{l}{  20336 }& 18684.0 & 3.9&8.30 $\pm$ 0.16 & 8.22 $\pm$ 0.15 \\
\multicolumn{1}{l}{  33328 }& 21150.0 & 3.6&7.89 $\pm$ 0.09 & 7.75 $\pm$ 0.09 \\
\multicolumn{1}{l}{  45725 }& 17800.0 & 3.9&7.92 $\pm$ 0.36 & 7.92 $\pm$ 0.28 \\
\multicolumn{1}{l}{  49567 }& 17270.0 & 3.3&8.42 $\pm$ 0.15 & 8.27 $\pm$ 0.14 \\
\multicolumn{1}{l}{  54309 }& 20859.0 & 3.4&7.49 $\pm$ 0.10 & 7.44 $\pm$ 0.08 \\
\multicolumn{1}{l}{  56139 }& 19537.0 & 3.4&7.21 $\pm$ 0.08 & 7.18 $\pm$ 0.08 \\
\multicolumn{1}{l}{  58050 }& 19961.0 & 3.8&7.48 $\pm$ 0.12 & 7.44 $\pm$ 0.12 \\
\multicolumn{1}{l}{  58343 }& 16530.0 & 3.6&7.57 $\pm$ 0.16 & 7.53 $\pm$ 0.16 \\
\multicolumn{1}{l}{  58978 }& 24445.0 & 3.4&7.84 $\pm$ 0.28 & 7.73 $\pm$ 0.24 \\
\multicolumn{1}{l}{  65875 }& 19900.0 & 3.3&7.35 $\pm$ 0.15 & 7.26 $\pm$ 0.14 \\
\multicolumn{1}{l}{  67698 }& 15500.0 & 3.7&8.38 $\pm$ 0.23 & 8.28 $\pm$ 0.21 \\
\multicolumn{1}{l}{ 120324 }& 20000.0 & 4.0&7.74 $\pm$ 0.11 & 7.68 $\pm$ 0.10 \\
\multicolumn{1}{l}{ 143275 }& 31478.0 & 3.5&8.47 $\pm$ 0.21 & 8.48 $\pm$ 0.25 \\
\multicolumn{1}{l}{ 174237 }& 17683.0 & 3.6&7.66 $\pm$ 0.12 & 7.62 $\pm$ 0.14 \\
\multicolumn{1}{l}{ 178175 }& 22000.0 & 3.5&7.40 $\pm$ 0.11 & 7.31 $\pm$ 0.10 \\
\multicolumn{1}{l}{ 187567 }& 23110.0 & 3.7&8.22 $\pm$ 0.10 & 8.07 $\pm$ 0.08 \\
\multicolumn{1}{l}{ 187811 }& 17800.0 & 3.8&7.26 $\pm$ 0.23 & 7.19 $\pm$ 0.19 \\
\multicolumn{1}{l}{ 189687 }& 18106.0 & 3.5&7.39 $\pm$ 0.23 & 7.33 $\pm$ 0.22 \\
\multicolumn{1}{l}{ 191610 }& 18350.0 & 3.7&7.76 $\pm$ 0.40 & 7.65 $\pm$ 0.32 \\
\multicolumn{1}{l}{ 192685 }& 18500.0 & 3.7&7.36 $\pm$ 0.24 & 7.35 $\pm$ 0.22 \\
\multicolumn{1}{l}{ 203467 }& 17087.0 & 3.4&7.35 $\pm$ 0.28 & 7.34 $\pm$ 0.27 \\
\multicolumn{1}{l}{ 205637 }& 17800.0 & 3.3& 9.29  $\pm$ 0.53 & 8.96  $\pm$ 0.49 \\
\multicolumn{1}{l}{ 212076 }& 19270.0 & 3.5&7.67 $\pm$ 0.09 & 7.57 $\pm$ 0.09 \\
\multicolumn{1}{l}{ 212571 }& 26061.0 & 3.7&8.31 $\pm$ 0.28 & 8.12 $\pm$ 0.25 \\
\multicolumn{1}{l}{ 217050 }& 17893.0 & 3.3&7.93 $\pm$ 0.18 & 7.84 $\pm$ 0.20 \\
\hline
\end{tabular}}
\end{center}
\end{table}

\begin{figure}
\centering
\vspace{0.5cm}
\hspace{-0.5cm}
\includegraphics[scale= 0.50]{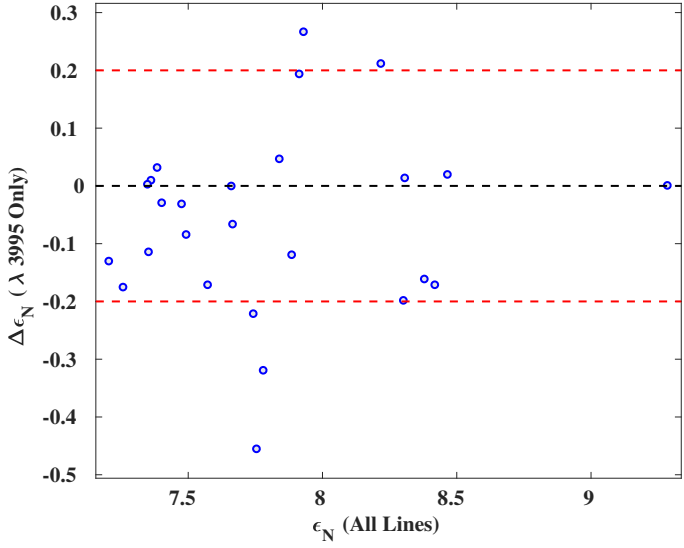}
\caption{Change in measured nitrogen abundance of the stars in the Be sample between an analysis using only N\,{\sc ii} $\lambda\,3995$ and an analysis including all available N\,{\sc ii} lines.}
\label{Nabund_3995}
\end{figure}

Because the N\,{\sc ii} $\lambda\,3995$ is less likely to be affected by line blends and is measured for all objects, it is the single most reliable line for nitrogen abundance determinations. For this reason, the abundance analysis was re-performed using only this line. Figure~\ref{Nabund_3995} shows the the nitrogen abundances based solely on $\lambda\,3995$ differ only by $\pm 0.2$ dex for most of the sample stars (22 out of 26 stars). The four Be stars that differ by more than 0.2~dex are HD\,11415, HD\,191610, HD\,205637 and HD\,217050.

\begin{table}
\centering{
\caption{Measured Nitrogen Abundances in the MiMeS Be Star Sample Based Only on N\,{\sc ii} $\lambda\,3995$.
\label{N_abundances_obser_NII3995}}
\begin{center}
{\small
\begin{tabular}{cccccc}
\hline\hline
HD &	$T_{\rm eff}\,(\rm K)$ & $\log g$	 & \multicolumn{2}{c}{$\epsilon_N\,\pm\,\Delta\epsilon_N$ (dex)}\\
	 &	& & \multicolumn{1}{c}{{\small $\xi_t=\,2.0\, \rm  km\,s^{-1}$}} & \multicolumn{1}{c}{{\small $\xi_t=\,5.0\,\rm  km\,s^{-1}$}}\\	
\hline
\multicolumn{1}{l}{  11415} &15600  & 3.50 &  7.46 $\pm$ 0.24 & 7.43$\pm$0.22\\
\multicolumn{1}{l}{  20336} &18684  & 3.90 &  8.10 $\pm$ 0.40 & 7.98$\pm$0.40\\
\multicolumn{1}{l}{  33328} &21150  & 3.60 &  7.77 $\pm$ 0.33 & 7.65$\pm$0.25\\
\multicolumn{1}{l}{  45725} &17800  & 3.90 &  8.11 $\pm$ 0.76 & 8.04$\pm$0.60\\
\multicolumn{1}{l}{  49567} &17270  & 3.30 &  8.25 $\pm$ 0.23 & 8.07$\pm$0.19\\
\multicolumn{1}{l}{  54309} &20859  & 3.40 &  7.41 $\pm$ 0.22 & 7.33$\pm$0.17\\
\multicolumn{1}{l}{  56139} &19537  & 3.40 &  7.08 $\pm$ 0.15 & 7.01$\pm$0.13 \\
\multicolumn{1}{l}{  58050} &19961  & 3.80 &  7.44 $\pm$ 0.31 & 7.38$\pm$0.24\\
\multicolumn{1}{l}{ 58343} &16530  & 3.60 &  7.40 $\pm$ 0.17 & 7.34$\pm$0.16\\
\multicolumn{1}{l}{  58978} &24500  & 3.40 &  7.89 $\pm$ 0.29 & 7.72$\pm$0.21\\
\multicolumn{1}{l}{ 65875} &19900  & 3.30 &  7.24 $\pm$ 0.38 & 7.15$\pm$0.31 \\
\multicolumn{1}{l}{ 67698} &15500  & 3.70 &  8.22 $\pm$ 0.30 & 8.05$\pm$0.28\\
\multicolumn{1}{l}{  120324} &20000  & 4.00 & 7.52 $\pm$ 0.26 & 7.44$\pm$0.29\\
\multicolumn{1}{l}{  143275} &31478  & 3.50 &  8.49 $\pm$ 0.66 & 8.64$\pm$0.46\\
\multicolumn{1}{l}{  174237} &17683  & 3.65 &  7.66 $\pm$ 0.31 & 7.57$\pm$0.28\\
\multicolumn{1}{l}{  178175} &22000  & 3.50 &  7.37 $\pm$ 0.14 & 7.27$\pm$0.15\\
\multicolumn{1}{l}{  187567} &23110  & 3.70 &  8.43 $\pm$ 0.17 & 8.20$\pm$0.14\\
\multicolumn{1}{l}{  187811} &17800  & 3.80 &  7.08 $\pm$ 0.50 & 7.01$\pm$0.45\\
\multicolumn{1}{l}{  189687} &18106  & 3.50 &  7.42 $\pm$ 0.52 & 7.31$\pm$0.47 \\
\multicolumn{1}{l}{  191610} &18350  & 3.70 &  7.30 $\pm$ 0.94 & 7.18$\pm$0.92 \\ 
\multicolumn{1}{l}{  192685} &18500  & 3.70 &  7.37 $\pm$ 0.36 & 7.27$\pm$0.37\\
\multicolumn{1}{l}{  203467} &17087  & 3.40 &  7.35 $\pm$ 0.49 & 7.23$\pm$0.44 \\
\multicolumn{1}{l}{ 205637} &17800  & 3.50 &  9.29 $\pm$ 0.53 & 8.96$\pm$0.49 \\
\multicolumn{1}{l}{  212076} &19270  & 3.50 &  7.60 $\pm$ 0.14 & 7.45$\pm$0.12 \\
\multicolumn{1}{l}{  212571} &26061  & 3.70 &  8.32 $\pm$ 0.32 & 8.11$\pm$0.27\\
\multicolumn{1}{l}{  217050} &17893  & 3.30 &  8.20 $\pm$ 0.28 & 7.99$\pm$0.22\\ 
\hline
\end{tabular}}
\end{center}}
\end{table}

\begin{figure}
\centering
\vspace{0.5cm}
\includegraphics[scale= 0.45]{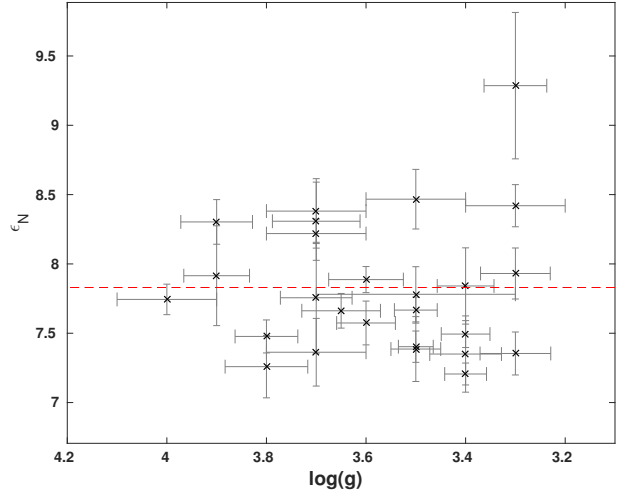}
\caption{Scatter-plot of nitrogen abundance versus stellar gravity $\log g$ for the MiMes Be star sample. The dashed red line is the solar nitrogen abundance, $\epsilon_{\rm N}^{\odot}=\,7.83$~dex.}
\label{Nabund_vs_rlg}
\end{figure}

\begin{figure}
\centering
\hspace{-0.5cm}
\includegraphics[scale= 0.40]{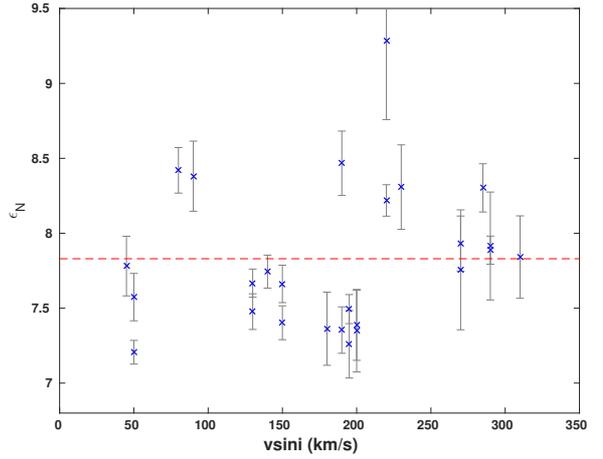}
\caption{Nitrogen abundance versus $v\sin i$ in the MiMeS Be star sample. The dashed red line represents the solar nitrogen abundance.}
\label{Nabund_vsini}
\end{figure}

Figures~\ref{Nabund_vs_rlg} and \ref{Nabund_vsini} show the measured nitrogen abundances of the Be star sample as a function stellar gravity and projected rotational velocity, respectively. Theoretical models of the evolution of rotating massive stars on the main-sequence predict the increase of the nitrogen abundance with age and velocity, and stellar age is correlated with surface gravity. As Figure~\ref{Nabund_vs_rlg} shows, there is no significant correlation, or anti-correlation, between the measured nitrogen abundances of the Be stars and their gravities. 

Figure~\ref{Nabund_vsini} does not show any clear trend of nitrogen abundance with $v\sin i$. It does show the increase in the uncertainty in the nitrogen abundance with increased $v\sin i$, as expected from the discussion above. There is a weak trend in that the Be stars with projected rotational velocities equal to or larger than 200 $\rm km\,s^{-1}$ are more likely to have nitrogen abundances larger than the solar value, while stars with smaller projected rotational velocities are likely to have nitrogen abundances equal to or lower than the solar value.

However, before the implications of the nitrogen abundances are discussed, the potential effects of circumstellar disk emission and gravitational darkening need to be carefully addressed. This is done in the next two sections. 

%

\subsection{Gravitational Darkening}
\label{gd_sec}

The surface gravity and effective temperature of a rotating star are dependent on latitude and the star's rotational velocity, a phenomena usually called gravitational darkening because the local effective temperature decreases towards the stellar equator \citep{von24}. The stellar surface also distorts with rotation, and the equatorial radius is predicted (in the Roche approximation) to be 50\% larger than the polar radius at critical rotation when the effective gravitational acceleration at the stellar equator vanishes. The combination of these effects can change photospheric spectrum of the star, and the observed spectrum will depend on the viewing inclination. The classic treatment of gravitational darkening by \citet{von24} predicts that the local effective temperature will depend on the local effective gravity, $g_{\rm eff}$, via $T_{\rm eff} \propto g_{\rm eff}^\beta$ where the exponent $\beta$ is $1/4$ \citep{Collins65}. However, interferometric studies of nearby, rapidly rotating stars suggest a lower value of $\beta$ gives a better fit to the intensity distribution over the resolved stellar surfaces \citep{bel12}. \citet{lar11} have reformulated the treatment of gravitational darkening in a way that naturally predicts variable $\beta$ values in better agreement with observations. The \citet{lar11} treatment generally predicts smaller gravitational darkening effects as the reduction in the stellar $T_{\rm eff}$ near the equator is less.

\begin{figure}
\centering
\includegraphics[scale= 1.]{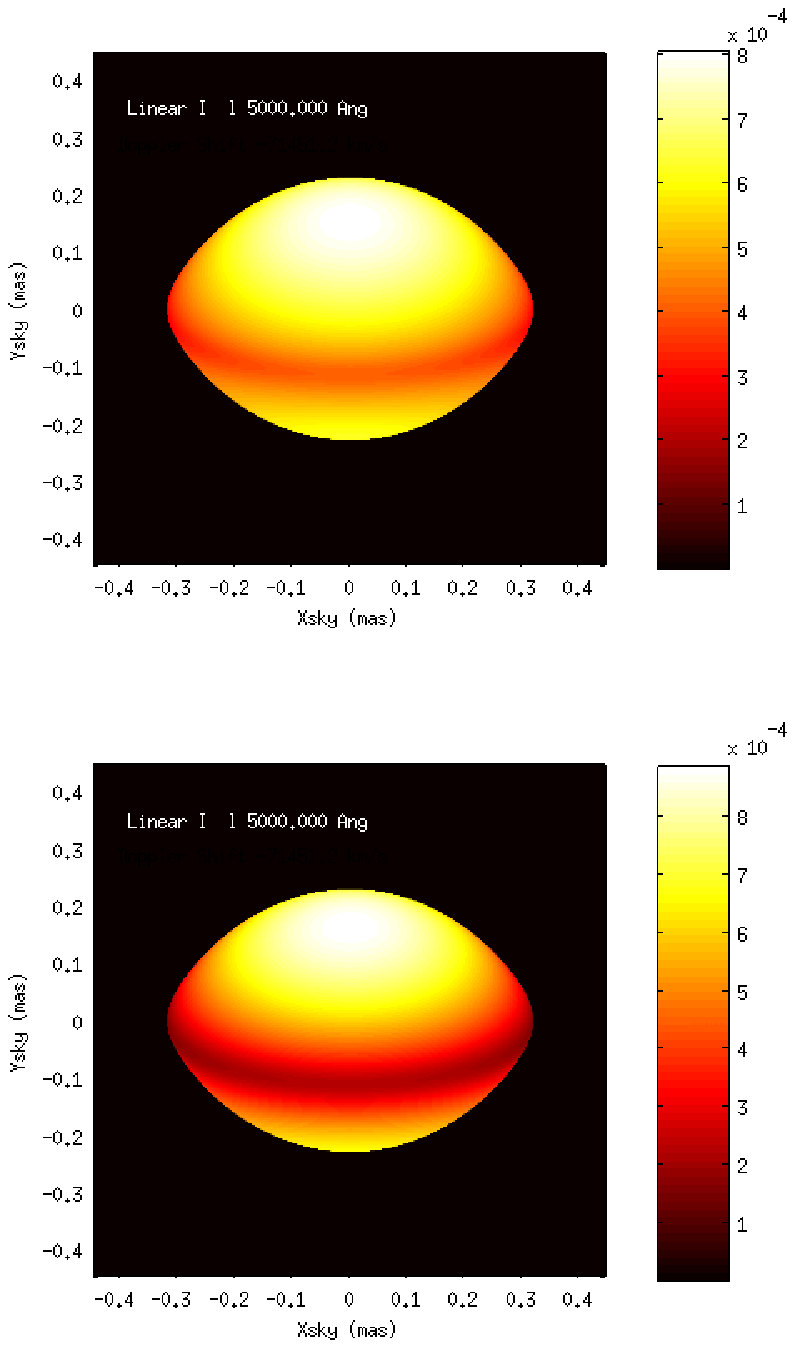}
\caption{The continuum intensity at $\lambda\,5000\;$\AA\ {\bf (in units of $\rm erg\,s^{-1}\,cm^{-2}\,Hz^{-1}\,Ster^{-1}$}) for a B3V star rotating at 95\% of its critical velocity using von Zeipel's relation for gravitational darkening (lower panel) and the treatment of \citet{lar11} (upper panel). The stars are assumed to be viewed at a distance of 100~pc, and the $x$ and $y$ axis are in milliarcseconds on the sky. The colourbars at the right give the linear intensity scales used for the figures.}
\label{GD_Ic}
\end{figure}

The effect of gravitational darkening on N\,{\sc ii} $\lambda\,3995$ was investigated using both the classical treatment of \citet{Collins65} and the newer formulation of \citet{lar11}.  To illustrate the differences between these two treatments, Figure~\ref{GD_Ic} compares two theoretical continuum images of a main sequence B3V star rotating at 95\% of its critical velocity. The star has an equatorial radius of $R_{eq}=6.87\,R_{\odot}$, {an equatorial-to-polar radius ratio of} $R_{eq}/R_p=1.43$, a critical velocity of $v_{\rm crit}=450\,\rm km\,s^{-1}$, and is viewed at an inclination of $i=70^{o}$. The continuum emission at $5000\,$\AA\ was computed assuming that each surface element radiates like a blackbody at its local $T_{\rm eff}$. The treatment of the geometry of the stellar surface as seen by an external observer at infinity is taken from \citet{Meghan13}. The darkening at the equator is weaker using the \citet{lar11} treatment: the local temperature varies between $\approx$ 11,000 K at the equator and $\approx$ 21,000 K at the pole using the classical formulation, and between $\approx$ 13,500 K at the equator and $\approx$ 20,000 K at the pole using the \citet{lar11} formalism.

\begin{figure}
\centering
\subfloat[]{													
\hspace*{-0.70cm}
\includegraphics[scale= 0.25]{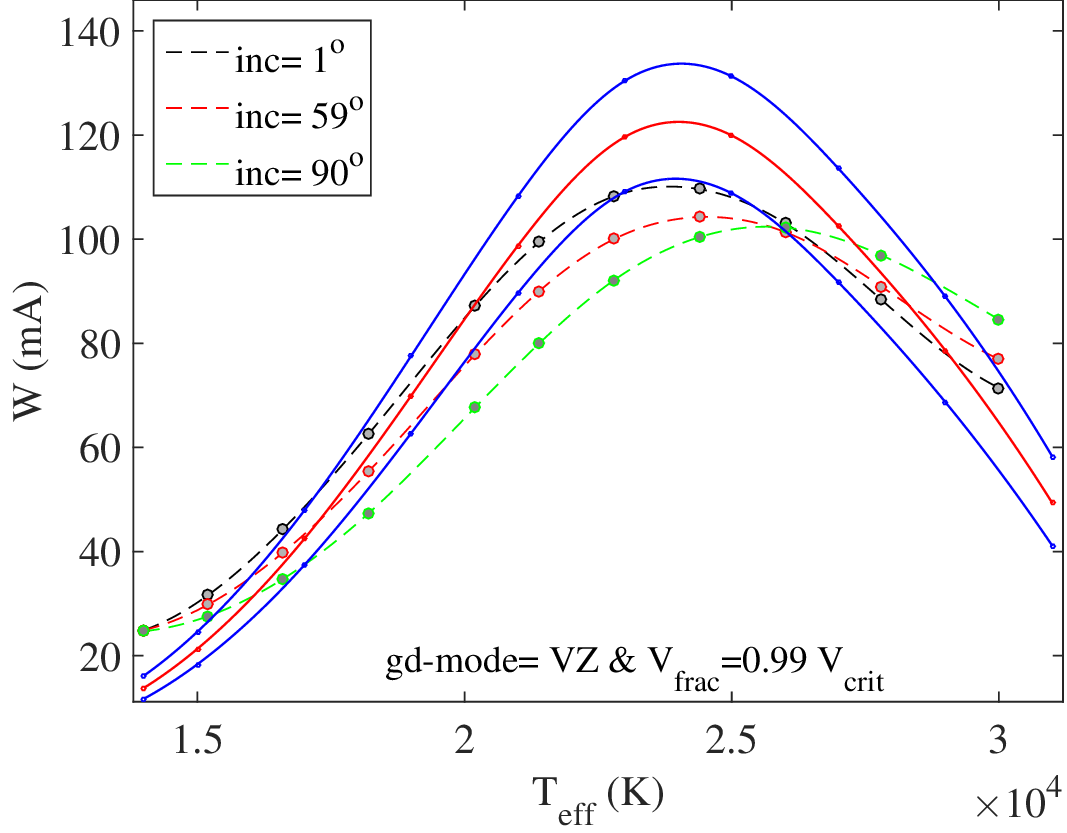}
\label{eqw_3995_gd_mode2_vz-0.99vc}
}
\hspace*{0.2cm}
\subfloat[]{
\includegraphics[scale= 0.25]{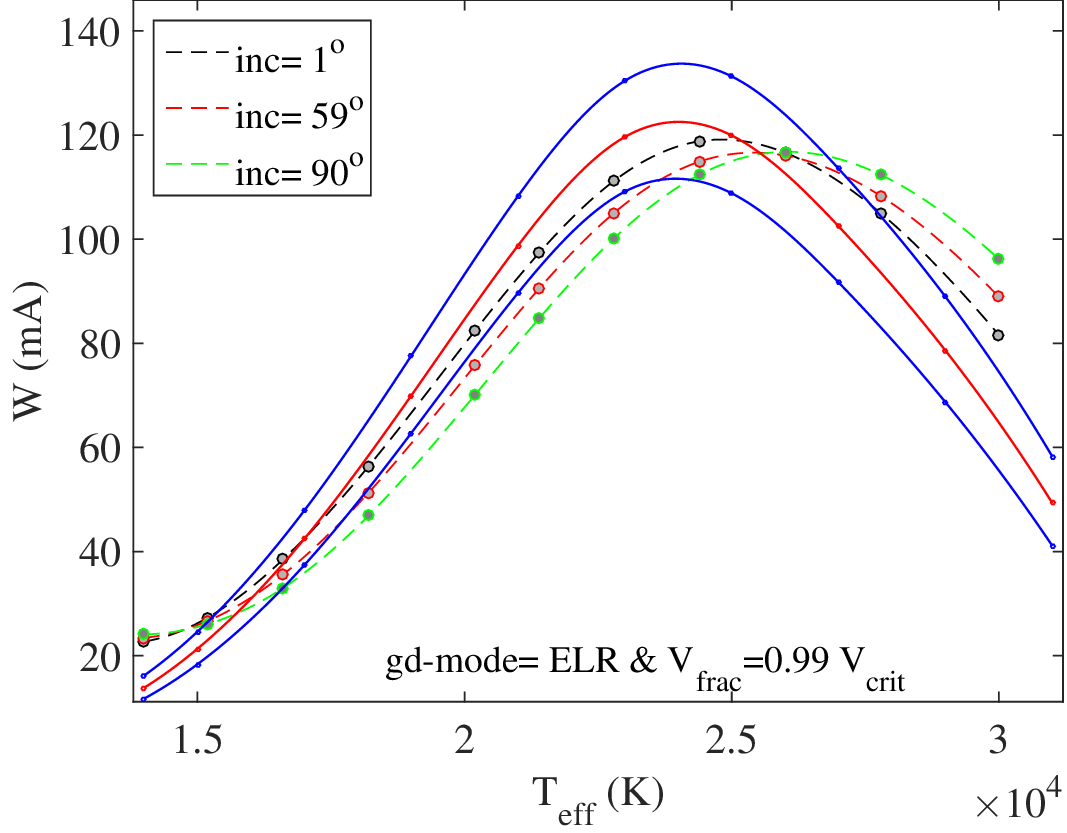}
\label{eqw_3995_gd_mode0_ELR-0.99vc}
}
\\
\subfloat[]{
\hspace*{-.7cm}
\includegraphics[scale= 0.25]{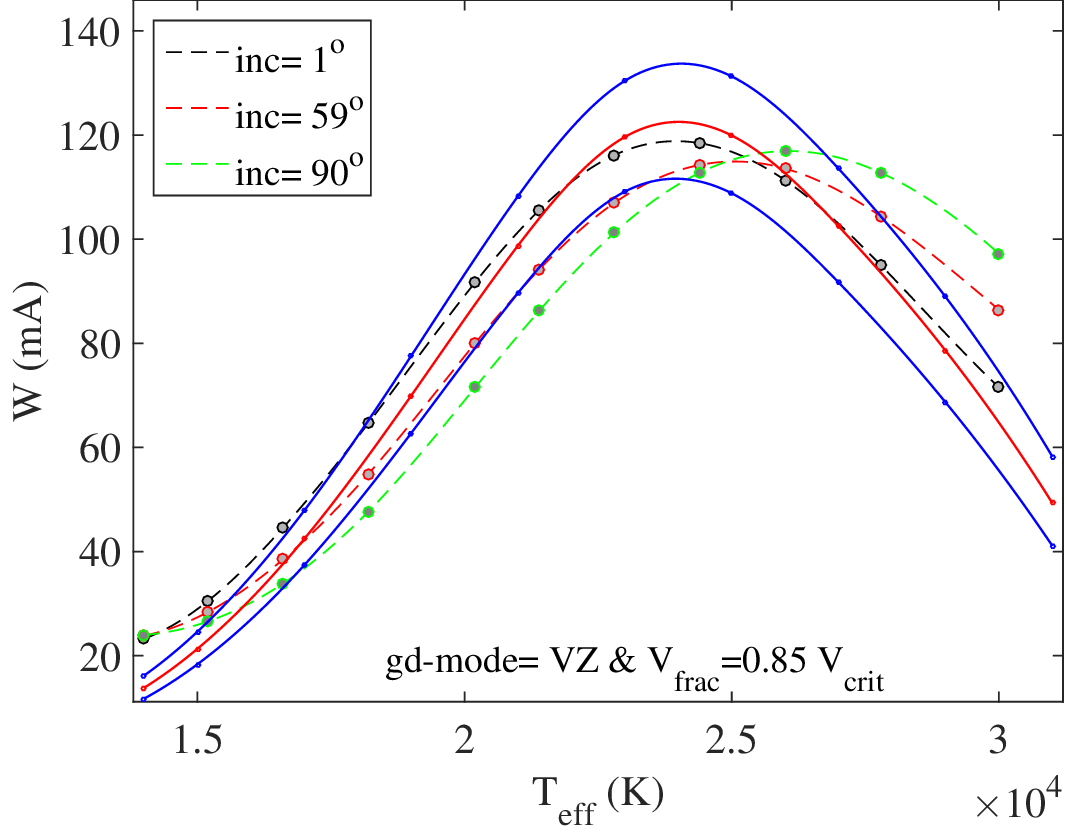}
\label{eqw_6482_gd_mode2_vz-0.99vc}
}
\hspace*{0.2cm}
\subfloat[]{
\includegraphics[scale= 0.25]{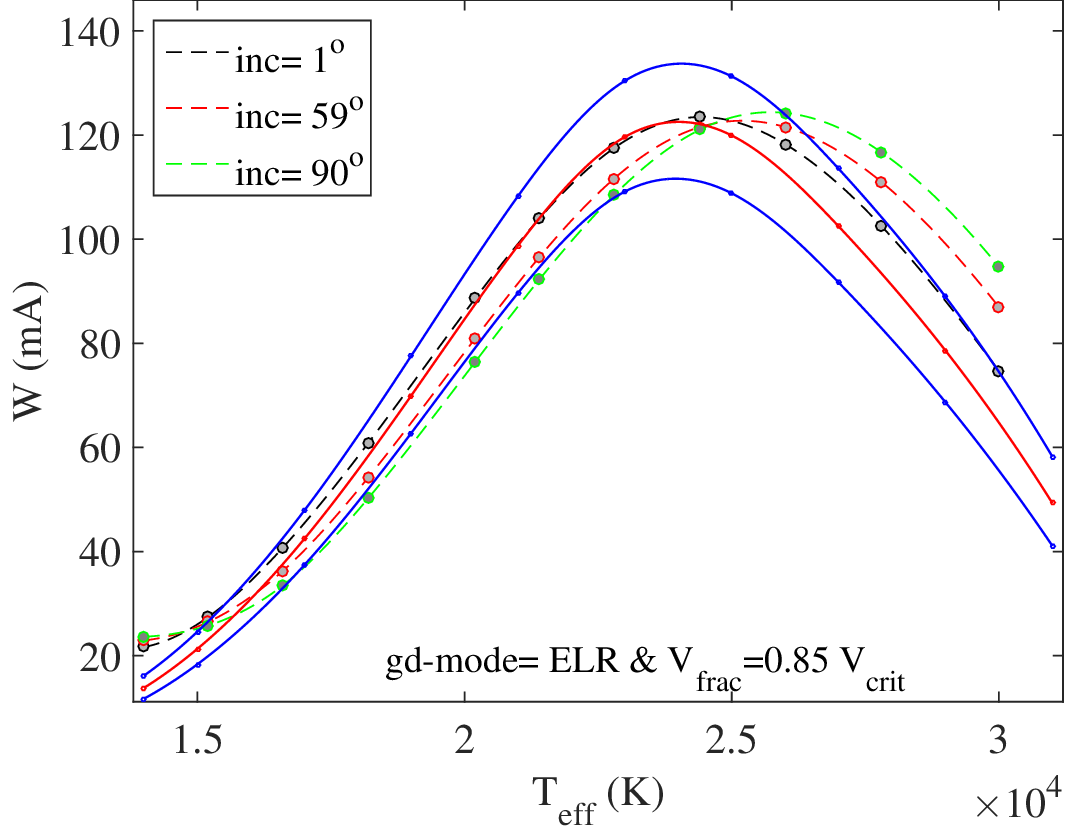}
\label{eqw_6482_gd_mode0_ELR-0.99vc}
}
\caption{Panels~\protect\subref{eqw_3995_gd_mode2_vz-0.99vc} and \protect\subref{eqw_3995_gd_mode0_ELR-0.99vc} show the predicted equivalent widths of N\,{\sc ii} $\lambda\,3995$ for the solar nitrogen abundance including gravitational darkening as a function of the $T_{\rm eff}$ of the non-rotating stars with the same mass. The left panel used the \citet{von24} formalism while the right panel uses the \citet{lar11} formalism. The rotational velocity was assumed to be 99\% of the critical velocity, and three inclinations angles are shown, $i=1^o$ (black dashed line), $i=59^o$ (red dashed line) and $i=90^o$ (green dashed line). Also show in each panel are the non-rotating equivalent widths for $\log g=4.0$ at the solar nitrogen abundance (red solid line) and $\pm\,0.1$ of the solar nitrogen abundance (blue solid lines). Panels \protect\subref{eqw_6482_gd_mode2_vz-0.99vc} and \protect\subref{eqw_6482_gd_mode0_ELR-0.99vc} are the same, but for a rotational velocity equal to 85\% of the critical velocity.
\label{eqw_gd}
}
\end{figure}

The effect of gravitational darkening on the N\,{\sc ii} $\lambda\,3995$ was computed with the same code that produced the continuum images, but with the calculations of \citet{AS15} used to provide non-LTE, photospheric line profiles as a function of local $T_{\rm eff}$ and $\log g$ that are integrated over the stellar surface. Calculations were done for a full range of B-type stars to gauge the effect of gravitational darkening on the predicted equivalent width of the line as shown in Figure~\ref{eqw_gd}. Rotation velocities of 85\% and 99\% of critical were considered, and the variation of the line's equivalent width as a function of $T_{\rm eff}$ of the underlying, non-rotating model is shown for three viewing inclinations, $i=0$, $59$ and $90^o$. The \citet{Collins65} and \citet{lar11} treatments are compared as before. 

As shown in Figure~\ref{eqw_gd}, the non-LTE equivalent of N\,{\sc ii} $\lambda\,3995$ width increases with $T_{\rm eff}$ until a maximum near $T_{\rm eff}\,\approx$24,000~K and then decreases for higher $T_{\rm eff}$'s because of the shift of the nitrogen ionization balance to N\,{\sc iii} \citep{AS15}. The equivalent widths for the rotating models depend on the viewing inclination. For $T_{\rm eff}$ well below the peak, the equivalent widths are larger for pole-on stars because the hotter stellar pole is viewed and the nitrogen line gets stronger with $T_{\rm eff}$ in this parameter range. Past the peak, the trend reverses with the equator-on models giving the largest equivalent widths because the nitrogen line gets stronger with decreasing $T_{\rm eff}$ in this parameter range. This basic behavior is seen in all cases considered.

However the most important conclusion from this figure is that in the case most applicable to the present work, panel~(d) with 85\% of critical rotation and the more realistic \citet{lar11} treatment, the effect of gravitational darkening is quite small. To put the predictions of the gravitationally-darkened models in perspective, also show in each figure panel are the predictions of non-rotating, spherical models as a function of $T_{\rm eff}$ for $\log g=4.0$ at the solar nitrogen abundance and $\pm0.1\;$dex around the solar nitrogen abundance. In panel (d), the gravitationally-darkened models fall within the $\pm0.1\;$dex envelope of the non-rotating model predictions for $T_{\rm eff}\leq 26,000\;$K, which includes the vast majority of our Be star sample. As noted by \citet{AS15}, the $\pm0.1\;$dex limit represents the achievable accuracy in nitrogen abundance determinations due to uncertainties in the basic atomic data used. For this reason, we do not expect the use of non-rotating models specified by a single $T_{\rm eff}$ and $\log g$ to be a dominate source of uncertainty in our abundance determinations.  A more subtle issue is the applicability of the adopted stellar parameters for the Be and B stars of Tables~\ref{parameters_table} and \ref{parameters_table_Bstars}.

\begin{figure}
\begin{center}
\includegraphics[scale= 0.4]{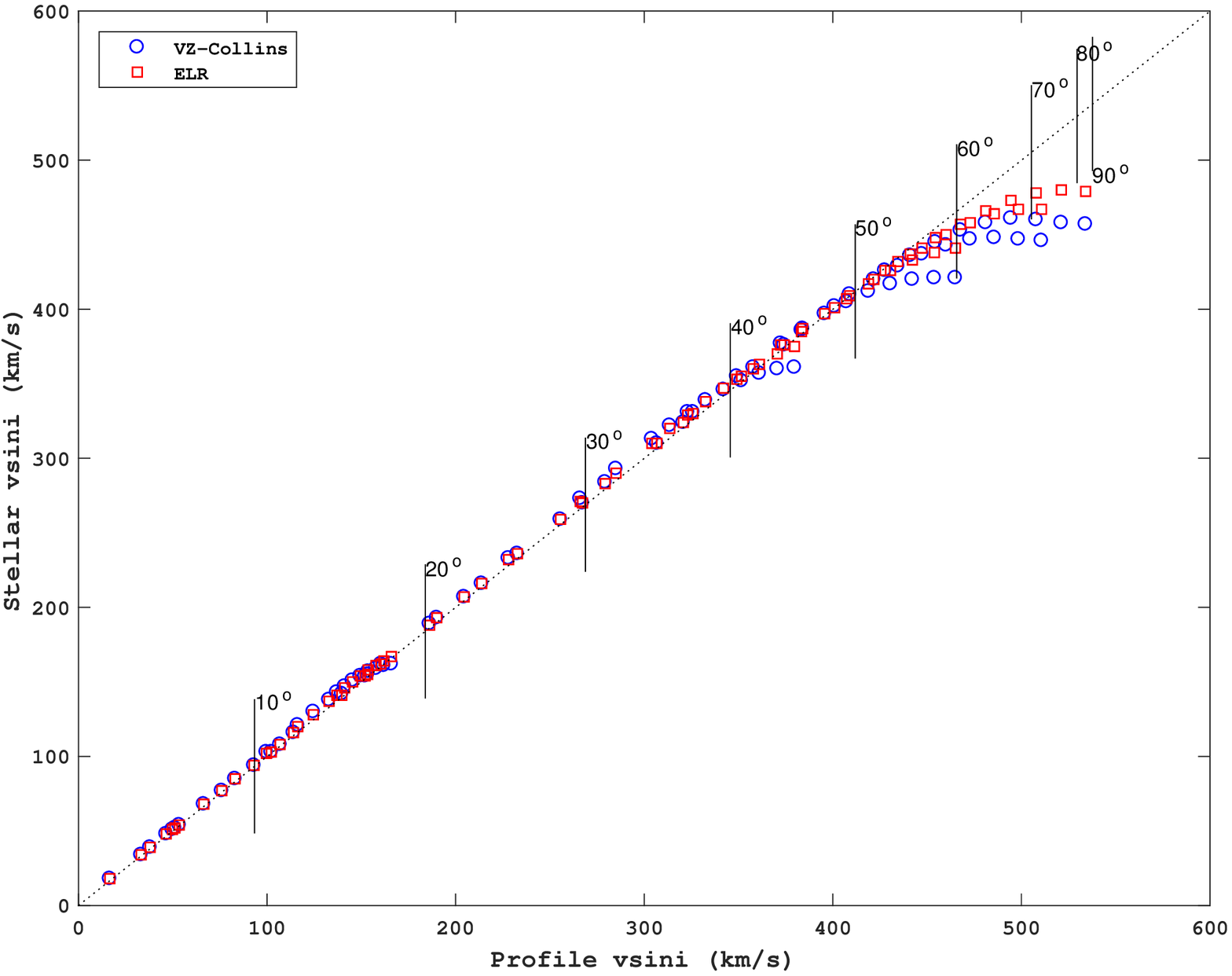}
\caption{Model $v\sin i$ values versus $v\sin i$ determined from fitting a pure rotation profile computed without gravitational darkening to the N\,{\sc ii} $\lambda\,3995$ line. The underlying stellar model was a B0V star with a critical velocity of $538\,\rm km\,s^{-1}$. Blue circles indicate results assuming the \citet{Collins65} treatment of gravitational darkening and red squares, the \citet{lar11} treatment. Vertical lines labelled by viewing inclination in degrees give the maximum possible $v\sin i$ at each model inclination. Profiles were computed for viewing inclination from 10 to $90^{\circ}$ with rotation rates, $v_{\rm frac}$, ranging from 10 to 100\% of the critical velocity. The black dotted line is the equality line of unit slope.}
\label{fig:vsini_from_NII}
\end{center}
\end{figure}

A final issue in connection with gravitational darkening is the impact on $v\sin i$ estimates. It has been known since the work of \citet{Sto68} that gravitational darkening can significantly impact $v\sin i$ measurements based on observed line widths. This bias was revived in context of the Be stars by \citet{tow04}, who demonstrated that $v\sin i$ values based on He\,{\sc i}~$\lambda\,4471$ can significantly underestimate rotation rates for stars near critical rotation because of the darkening of the equatorial regions (which are highest velocity portions of the stellar surface). However, this conclusion was based on the \citet{Collins65} formulation of gravitational darkening, which overestimates the equatorial darkening compared to the treatment of \citet{lar11}. To gauge the effect of gravitational darkening on the N\,{\sc ii} $\lambda\,3995$ line width, we have computed, via disk integration as before, synthetic, rotationally broadened lines in a B0V star (parameters in Table~\ref{central_stars_parameters_table}, implying a critical velocity of $538\,\rm km\,s^{-1}$) and measured the apparent $v\sin i$ by fitting pure rotation profiles uncorrected for gravitational darkening. The B0V spectral type was chosen as the nitrogen line will strongly increase in strength towards the cooler and higher velocity equatorial region. The results are shown in Figure~\ref{fig:vsini_from_NII} which plots the $v\sin i$ recovered from the profile versus the known model $v\sin i$. Viewing inclinations from $10$ to $90^{\circ}$ were considered in models with rotation rates from 10\% to 100\% of $v_{\rm crit}$. Note that at any  given viewing inclination, the maximum possible $v\sin i$ is $v_{\rm crit} \sin i$. While the effect of gravitational darkening is clearly seen at high rotational rates in stars seen more equator-on (i.e.\ at high inclination), the effect is not large and the treatment of \citet{lar11} produces a noticeably smaller effect. Given this, we do not consider the neglect of gravitational darkening on the measurement of $v\,\sin i$ from N\,{\sc ii}~$\lambda\,3995$ to be a large contributor to the uncertainty in the $v\,\sin i$ values; much more significant is the uncertainty in fitting the rotational profiles to the observed spectra, given their $S/N$ values, and the uncertainty associated with the continuum placement.   

\subsection{Circumstellar Disk Emission}
\label{Disk_paras}

\begin{table}
\centering{
\caption{Adopted Central Star Parameters for the Be Star Modelling. 
\label{central_stars_parameters_table}}}
\begin{center}
{\small
\begin{tabular}{c|c|c|c|c}
\hline\hline
\multicolumn{1}{l}{Spectral Type}	&	\multicolumn{1}{|c|}{$T_{\rm eff} (K)$}	&	$\log g$	& $M\,(M_{\odot}$) & $R\,(R_{\odot}$)\\  
\hline\hline
\multicolumn{1}{l|}{B0V}	 & 30000 &  4.0 & 17.5 & 7.4\\
\multicolumn{1}{l|}{B0.5V}	 & 28000 &  4.0 & 15.4 & 6.9\\
\multicolumn{1}{l|}{B1V}	 & 25000 &  4.0 & 13.2 & 6.4\\
\multicolumn{1}{l|}{B1.5V}	 & 23000 &  4.0	& 11.0 & 5.9\\
\multicolumn{1}{l|}{B2V}	 & 21000 &  4.0	& 9.1  & 5.3\\
\multicolumn{1}{l|}{B3V}	 & 19000 &  4.0	& 7.6  & 4.8\\
\multicolumn{1}{l|}{B4V}	 & 17000 &  4.0	& 6.6  & 4.3\\
\multicolumn{1}{l|}{B5V}	 & 15000 &  4.0	& 5.9  & 3.9\\
\hline
\end{tabular}}
\end{center}
\noindent{\it Source: \citet{Cox00} based on \citet{SK82}.} 
\end{table}

In the Be star sample, an important consideration is the potential contamination of the photospheric spectrum by the circumstellar disk. While direct line emission is not observed in the N\,{\sc ii} spectrum in the MiMeS Be star sample (unlike the hydrogen Balmer series, which is the defining characteristic of Be stars), the circumstellar disk can also give rise to continuum emission. This continuum emission usually first becomes apparent in the near-IR, and Be stars as a class are well-known to have an IR excess compared to normal B stars of the same spectral type \citep{Riv13}. However, continuum emission is also present, and expected, in the optical. In their abundance study of Magellanic Cloud Be stars, \cite{dun11} added a featureless continuum to their model spectra for each star, diluting the photospheric spectrum by the amount required for the measured silicon equivalent widths to recover the reference silicon abundance for either the LMC or SMC. With this method, sometimes large dilutions were found.

In the present work, the {\sc bedisk} \citep{sig07} and {\sc beray} \citep{sig11} codes are used to explicitly model the observed H$\alpha$ emission line in each Be star spectrum, and the disk parameters (see below) are found by matching the observed H$\alpha$ emission line. Using the best-fit disk density model, the effect of the circumstellar disk emission on the N\,{\sc ii} spectrum was explicitly computed. This technique has the advantage that it does not rely on an assumption like that used in the \citet{dun11}. In addition, the full range of behavior of the disk's effect on the spectrum is possible; for example at low inclinations (pole-on star, face-on disk), it is reasonable to expect the disk contribution to be added continuum emission, like that assumed by \cite{dun11}; however, for higher inclination systems, the photosphere can be viewed through the disk, and there is the possibility of a reduction in the photospheric spectrum due to this obscuration \citep[see][]{sp13}. Another advantage of direct H$\alpha$ modelling is that an estimate of the system inclination is obtained as the morphology of the H$\alpha$ line (singly-peaked, doubly-peaked, shell absorption) depends strongly on the inclination.

To represent the Be star circumstellar disk, a geometrically thin, disk density model was adopted with the functional form
\begin{equation}
\label{disk_eqn}
\rho(R,Z)=\rho_0 \left( R / R_* \right)^{-n} e^{-(z/H)^2}\;,
\end{equation}
where $(R,\,Z)$ are the cylindrical co-ordinates on the disk, $R$ being the distance from the star's rotation axis, $Z$, the height above or below the equatorial plane, and $R_*$, the stellar radius. Thus the disk density is specified by the power-law index $n$ and the density at the base of the disk, $\rho_0$. The disk is assumed to follow this density law until an outer radius of $R=R_{\rm disk}$. The disk scale height, $H$, is of the form $H(R)=\,H_0\,(R/R_*)^{3/2}$, where the scale height at the stellar surface is given by
\begin{equation}
H_0 = \left( \frac{2 R_{*}^{3} k T_{0}}{G M_{*} \mu m_{H}}\right)^{1/2} \, ,
\end{equation}
where $M_*$ and $R_*$ are the stellar mass and radius and $\mu$ is the mean-molecular weight of the disk gas. This expression follows from the assumption of vertical hydrostatic equilibrium, and the temperature $T_0$ is a parameter set to $T_0=0.6\,T_{\rm eff}$ and is used only to set the scale height of the disk \citep[see][for details]{sig09}. This density distribution has been used extensively to model the spectra of Be stars \citep{sil10,sil14,arc17}.

Values for the stellar parameters of the central B stars are listed in Table~\ref{central_stars_parameters_table}. For the purpose of the circumstellar disk calculations, we have assumed that the central stars are spherical with radius $R_*$ and have neglected gravitational darkening. The influence of gravitational darkening on the thermal structure of Be stars disks has been studied by \citet{MJS11} and \citet{MJS13} who find the thermal effects to be relatively small, particularly with \citet{lar11} formulation of gravitational darkening. In the context of computed H$\alpha$ emission profiles, one consequence of non-spherical models would be the expansion of the stellar equatorial radius (and hence the inner edge of the disk) with increased rotation. For a fixed mass, this decreases the overall scale of the disk's Keplerian rotation and acts to narrow the width of the predicted H$\alpha$ emission, which may translate into systematically higher viewing inclinations derived from observed spectra. Of course, the stellar parameters in Table \ref{central_stars_parameters_table} themselves set the scale of the disk's Keplerian rotation. We have adopted the mass and radius calibration of the main sequence B stars used by \citet{Cox00}, which is ultimately based on the calibration of \citet{SK82}. We note that \citet{NP14} find that the available evidence from nearby B stars suggests this calibration may be systematically too large by 10-20\% in mass and 25\% in radius.   

Given this density distribution and a spectral type for the central star (which fixes the star's mass, radius and luminosity), the {\sc bedisk} code of \cite{sig07} was used to compute the thermal structure of the disk. The disk temperatures are found by enforcing radiative equilibrium in a gas of solar composition, and the {\sc bedisk} solution provides both the thermodynamic state of the gas $(T,\rho)$ and all of the atomic level populations required to construct the opacity and emissivity required by radiative transfer calculations.  This solution is then input into the {\sc beray} code of \cite{sig11} which solves the radiative transfer equation along a series of rays threading the star+disk system directed at the observer. For rays that terminate on the stellar surface, a photospheric boundary condition of an appropriately Doppler-shifted, photospheric, LTE H$\alpha$ line is used. Hence the composite nature of the spectrum, stellar photosphere plus disk, is consistently modeled, and the full range of Be H$\alpha$ profiles can be reproduced, from singly-peaked emission to deep shell absorption. The {\sc beray} modelling adds an addition parameter, namely the viewing inclination angle, $i$. Typically low inclination systems produce singly-peaked H$\alpha$ emission lines, intermediate inclinations give doubly-peaked profiles, and high inclination systems, where the star is viewed through the disk, give shell absorption. For this reason, the H$\alpha$ modeling can yield an estimate of the system inclination. 

Grids of synthetic H$\alpha$ line profiles were computed for all combinations of the following disk parameters for central stars of spectral types B0 through B5. The selected values of $\rho_0$ were $1.0\times10^{-12}$, $2.5\times10^{-12}$,	$5.0\times10^{-12}$,	$7.5\times10^{-12}$,	$1.0\times10^{-11}$,	$2.5\times10^{-11}$,	$5.0\times10^{-11}$, $7.5\times10^{-11}$, $1.0\times10^{-10}$ and $2.5\times10^{-10}\;\rm g\,cm^{-3}$.  The selected values of the power-law index $n$ were 2.0, 2.5, 3.0, 3.5 and 4.0. The selected values of disk radii were 6.0, 12.5, 25.0 and 50.0 stellar radii. Finally, 13  inclination angles, 10, 18, 20, 30, 40, 45, 50, 60, 70, 72, 80,  84, and 89 degrees, were considered. All combinations of these parameters makes a library of 2600 model H$\alpha$ line profiles for each spectral type. The physical parameters of the central stars, i.e. their masses, radii and temperatures, are listed in Table~\ref{central_stars_parameters_table}. Each star in the Be sample was assigned to the spectral type with the closest match in $T_{\rm eff}$ as listed in Table~\ref{Halpha_MATCHING_RESULTS}. Note that this assigned spectral type can differ from the spectral type listed in Table~\ref{parameters_table}, taken from the SIMBAD database.

The disk parameters for each star of the sample were taken to be those of the synthetic H$\alpha$ profile that best matched the observed H$\alpha$ line profile from the library computed for that spectral type. To make the match, the figure of merit ${\cal F}$ of Equation~(\ref{eq:fom}) was computed for each library profile, and the disk parameters of the model corresponding to the minimum in ${\cal F}$ were adopted. We note that in the Be sample, two stars, HD~11415 and HD~49567, had no sign of emission in the H$\alpha$ line, and hence the disk contamination was assumed to be zero. The absence of H$\alpha$ emission is not inconsistent with a Be designation as it is well-known that Be stars transition between Be and B star states as the disk forms and dissipates \citep{Riv13}. 

\begin{figure}
\begin{center}
\includegraphics[scale= 0.6]{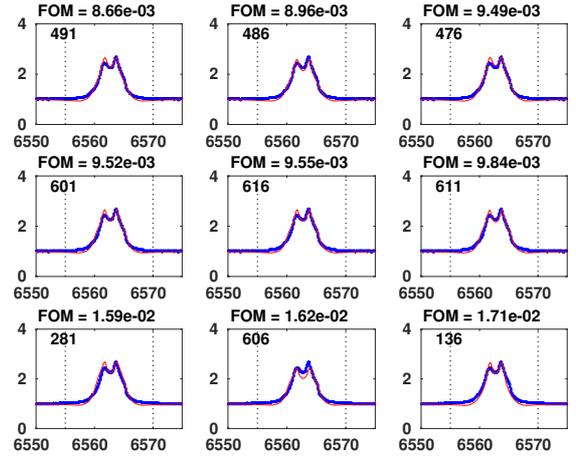}
\caption{The top nine H$\alpha$ library fits (red lines) for the Be star HD~143275, a star classified with a good H$\alpha$ fit. The observed H$\alpha$ line, the same in each panel, is shown in blue. The figure-of-merit, ${\cal F}$, is shown at the top of each panel, and the disk parameters for the best-fit profile (top left) are $\rho_0 = 7.5\cdot\,10^{-12}\,\rm g\,cm^{-3}$, $n=2.5$, $R_{\rm disk}=50\,R_{*}$, and $i=18^{\circ}$.}
\label{Halpha_fitting_HD143275}
\end{center}
\end{figure}

\begin{figure}
\begin{center}
\includegraphics[scale= 0.6]{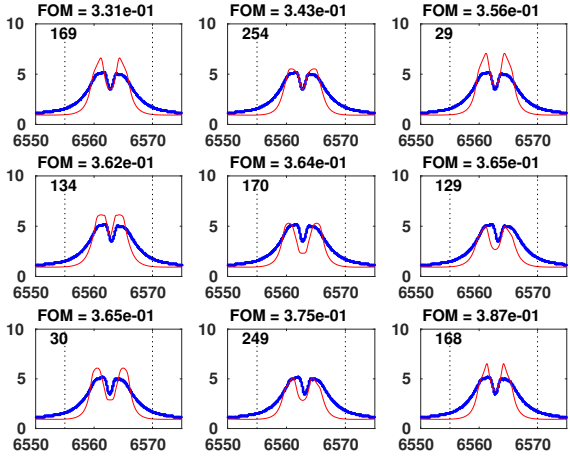}
\caption{The same as Figure (\ref{Halpha_fitting_HD143275}) but for the Be star HD~45725. The fit {\bf is} classified as acceptable.}
\label{Halpha_fitting_HD45725}
\end{center}
\end{figure}

\begin{figure}
\begin{center}
\includegraphics[scale= 0.6]{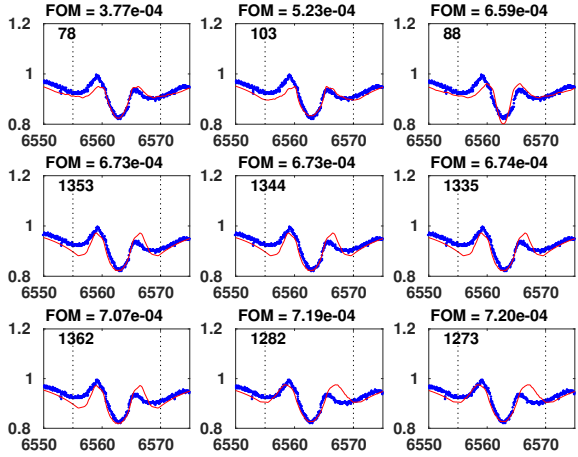}
\caption{The same as Figure (\ref{Halpha_fitting_HD143275}) but for the Be star HD~189687 with an asymmetric H$\alpha$ line. The fit is classified as poor, and only the blue half of the line was used in the fit.}
\label{Halpha_fitting_HD189687}
\end{center}
\end{figure}

The rest of the Be stars in the sample were classified into three groups based on the quality of the H$\alpha$ fits: good matching, acceptable matching, and poor matching. In the first group, the adopted disk models reproduce the observed profiles well, especially the peaks. They also have figure of merit ${\cal F}<0.1$. An example of a good fit it given in Figure~\ref{Halpha_fitting_HD143275} where the H$\alpha$ line is reproduced well in peak height, shape, and overall line width. In the second group of acceptable matching, the adopted models do not reproduce the observed profiles as well as the first group, usually failing to simultaneously match both the peak heights and overall line width. These fits typically have ${\cal F}\,>0.1$ and an example for the Be star HD~45725 is given in Figure~\ref{Halpha_fitting_HD45725}. Two stars, HD174237 and HD189687, were classified as poor matches with the observed H$\alpha$ profile having asymmetric, doubly-peaked profiles. To extract disk parameters, only the blue half of the H$\alpha$ profile was modelled. Figure~\ref{Halpha_fitting_HD189687} illustrates the case of HD~189687. HD~174237 (CX Dra) is known to be an interacting binary \citep{RICH00}, and its disk density distribution is likely poorly described by a simple power law.

The acceptable and poor fits may have their origin in the $(M_*,R_*)$ calibration of Table~\ref{central_stars_parameters_table} as previously discussed. In addition, the assumed power-law form for the equatorial disk density may not be sufficiently general for all Be star disks. Finally, the photospheric H$\alpha$ profiles assumed $\log\,g=4.0$ for the central stars, whereas many of the observed sample of Be stars have lower $\log\,g$ values. Despite these uncertainties, we feel that the disk density parameters are extracted reliably enough for useful estimates of the disk contamination at the wavelength positions of the N\,{\sc ii} lines used in the abundance analysis, as discussed in the next section. 

Table~\ref{Halpha_MATCHING_RESULTS} lists all of the derived disk parameters for the Be stars in the sample, and classifies them, as discussed above, into good, acceptable and poor matching. Of the 26 stars, 14 match well, 10 match acceptably, two have poor matches, and two Be stars are diskless. The derived disk based densities fall in the range of $1.0\,\cdot\,10^{-12}\,\rm g\,cm^{-3}$ to $1.0\,\cdot\,10^{-10}\,\rm g\,cm^{-3}$ and the power-law index, in the range of $2.0$ to $3.5$. These values are very typical of disk density parameters found for the Be stars \citep{sil10,sil14,arc17}.

\begin{table}
\centering{
\caption{Disk Density Parameters for the MiMeS Be Star Sample.
\label{Halpha_MATCHING_RESULTS}}}
\begin{center}
{\scriptsize
\begin{tabular*}{0.45\textwidth}{l@{\hskip 0.09in}c@{\hskip 0.09in}c@{\hskip 0.1in}c@{\hskip  0.0in}c@{\hskip  0.09in}c@{\hskip  0.09in}c@{\hskip  -0.15in}c@{\hskip  -0.05in}c@{\hskip  -0.05in}c}
\hline\hline
HD				&	\multicolumn{1}{c}{Spectral}  & $\rho_0$	& n 	& $R_{disk}$& $i$ & $i_{\rm Fre05}$ & \\
				& 		Type					& $\rm (g\,cm^{-3})$	& 		&	$(R_{*})$	& $(^\circ)$ & \multicolumn{1}{c}{$(^\circ)$} & \\
\hline 				
\multicolumn{8}{c}{Good Match}\\
\multicolumn{1}{l}{33328}	&	\multicolumn{1}{l}{B2V}	&		1.00$^{-12}$	&	3.5	&	6.0	& 	50.0 	&	73.0	&	\\
\multicolumn{1}{l}{56139}	&	\multicolumn{1}{l}{B3V}	&	2.50$^{-11}$	& 	2.5	&	25.0	&	30.0 	& 	16.6 	&	\\  
\multicolumn{1}{l}{58050} 	&	\multicolumn{1}{l}{B3V}	&	1.00$^{-12}$	&	2.5 	& 6.0		&	20.0	& 	21.9	&	\\	
\multicolumn{1}{l}{58343} 	&	\multicolumn{1}{l}{B4V}	&	1.00$^{-11}$ 	& 	2.0 &	25.0	&	10.0	&	9.7	& \\
\multicolumn{1}{l}{67698}	&	\multicolumn{1}{l}{B5V}	&		2.50$^{-11}$	&	3.0	& 	12.5	& 	20.0 	&		 	&	\\
%
\multicolumn{1}{l}{120324} & 	\multicolumn{1}{l}{B2V}	&		7.50$^{-12}$ &	2.5  	& 	12.5	&	20.0	& 			&	\\
\multicolumn{1}{l}{143275}	& 	\multicolumn{1}{l}{B0V}	&		7.50$^{-12}$ &	2.5	& 	50.0 	&	18.0	& 			& 	\\
%
\multicolumn{1}{l}{187811} &	\multicolumn{1}{l}{B4V}	&	7.50$^{-12}$	&	2.5 	&	6.0	&	50.0 	&	48.9	&	\\
\multicolumn{1}{l}{191610}	& 	\multicolumn{1}{l}{B4V}	&		7.50$^{-12}$ &	3.0  	&	6.0 	& 	45.0	&	63.7 	&	\\
%
\multicolumn{1}{l}{192685} &	\multicolumn{1}{l}{B4V}	&	1.00$^{-11}$	&	3.0 	&	6.0	&	50.0	&			&	\\
\multicolumn{1}{l}{205637}  & \multicolumn{1}{l}{B4V}	&	5.00$^{-11}$ 	&	3.5 	&	25.0 	&	80.0 	&	54.7	&	\\
\multicolumn{1}{l}{212571} &	\multicolumn{1}{l}{B1V}	&		2.50$^{-11}$	& 	3.0	&	 12.5	& 	60.0	& 33.6	&	\\
\hline
\multicolumn{8}{c}{Acceptable Match}\\
\multicolumn{1}{l}{20336}	&	\multicolumn{1}{l}{B4V}	&	5.00$^{-11}$	&	3.0	&	25.0	&	70.0	& 66.7 \\
\multicolumn{1}{l}{45725}	&	\multicolumn{1}{l}{B4V}	&	7.50$^{-12}$	&	2.0 	&	25.0	&	72.0	&	66.6 	&	\\
\multicolumn{1}{l}{54309}	&	\multicolumn{1}{l}{B3V}	&	1.00$^{-10}$	& 	3.0	&	25.0	& 	60.0	&	38.4 	& \\
\multicolumn{1}{l}{58978}	&	\multicolumn{1}{l}{B1.5V}&	5.00$^{-11}$	&	3.5 	&	50.0	&	60.0	&	55.2	&	\\
\multicolumn{1}{l}{65875}	&	\multicolumn{1}{l}{B3V}	&	7.50$^{-12}$	& 	2.0	&	50.0	&	50.0	& 	27.8 	&	\\
\multicolumn{1}{l}{178175} &	\multicolumn{1}{l}{B2V}	&	5.00$^{-12}$	&	2.0 	&	12.5	&	30.0 	&	22.4	&	\\
\multicolumn{1}{l}{187567} &	\multicolumn{1}{l}{B2V}	& 	5.00$^{-12}$	&	2.0 	&	25.0	&	40.0	& 			& \\ 
\multicolumn{1}{l}{203467} &	\multicolumn{1}{l}{B5V}	&	7.50$^{-11}$	&	2.5 	&	50.0	&	60.0	&			&	\\
%
\multicolumn{1}{l}{212076} &	\multicolumn{1}{l}{B3V}	&	5.00$^{-12}$	&	2.0 	&	50.0	&	40.0	&	18.6	& 	\\
\multicolumn{1}{l}{217050$^B$} &	\multicolumn{1}{l}{B4V}	&	5.00$^{-12}$	&	2.0 	&	25.0	& 	84.0	&	78.5	& \\
\multicolumn{1}{l}{217050$^R$} &\multicolumn{1}{l}{B4V}		&	2.50$^{-11}$ 	&  2.5 	&	50.0	&	84.0	&			&\\
\hline
\multicolumn{8}{c}{Poor Match}\\	
\multicolumn{1}{l}{174237} &	\multicolumn{1}{l}{B4V}	&	1.00$^{-12}$	&	2.0	&  6.0	& 	50.0	& 	33.2	&  \\ 
\multicolumn{1}{l}{189687}	& 	\multicolumn{1}{l}{B4V}	&		2.50$^{-12}$ &	3.0 	& 	6.0	&	30.0	& 46.6 	& 	\\	
\hline
\multicolumn{8}{c}{No Disk}\\	
\multicolumn{1}{l}{11415} 	& 	\multicolumn{1}{l}{B5V} & \multicolumn{1}{c}{$-$}& \multicolumn{1}{c}{$-$}& \multicolumn{1}{c}{$-$}& \multicolumn{1}{c}{$-$}& \multicolumn{1}{c}{$-$}&	\\	
%
\multicolumn{1}{l}{49567} 	&	\multicolumn{1}{l}{B4V}	& \multicolumn{1}{c}{$-$}& \multicolumn{1}{c}{$-$}& \multicolumn{1}{c}{$-$}& \multicolumn{1}{c}{$-$}& \multicolumn{1}{c}{$-$}&	\\		
\hline
\end{tabular*}}
\end{center}
\noindent{\scriptsize Notes: The two entries for HD\,217050 are separate fits to the red (R) and blue (B) wings of the H$\alpha$ profile. Entries in the seventh column ($i_{\rm Fre05}$) are the estimated inclinations of \protect\citet{fre05}.}
\end{table}

\begin{figure}
\begin{center}
\includegraphics[scale=0.40]{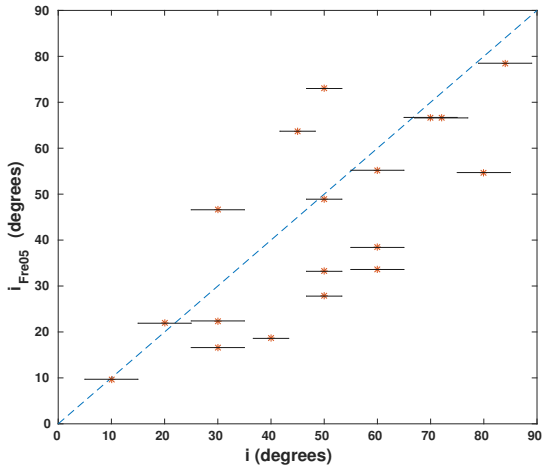}
\caption{Estimated inclinations of the MiMeS Be star sample compared with \citet{fre05} for the 18 stars in common.}
\label{inc_estimations_comp}
\end{center}
\end{figure}

As noted above, the best-fit H$\alpha$ profile for each star yields an estimate of the system inclination, and these inclinations are also listed in Table~\ref{Halpha_MATCHING_RESULTS}. We compare our inclination estimates with the inclinations of \citet{fre05} for the 18 stars in common stars in Figure~\ref{inc_estimations_comp}. The \citet{fre05} inclinations are based on the gravitational darkening effect in the Mg\,{\sc ii} and He\,{\sc i} spectral lines. While the two estimates are correlated, there is a large scatter. In general, our estimates tend to be higher than those of \citet{fre05}. We have assigned the bin size in inclination, $\pm 5^{\circ}$, as an error, but this is quite likely a significant underestimate. \citet{fre05} cite very small errors on their inclinations, typically $\pm 1^{\circ}$, but we do not fully understand how these errors were determined.

A one-tailed Kolmogorov-Smirnov test was applied to the cumulative distribution of inclinations for our entire Be star sample (26 stars) and compared to the $(1-\cos\,i)$ distribution expected for random orientations. The sample inclinations agree with the expected random distribution at the 1\% level.  

\subsection{Corrected Nitrogen Abundances}
\label{Nabund_corr}

The disk parameters obtained in Section~\ref{Disk_paras} were used to compute the expected disk contamination using the {\sc beray} code.  For each model, we compute the ratio of the star+disk continuun flux and the stellar photospheric flux alone, which we call the disk contamination, denoted $\Delta_{\rm D}$. This is a wavelength dependent quantity and as discussed above, one niavely expects an excess with $\Delta_{\rm D}\geq 1$; however, this need not be the case for higher inclinations where the disk can obscure the photospheric flux.

\begin{figure}
\centering
\includegraphics[scale= 0.40]{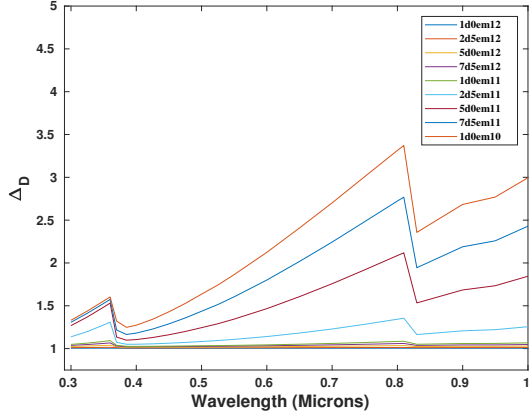}
\caption{Disk contamination as function of the wavelength for a B3V central star. The model viewing inclination was $i=30^o$, and curves are shown for nine values of $\rho_0$ as given in the legend. The other disk density parameters were set to $n=\,2.5$ and $R_{disk}=\,25\,R_*$.}
\label{DiskCont1}
\end{figure}

Figures~\ref{DiskCont1} shows the wavelength-dependent disk contamination for a B3V central star seen at $i=30^\circ$ (nearly face-on disk) with disk parameters $n=\,2.5$, and $R_{disk}=\,25\,R_*$ for nine selected values of $\rho_0$ between $1.0\times 10^{-12}$ and $1.0\times 10^{-10} \,(\rm g\,cm^{-3})$. The disk model with the lowest $\rho_0$ value has no significant contribution to the total flux, and the disk contamination increases with the increase in the disk density. In this case, as the disk is seen nearly face-on, and the disk adds continuum flux to the photospheric flux; however, the contamination for $\lambda<5000\,$\AA\ is less than 50\% even for the densest disks considered. Finally, emission edges corresponding to the Balmer ($\lambda_{th}=3646\,$\AA) and Paschen ($\lambda_{th}=8204\,$\AA) continua are evident at the highest disk densities. 

\begin{figure}
\centering
\includegraphics[scale= 0.40]{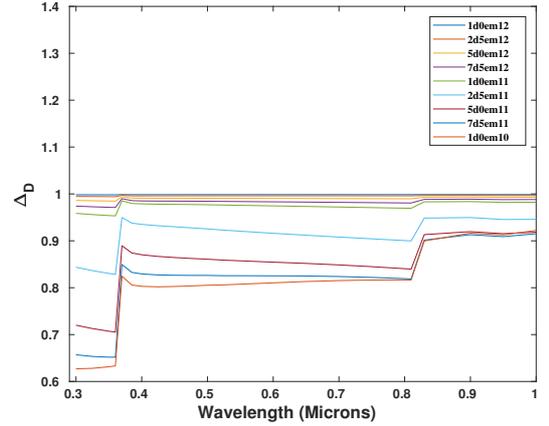}
\caption{The same as Figure (\ref{DiskCont1}) but for B4V Be central star seen at $i=\,80^o$.}
\label{DiskCont2}
\end{figure}

On the other hand, the expected disk contamination for a model with a B4V central star and disk parameters $i=\,80^\circ$ (nearly edge-on disk), $n=\,3.5$, and $R_{disk}=\,25\,R_*$ and for the same $\rho_0$ values as Figure~\ref{DiskCont1} is shown in Figure~\ref{DiskCont2}. Again, the disk model with the lowest $\rho_0$ value has no effect on the observed stellar radiation. However in this case of a nearly edge-on disk, as the disk density increases, the disk acts to reduce the stellar flux and $\Delta_{\rm D}<1$. However, in the optical region, the disk contamination parameter stays above 0.8 even for the densest disks.

Table~\ref{Disk_Cont_results} lists the predicted disk contamination, $\Delta_{\rm D}$, at three optical wavelengths corresponding to N\,{\sc ii} transitions computed for stellar models with disk parameters equal to those of the best fit for each Be star in the sample. Only 24 or the 26 sample Be stars appear in this table as two sample Be stars, HD\,11415 and 49567, were found to be diskless (see Table~\ref{Halpha_MATCHING_RESULTS}). The effects of disk contamination are predicted to be quite small, with $\Delta_{\rm D}$ below $1.1$ in all except two cases at the longest wavelength considered ($\lambda\,5679.6\,$\AA). \cite{dun11} found disk correction factors of up to 1.60, with a median of $1.21$ and half of the values greater than $1.1$, in their analysis of the nitrogen abundances of 30 Magellanic Cloud Be stars. In our models, such large disk contaminations would be associated with very large emission equivalent widths for H$\alpha$. While Be stars disks at the low LMC/SMC abundances are likely hotter, the effect on the H$\alpha$ emission equivalent width from the disk is small \citep{AS12}.

\begin{table}
\centering{
\caption{Disk Contamination $\Delta_{D}$ for the MiMeS Be Star Sample at Three Wavelengths.
\label{Disk_Cont_results}}}
\begin{center}
\begin{tabular*}{0.35\textwidth}{crrrrr|rrrrrr}
\hline\hline
HD			   & \multicolumn{3}{c}{$\Delta_{\rm D}\equiv\rm F_{star+disk}/F_{star}$}	&\\
				& \multicolumn{1}{l}{{$\lambda$ 3995}} & \multicolumn{1}{l}{{$\lambda$ 4481}} & \multicolumn{1}{l}{{$\lambda$ 5679.6 (\AA)}}	&\\
\hline 				
\multicolumn{5}{c}{Good Match}\\
%
\multicolumn{1}{l}{33328}	&	\multicolumn{1}{l}{1.001}&	\multicolumn{1}{l}{1.001}&	\multicolumn{1}{l}{1.001}\\
\multicolumn{1}{l}{56139}	&	\multicolumn{1}{l}{1.051}& 	\multicolumn{1}{l}{1.062}& 	\multicolumn{1}{l}{1.120}\\  
\multicolumn{1}{l}{58050} 	&	\multicolumn{1}{l}{1.001}& 	\multicolumn{1}{l}{1.001}& 	\multicolumn{1}{l}{1.002}\\
\multicolumn{1}{l}{58343}	&	\multicolumn{1}{l}{1.042}&	\multicolumn{1}{l}{1.048}&	\multicolumn{1}{l}{1.071}\\	
\multicolumn{1}{l}{67698}	&	\multicolumn{1}{l}{1.036}&	\multicolumn{1}{l}{1.044}	&	\multicolumn{1}{l}{1.083}\\
%
\multicolumn{1}{l}{120324}	&  \multicolumn{1}{l}{1.026}& \multicolumn{1}{l}{1.029}& 	\multicolumn{1}{l}{1.038}\\
\multicolumn{1}{l}{143275}	&  \multicolumn{1}{l}{1.038}&	\multicolumn{1}{l}{1.040}& 	\multicolumn{1}{l}{1.047}\\
\multicolumn{1}{l}{187811} &	\multicolumn{1}{l}{1.006}&	\multicolumn{1}{l}{1.006}&	\multicolumn{1}{l}{1.010}\\
\multicolumn{1}{l}{191610}	&  \multicolumn{1}{l}{1.006}& 	\multicolumn{1}{l}{1.007}& 	\multicolumn{1}{l}{1.01}\\
\multicolumn{1}{l}{192685} &	\multicolumn{1}{l}{1.004}&	\multicolumn{1}{l}{1.004}&	\multicolumn{1}{l}{1.007}\\
\multicolumn{1}{l}{205637}  & \multicolumn{1}{l}{0.870}&	\multicolumn{1}{l}{0.865}&	\multicolumn{1}{l}{0.856}\\
\multicolumn{1}{l}{212571} &	\multicolumn{1}{l}{0.973}& 	\multicolumn{1}{l}{0.975}& 	\multicolumn{1}{l}{0.989}\\
\hline
\multicolumn{5}{c}{Acceptable Match}\\
\multicolumn{1}{l}{20336}	&	\multicolumn{1}{l}{0.886}& 	\multicolumn{1}{l}{0.884}& 	\multicolumn{1}{l}{0.903}\\
\multicolumn{1}{l}{45725}	&	\multicolumn{1}{l}{0.992}& 	\multicolumn{1}{l}{0.992}& 	\multicolumn{1}{l}{0.993}\\
\multicolumn{1}{l}{54309}	&	\multicolumn{1}{l}{0.915}& 	\multicolumn{1}{l}{0.955}& 	\multicolumn{1}{l}{1.109}\\
\multicolumn{1}{l}{58978}	&	\multicolumn{1}{l}{0.935}&	\multicolumn{1}{l}{0.939}&	\multicolumn{1}{l}{0.971}	\\
\multicolumn{1}{l}{65875}	&	\multicolumn{1}{l}{1.027}& 	\multicolumn{1}{l}{1.030}& 	\multicolumn{1}{l}{1.046}\\
\multicolumn{1}{l}{178175} &	\multicolumn{1}{l}{1.019}&	\multicolumn{1}{l}{1.021}&	\multicolumn{1}{l}{1.027}\\
\multicolumn{1}{l}{187567} &	\multicolumn{1}{l}{1.026}& 	\multicolumn{1}{l}{1.029}& 	\multicolumn{1}{l}{1.039}	&	\\ 
\multicolumn{1}{l}{203467} &	\multicolumn{1}{l}{0.958}&	\multicolumn{1}{l}{1.010} &	\multicolumn{1}{l}{1.231}\\
\multicolumn{1}{l}{212076} &	\multicolumn{1}{l}{1.024}&	\multicolumn{1}{l}{1.027}&	\multicolumn{1}{l}{1.037}\\
\multicolumn{1}{l}{217050$^B$} &	\multicolumn{1}{l}{0.941}&	\multicolumn{1}{l}{0.938} &	\multicolumn{1}{l}{0.929}\\ 
\multicolumn{1}{l}{217050$^R$}	&	\multicolumn{1}{l}{0.892}&	\multicolumn{1}{l}{0.886} &	\multicolumn{1}{l}{0.870}\\ 
\hline
\multicolumn{5}{c}{Poor Match}\\
\multicolumn{1}{l}{174237} & 	\multicolumn{1}{l}{1.002}& 	\multicolumn{1}{l}{1.003}& 	\multicolumn{1}{l}{1.003}\\
\multicolumn{1}{l}{189687}	&	 \multicolumn{1}{l}{1.004}& 	\multicolumn{1}{l}{1.005}& 	\multicolumn{1}{l}{1.006}\\
\hline
\end{tabular*}
\end{center}
\end{table}

\begin{figure}
\centering
\includegraphics[scale= 0.45]{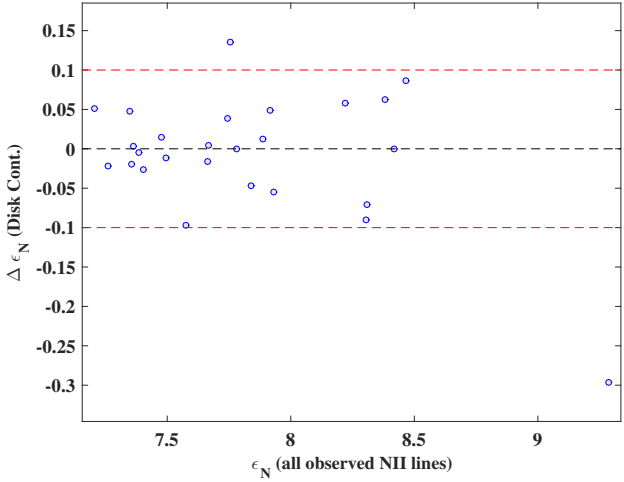}
\caption{Change in the measured nitrogen abundances after including corrections for disk contamination for the MiMeS Be star sample.}
\label{Nabund_DiskCont_Corr}
\end{figure}

To derive nitrogen abundances corrected for disk emissiom, we correct the measured line equivalent width for the $\Delta_{\rm D}$ factor listed in Table~\ref{Disk_Cont_results} as follows: over the narrow width of the N\,{\sc ii} spectral lines, the disk contributes a wavelength independent flux, ${F_D}$, making the observed flux at wavelength $\lambda$
\begin{equation}
F_{\lambda} = F_{\lambda}^* + F_D \,,
\end{equation}
where $F_{\lambda}^*$ is the stellar flux in the absence of the disk.
Choosing a wavelength outside of the line, we have the continuum flux
\begin{equation}
F_{C} = F_{C}^* + F_D \,,
\end{equation}
where $F_{C}^*$ is the stellar continuum flux in the absence of the disk. The observed equivalent width is
\begin{equation}
W_{\lambda}^{obs} \equiv \int_{\rm line} \left(\frac{F_{\lambda}-F_{C}}{F_{C}}\right)\,d\lambda \,,
\end{equation}
which includes the contribution of the disk and hence is not equal to the photospheric equivalent width.
Inserting the above relations, and using the definition of the photospheric equivalent width as
\begin{equation}
W_{\lambda}^{photo} \equiv \int_{\rm line}  \left(\frac{F_{\lambda}^*-F_{C}^*}{F_{C}^*}\right)\,d\lambda \,,
\end{equation}
the relation between $W_{\lambda}^{obs}$ and $W_{\lambda}^{cor}$ is
\begin{equation}
W_{\lambda}^{obs} = \frac{F_C^*}{F_C^*+F_D}\, W_{\lambda}^{photo} \,.
\end{equation}
As noted above, model calculations with {\sc bedisk} and {\sc beray} provide the quantity $\Delta_D\equiv (F_C^*+F_D)/F_C^*$
at the wavelength of each line for the appropriate disk density model found by matching the H$\alpha$
profile. To obtain the photospheric equivalent, we correct the observed equivalent widths as
\begin{equation}
W_{\lambda}^{photo} = \Delta_D\, W_{\lambda}^{obs} \,.
\end{equation}
While one may expect $\Delta_D>1$, making the corrected equivalent widths larger than the observed ones,
this is not necessarily the case. In shell stars, for example, the stellar continuum
is seen through the disk and there is a reduction of the stellar flux associated with the optical depth
of the disk. In this case, $F_D$ will be negative, making $F_C^*+F_D < F_C^*$ and $\Delta_D<1$.

Finally, we note that for all of the N\,{\sc ii} lines, we have neglected line emission from the disk in the
N\,{\sc ii} transitions themselves. We have explicitly calculated disk contributions to the N\,{\sc ii} $\lambda\,3995$ line over the range of disk models considered and find negligible line emission contribution of the disk to the line profile for disk density models
found by H$\alpha$ fitting. In very dense disks, $n=2.0$ and $\rho_0=1.0\cdot 10^{-10}\;\rm g\,cm^{-3}$, some small emission is predicted in the far wings of the lines (i.e.\ at high velocity) as the N\,{\sc ii} lines are formed in the inner portion of the disk close to the star. We note that none of our observed N\,{\sc ii} lines, including the strongest transition at $\lambda\,3995$, show any trace of emission in the line wings.  

After correcting the observed equivalent widths in this fashion, the abundances analysis was re-performed with the corrected equivalent widths. Figure~\ref{Nabund_DiskCont_Corr} shows the change in the estimated nitrogen abundances after applying corrections for the disk contamination. As the figure shows, the corrections of the nitrogen abundance for the disk contamination lie within $\pm\,0.1$~dex for most of the sample. There are a few stars with larger differences; however, all of these differences are smaller than the estimated uncertainties of the nitrogen abundances due to errors in the continuum normalization. 

\vspace{0.5cm}
\begin{figure}
\begin{center}
\hspace{-0.5cm}
\includegraphics[scale=0.40]{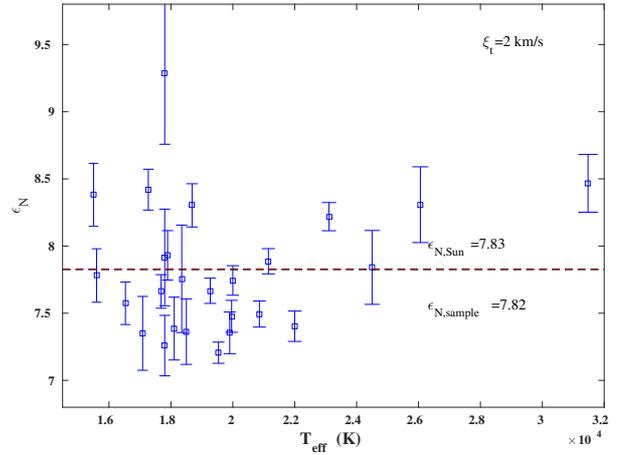}
\caption{Nitrogen abundance corrected for disk contamination versus $T_{\rm eff}$ for the MiMeS Be star sample.}
\label{N_abund_corr_Teff}
\end{center}
\end{figure}

Figure~\ref{N_abund_corr_Teff} shows the nitrogen abundances corrected for the disk contamination effect as a function of the stellar effective temperature. The observed pattern does not change significantly from that seen in Figure~\ref{N_abund_Teff}. Consequently, we can conclude that the observed nitrogen enrichments in the atmospheres of a number of Be stars are real. 

\begin{table}
\centering{
\caption{Nitrogen Abundances of the MiMeS Be Star Sample Corrected for Disk Contamination.
\label{N_abundances_corrected}}}
\begin{center}
{\small
\begin{tabular}{cccccc}
\hline\hline
HD &	$T_{\rm eff}\,(\rm K)$ & $\log g$	 & \multicolumn{2}{c}{$\epsilon_N\,\pm\,\Delta\epsilon_N$ (dex)}\\
	 &	& & \multicolumn{1}{c}{{\small $\xi_t=\,2.0\, \rm  km\,s^{-1}$}} & \multicolumn{1}{c}{{\small $\xi_t=\,5.0\,\rm  km\,s^{-1}$}}\\	
\hline
\multicolumn{1}{l}{ 11415} &15600 & 3.50  &   7.78 $\pm$  0.20 &  7.75$\pm$0.15\\
\multicolumn{1}{l}{ 20336} &18684 & 3.90  &   8.21 $\pm$  0.18 &  8.12$\pm$0.15\\
\multicolumn{1}{l}{ 33328} &21150 & 3.60  &   7.90 $\pm$  0.09 &  7.76$\pm$0.09\\
\multicolumn{1}{l}{ 45725} &17800 & 3.90  &   7.96 $\pm$  0.31 &  7.89$\pm$0.28\\
\multicolumn{1}{l}{ 49567} &17270 & 3.30  &   8.42 $\pm$  0.15 &  8.27$\pm$0.14\\
\multicolumn{1}{l}{ 54309} &20859 & 3.40  &   7.48 $\pm$  0.07 &  7.43$\pm$0.08\\
\multicolumn{1}{l}{ 56139} &19537 & 3.40  &   7.26 $\pm$  0.10 &  7.22$\pm$0.08\\
\multicolumn{1}{l}{ 58050} &19961 & 3.80  &   7.49 $\pm$  0.13 &  7.44$\pm$0.10\\
\multicolumn{1}{l}{ 58343} &16530 & 3.60  &   7.48 $\pm$  0.13 &  7.43$\pm$0.13\\
\multicolumn{1}{l}{ 58978} &24500 & 3.40  &   7.79 $\pm$  0.26 &  7.67$\pm$0.25\\
\multicolumn{1}{l}{ 65875} &19900 & 3.30  &   7.33 $\pm$  0.17 &  7.32$\pm$0.14\\
\multicolumn{1}{l}{ 67698} &15500 & 3.70  &   8.44 $\pm$  0.24 &  8.31$\pm$0.23\\
\multicolumn{1}{l}{120324} &20000 & 4.00  &   7.78 $\pm$  0.12 &  7.70$\pm$0.09 \\
\multicolumn{1}{l}{ 143275} &31478 & 3.50  &   8.55 $\pm$  0.32 &  8.54$\pm$0.31\\
\multicolumn{1}{l}{ 174237} &17683 & 3.65  &   7.65 $\pm$  0.14 &  7.61$\pm$0.13\\
\multicolumn{1}{l}{ 178175} &22000 & 3.50  &   7.38 $\pm$  0.12 &  7.33$\pm$0.10\\
\multicolumn{1}{l}{ 187567} &23110 & 3.70  &   8.28 $\pm$  0.11 &  8.10$\pm$0.08\\
\multicolumn{1}{l}{ 187811} &17800 & 3.80  &   7.24 $\pm$  0.24 &  7.19$\pm$0.19\\
\multicolumn{1}{l}{ 189687} &18106 & 3.50  &   7.38 $\pm$  0.23 &  7.40$\pm$0.23\\
\multicolumn{1}{l}{ 191610} &18350 & 3.70  &   7.89 $\pm$  0.34 &  7.73$\pm$0.32\\
\multicolumn{1}{l}{ 192685} &18500 & 3.70  &   7.37 $\pm$  0.24 &  7.34$\pm$0.26\\
\multicolumn{1}{l}{ 203467} &17087 & 3.38  &   7.40 $\pm$  0.30 &  7.35$\pm$ 0.28\\
\multicolumn{1}{l}{205637} &17800 & 3.50  &   8.99 $\pm$ 0.54 & 9.42$\pm$0.17 \\
\multicolumn{1}{l}{ 212076} &19270 & 3.50  &   7.67 $\pm$  0.10 &  7.61$\pm$0.08\\
\multicolumn{1}{l}{ 212571} &26061 & 3.70  &   8.24 $\pm$  0.25 &  8.06$\pm$0.23\\
\multicolumn{1}{l}{ 217050} &17893 & 3.30  &   7.88 $\pm$  0.19 &  7.78$\pm$0.16\\
\hline
\end{tabular}}
\end{center}
\end{table}
\section{Nitrogen Abundances for the normal B star sample}
\label{Nabund_res_B_stars}

\begin{table}
\centering{
\caption{Nitrogen Abundances of the MiMeS Normal B Star Sample. 
\label{N_abundances_obser_BStars}}}
\begin{center}
{\small
\begin{tabular}{cccccc}
\hline\hline
HD &	$T_{\rm eff}\,(\rm K)$ & $\log g$	 & \multicolumn{2}{c}{$\epsilon_N\,\pm\,\Delta\epsilon_N$ (dex)}\\
	 &	& & \multicolumn{1}{c}{{\small $\xi_t=\,2.0\, \rm  km\,s^{-1}$}} & \multicolumn{1}{c}{{\small $\xi_t$ is not fixed}}\\ 
\hline
\multicolumn{1}{l}{   3360 }& 20750.0 & 3.8&8.14 $\pm$ 0.03& 8.23 $\pm$ 0.05 \\
\multicolumn{1}{l}{  30836 }& 21874.0 & 3.5&7.79 $\pm$ 0.04& 7.69 $\pm$ 0.08 \\
\multicolumn{1}{l}{  35468 }& 22000.0 & 3.6&8.29 $\pm$ 0.05& 8.33 $\pm$ 0.14 \\
\multicolumn{1}{l}{  35708 }& 20700.0 & 4.2&8.13 $\pm$ 0.05& 8.19 $\pm$ 0.09 \\
\multicolumn{1}{l}{  36629 }& 20300.0 & 4.2&7.70 $\pm$ 0.04& 7.70 $\pm$ 0.07 \\
\multicolumn{1}{l}{  36822 }& 30000.0 & 4.0&8.13 $\pm$ 0.06& 8.05 $\pm$ 0.13 \\
\multicolumn{1}{l}{  36959 }& 26100.0 & 4.2&7.77 $\pm$ 0.03& 7.82 $\pm$ 0.02 \\
\multicolumn{1}{l}{  36960 }& 29000.0 & 4.1&7.79 $\pm$ 0.06& 7.77 $\pm$ 0.10 \\
\multicolumn{1}{l}{  46328 }& 27000.0 & 3.8&8.12 $\pm$ 0.03& 8.18 $\pm$ 0.04 \\
\multicolumn{1}{l}{  48977 }& 20000.0 & 4.2&7.45 $\pm$ 0.09& 7.42 $\pm$ 0.10 \\
\multicolumn{1}{l}{  61068 }& 23800.0 & 4.0&8.00 $\pm$ 0.04& 8.07 $\pm$ 0.04 \\
\multicolumn{1}{l}{  66665 }& 28500.0 & 3.9&8.07 $\pm$ 0.05& 8.11 $\pm$ 0.05 \\
\multicolumn{1}{l}{  74560 }& 17000.0 & 4.0&7.52 $\pm$ 0.13& 7.44 $\pm$ 0.14 \\
\multicolumn{1}{l}{  74575 }& 22900.0 & 3.6&8.38 $\pm$ 0.04& 8.17 $\pm$ 0.10 \\
\multicolumn{1}{l}{  85953 }& 18600.0 & 3.9&7.88 $\pm$ 0.07& 7.86 $\pm$ 0.12 \\
\multicolumn{1}{l}{ 169467 }& 16700.0 & 4.1&7.85 $\pm$ 0.13& 7.81 $\pm$ 0.12 \\
\hline
\end{tabular}}
\end{center}
\end{table}

Estimates of the nitrogen abundances of the normal B-type star sample were obtained following the same procedure as described in previous sections. As there are several, recent, high-precision abundance analysis of Galactic, main sequence B, stars available \citep{nie12,lyu12}, comparison to these studies will shed some light on the accuracy of the analysis done in this work. 

\begin{figure}
\centering
\hspace{-0.5cm}
\includegraphics[scale= 0.35]{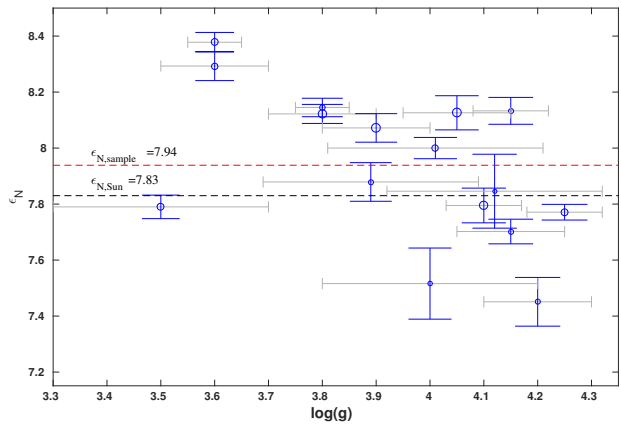}
\caption{Nitrogen abundances of the normal B star sample as function of stellar surface gravity. The dashed black line represent the solar nitrogen abundance, and the red dashed line, the mean abundance of the sample. The symbol sizes increase with stellar masses. \label{fig:Normal_BStars_Nabund_vs_logg}}  
\end{figure}

The derived nitrogen abundances for the B star sample are listed in Table~\ref{N_abundances_obser_BStars}.  Two cases are presented: in the first case, the microturbulent velocity was kept fixed at $2\,\rm km\,s^{-1}$, while in the second case, the microturbulent velocity was allowed to vary and set by forcing weak and strong N\,{\sc ii} lines to yield the same abundance. This was possible because the B-star sample has low $v\sin i$ values, as noted before. The mean abundance and 1$\,\sigma$ variations of the sample are $\epsilon_{\rm N}=7.92\,\pm\,0.26$ for the fixed $2\,\rm km\,s^{-1}$ microturbulent case, and $\epsilon_{\rm N}=7.93\,\pm\,0.28$ for the variable microturbulence case. Thus in both case, the nitrogen abundance of the MiMeS B star sample coincides with the solar abundance, and agrees well with the mean nitrogen abundance determined by \citet{nie12} and \citet{lyu12} for Galactic B~stars. Figure~\ref{fig:Normal_BStars_Nabund_vs_logg} plots the nitrogen abundances of the B star sample as a function of the adopted stellar gravities. There is a suggestion of an anti-correlation, with the lowest $\log g$ stars having the highest measured nitrogen abundances.

\begin{figure}
\centering
\vspace{0.5cm}
\includegraphics[scale= 0.5]{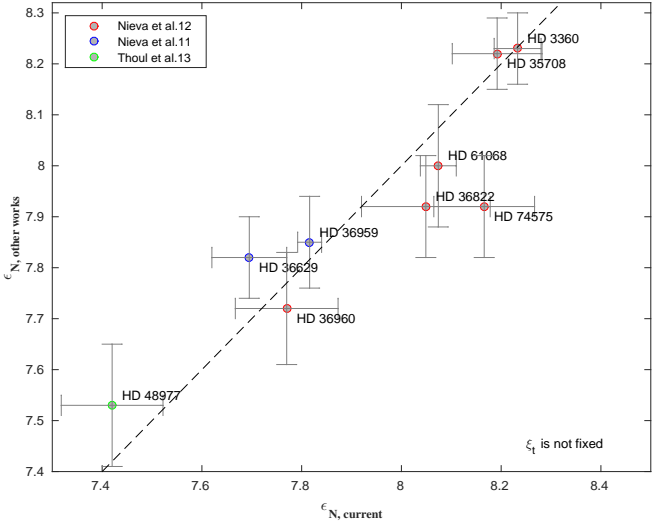}
\caption{Comparison of the nitrogen abundances derived in this work for normal B stars with other, recent, high-precision nitrogen determinations. Each point is identified by its stellar HD number. The comparison abundance sources are \protect\citet{nie11} (blue), \protect\citet{nie12} (red) and \protect\citet{thou13} (green). \label{fig:Normal_BStars_Nabund_Comp}}  
\end{figure}

A more detailed comparison of the nitrogen abundance of our MiMeS sample is possible as there are nine stars in common with other studies. Figure~\ref{fig:Normal_BStars_Nabund_Comp} compares our results with those available in the literature. Agreement is quite good, with most estimates agreeing withing $\pm0.1$~dex and within the respective errors. A single discordant case is the B star HD74575 where our nitrogen abundance is significantly larger, although the same stellar parameters were used in the current analysis and in that of \citet{nie12}. Finally we note one star, HD~48977, with a significant underabundance in nitrogen compared to the solar value. However, this result agrees within the uncertainties with the abundance of \cite{thou13}. 

Given that our abundance analysis for the MiMeS normal B star sample seems reliable and that the errors are consistently estimated, we return to the apparent correlation between nitrogen abundance and $\log g$ noted above. We estimate the stellar masses, radii, and ages for the MiMeS sample by interpolation in the evolutionary grids of rotating, massive stars of \citet{Eks12}. For this purpose, we followed the spectroscopic Hertzsprung-Russell (sHR) diagram procedure of \citet{LK14}. 
In Figure~\ref{fig:Normal_BStars_Nabund_vs_logg}, the stellar mass is represented by the size of the symbol such that the symbol size increases with the increase of the stellar mass. There is a weak correlation between the stellar masses and the measured abundances. Stars with masses equal to or less than seven solar masses do not show nitrogen enrichment, while many of the more massive stars do show significant nitrogen enrichment.  This trend agrees with the prediction of the theoretical models of the evolution of massive stars that show that the efficiency of matter transport by rotational mixing increases with stellar mass \citep{MM12}.

Returning to the Be star MiMeS sample and Figure~\ref{N_abund_corr_Teff}, there is an unsettling aspect to the Be star results. A significant number of Be stars have nitrogen abundances $\epsilon_{\rm N} < 7.6\;$dex whereas the above B star sample has only a few such objects. It is difficult to easily reconcile a significant population of sub-solar abundance Be stars when no corresponding main sequence population of B stars exists. One source of uncertainty is the neglect of line emission in the N\,{\sc ii} transitions by the disk which would fill in the lines. However, we explicitly checked this possibility with direct calculations based on our sample of disk density parameters and do not find significant N\,{\sc ii} emission. Another possibility is dilution of the primary Be star spectrum by an unresolved binary companion. We return to this point in the discussion.
In addition, there are a small number of high nitrogen abundance objects in the Be star sample. The highest abundance object, $\epsilon_{\rm N} > 9$, has a very large uncertainty. The next highest nitrogen abundance, HD~49567, with $\epsilon_{\rm N} = 8.42\pm 0.15$, is an interesting case. With sharp lines due to its lower $v\,\sin i$, the abundance seems reliably determined. This star is a member of a HMXB binary star system \citep{SM96}, and hence its high nitrogen abundance may be due to mass transfer.

\section{Discussion}
\label{disc}

\begin{figure}
\centering
\includegraphics[scale=0.65]{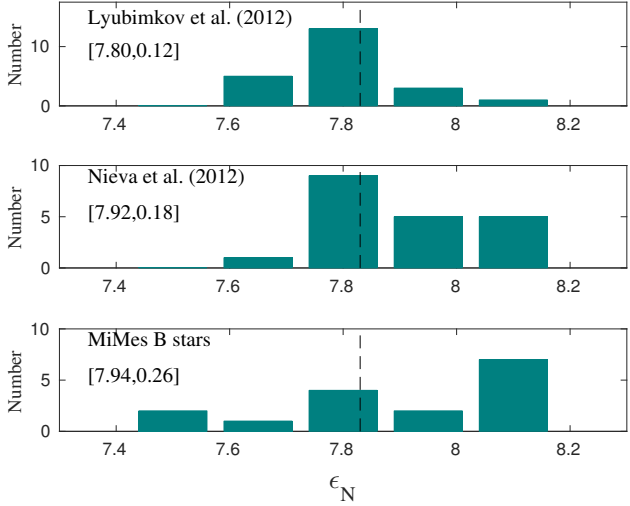}
\caption{Histograms of the measured nitrogen abundances in B stars. The top panel is the B star sample of \citet{lyu12}, the middle panel is the B star sample of \citet{nie12}, and the bottom panel is the current MiMeS normal B star sample. The mean and standard deviation of each sample are given in the panels. The solar nitrogen abundance is indicated by the dashed vertical line in each panel. 
\label{fig:nitrogen_hist}}  
\end{figure}

Figure~\ref{fig:nitrogen_hist} compares our MiMeS sample B star nitrogen abundances to \citet{lyu12} and \citet{nie12}. The mean nitrogen abundances are in excellent agreement, and all are close to the solar value. The dispersion in the current results, however, is somewhat lager, particularly compared to \citet{lyu12}, mostly due to a small number of stars with decidely sub-solar abundances, $\epsilon_{\rm N}<7.6$. We also note that our sample finds a significant fraction of nitrogen-enriched objects, agreeing roughly with \citet{nie12}. This figure demonstrates that our method and analysis can approach the accuracy of current high-precision abundance determinations. 

\begin{figure}
\centering
\includegraphics[scale=0.65]{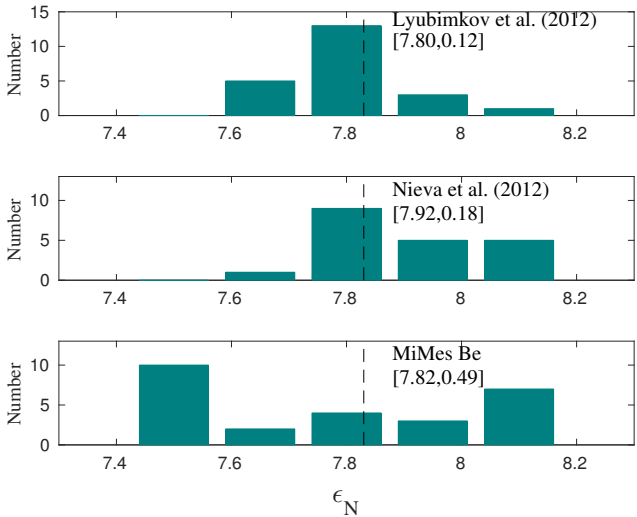}
\caption {The same as Figure~\ref{fig:nitrogen_hist} but the results of \citet{lyu12} (top panel) and \citet{nie12} (middle panel) are now compared with the measured nitrogen abundances (corrected for disk emission) of the MiMeS Be star sample.
\label{fig:nitrogen_hist2}}  
\end{figure}

Figure~\ref{fig:nitrogen_hist2} compares the results of our MiMeS Be star sample, corrected for disk emission, with the same two samples of Galactic B stars as above, \citet{lyu12} and \citet{nie12}. The mean nitrogen abundance again coincides with the solar nitrogen abundance, but this time the dispersion in the abundances in the Be star sample is much larger, $\sigma_{\rm N}=0.49$, versus $0.12$ to $0.18$ for the B star samples. This large dispersion is a result of a significant number of Be stars, about $10$ out of $26$, with decidedly sub-solar nitrogen abundances. This population of objects is not present in the galactic B star samples of \citet{lyu12} or \citet{nie12}. Also in the Be star sample, there is a significant fraction of Be stars with mildly enhanced nitrogen abundances. In fact, both our B and Be star samples are consistent with the finding of \citet{nie12} with about one third of the sample giving enhanced nitrogren abundances.  

\begin{figure}
\centering
\includegraphics[scale=0.40]{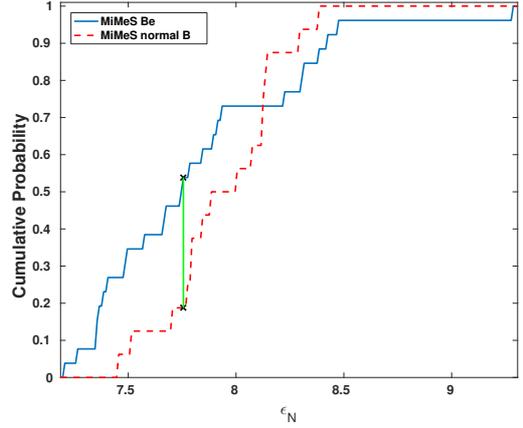}
\caption {The cumulative distribution of nitrogen abundances for the MiMeS normal B (solid line) and Be star (dotted line) samples. A two-tailed K-S test reveals that the two distributions are statistically different. The vertical green line highlights the maximum difference between the two cumulative distributions.
\label{KS_test_MiMeS_sample}}  
\end{figure}

Finally, in Figure~\ref{KS_test_MiMeS_sample}, we compare the cumulative distribution of nitrogen abundance in the B and Be star samples. They are statistically different at the 5\% level using a two-tailed K-S test.

About half of the stars in the normal B-type sample and one third of the stars in the Be sample have nitrogen abundances higher than the solar value. This results agrees with previous works on normal B stars, such as \citet{Mae14} and \citet{nie12},  who found nitrogen enrichment in about one third of their samples.The largest source of uncertainties in the current analysis is the continuum normalization of the observed spectra, which is a consequence of the rapid rotation of the stars included in our samples. 

The order of the expected corrections for gravitational darkening lies within $\pm 0.1\;$dex for most of the stars included in this analysis. This is smaller than the estimated uncertainties of all Be stars in the sample. As a result, it is not expected that correcting for gravitational darkening would significantly change the estimated nitrogen abundances. The disk contamination, determined by extracting the disk density parameters for the Be stars from H$\alpha$ line profile fitting, is less than $5\%$ for most of the cases, although it can reach as high as $\approx\,25\%$. This is smaller than the contamination found by \citet{dun11}, where values as large as $60\,\%$ were found; however, the \citet{dun11} method is quite different from the direct modelling approach of the current work. In the current Be star sample, including disk contamination corrections changes the estimated nitrogen abundance by $\approx\,\pm\,0.1$ dex in most cases. These corrections change the estimated nitrogen abundances of a few stars, but they do not significantly change the overall results. In conclusion, corrections for gravitational darkening and disk contamination do not significantly change the measured nitrogen abundance for the Be stars from values based solely on the observed equivalent widths.

The large fraction of decidedly sub-solar nitrogen abundances among the Be stars is a puzzle. While such lower abundances are found among the B star sample also, the fraction of low abundance stars in the B star sample is much smaller. Our detailed error analysis indicates suggests that the low abundances among the Be stars is statistically significant, but error budget only accounts for random errors in the curves-of-growth, the observed equivalent widths, and the adopted stellar parameters. We may have underestimated the systematic effects of either gravitational darkening, disk contamination, or systematic errors in the observed line profiles and determined equivalent widths. Another possibility is dilution of the spectra by unresolved binary companions that would lead to lower observed nitrogen equivalent widths. A SIMBAD \citep{W00} search of the stars in the Be sample reveals only a single known spectroscopic binary, HD\,147235 ($\delta\,$Soc), although others may certainly be present. Among the B star sample, there are four known spectroscopic binaries: HD\,30836, HD\,36822, HD\,48977, and HD\,74560. 

Finally we note that, in general, Be star samples, consisting of large $v\,\sin i$ objects, are difficult cases for precision abundance analysis, and this is reflected in the larger dispersion among the Be star abundances, approximately twice that of the B star sample.

\section{Future Directions}
\label{future_plan}

Perhaps the most important theoretical limitation of the present work is the treatment of gravitational darkening as a perturbation. While the current work carefully quantified the effect of gravitational darkening on the nitrogen abundances, gravitational darkening was not included in the underlying, fundamental stellar parameters. Although the more recent treatment of gravitational darkening of \cite{lar11} suggests that the amount of gravitational darkening with rotation is significantly less than previously assumed, it would be preferable to close this consistency loop. It is certainly possible to do this, although the amount of calculation is considerable as non-LTE treatments of hydrogen and helium, and other metals, are required to fix the stellar parameters.

Another limitation of the current work is the exclusive focus on nitrogen. As noted by \cite{prz2010} and \cite{Mae14}, rotational mixing changes the surface CNO abundances in systematic and correlated ways that can be understood with fairly a straight-forward analysis. Enlarging the determined abundances for the Be stars to include both carbon and oxygen would allow for an important consistency check on the results, ensuring that the determined abundances, within their errors, change in the predicted, systematic way.

We are currently working to self-consistently determine Be star stellar parameters with the effects of gravitational darkening and disk contamination. We are also extending the non-LTE abundance analysis of both the B and Be stars of the MiMeS survey to carbon and oxygen. Finally, including more accurate stellar masses and radii using grids of stellar evolutionary models in the computation of gravitational darkening and disk contamination \cite[e.g.\/][]{geo2013,gra2013}, and a more detailed treatment of the possible effects of binarity, may improve the results of the Be star abundance analysis. 

\section*{Acknowledgements}

\vspace{0.3in}
We would like to thank the anonymous referee for many helpful comments. This work is supported by the Canadian Natural Sciences and Engineering Research Council (NSERC). TAAS also greatfully acknowledges support from Western University. This research made use of the SIMBAD database, operated at CDS, Strasbourg, France.





\begin{thebibliography}{99}

\bibitem[\protect\citeauthoryear{Arcos et al.}{2017}]{arc17} Arcos, C., Jones, C.\ E., Sigut, T.\ A.\ A., Kanaan, S., \& Cur\'{e}, M., 2017, ApJ, 842, 48

\bibitem[\protect\citeauthoryear{Aerts et al.}{1999}]{aer99} Aerts, C., et al., 1999, A\&A, 343, 872

\bibitem[\protect\citeauthoryear{Ahmed \& Sigut}{2016}]{AS15} Ahmed, A. \& Sigut, T.A.A.\ 2016, MNRAS, 455, 1099

\bibitem[\protect\citeauthoryear{Ahmed \& Sigut}{2012}]{AS12} Ahmed, A. \& Sigut, T.A.A.\ 2012, ApJ, 744, 191

\bibitem[\protect\citeauthoryear{Brott er al.}{2011}]{brott11}Brott, I., et al., 2011, A\&A, 530, A115	


\bibitem[\protect\citeauthoryear{Catanzaro}{2013}]{Cat13}Catanzaro, G.\ 2013, A\&A, 550, A79

\bibitem[\protect\citeauthoryear{Collins}{1965}]{Collins65} Collins, G. W., II.\ 1965, ApJ, 142, 265


\bibitem[\protect\citeauthoryear{Cot\'{e} et al}{1996}]{cote96}Cot\'{e}, J. \& Waters, L.B.F.M. \& Marlborough, J.M.\ 1996, A\&A, 307, 184

\bibitem[\protect\citeauthoryear{Cox}{2000}]{Cox00} Cox, Arthur, N., 2000, Allen's Astrophysical Quantities, Fourth Edition, Springer-Verlag, Berlin

\bibitem[\protect\citeauthoryear{Cranmer}{2005}]{cra05} Cranmer. Steven R.\ 2005, ApJ, 634, 585 

\bibitem[\protect\citeauthoryear{Dunstall et al.}{2011}]{dun11} Duntsall, P. R., et al., 2011, A\&A, 536, A65 

\bibitem[\protect\citeauthoryear{Ekstr\"{o}m et al.}{2012}]{Eks12}Ekstr\"{o}m, S., et al., 2012, A\&A, 537, A146

\bibitem[\protect\citeauthoryear{Espinosa Lara \& Rieutord}{2011}]{lar11} Espinosa Lara, F. \& Rieutord, M.\ 2011, A\&A, 533, A43

\bibitem[\protect\citeauthoryear{Evans et al.}{2005}]{evans05} Evans, C.J., et al., 2005, \aap, 437, 467

\bibitem[\protect\citeauthoryear{Fr\'{e}mat et al.}{2005}]{fre05} Fr\'{e}mat, Y., Zorec, J., Hubert, A.-M. \& Floquet, M.\ 2005, A\&A, 440, 305

\bibitem[\protect\citeauthoryear{Fukuda}{1982}]{Fuk82}Fukuda, Ichiro\ 1982, PASP, 94, 271	

\bibitem[\protect\citeauthoryear{Gies \& Lambert}{1992}]{gie92} Gies, Douglas R., \& Lambert, David L.\ 1992, ApJ, 387, 673

\bibitem[\protect\citeauthoryear{Gray}{2005}]{gray05} Gray, David F., 2005, The Observation and Analysis of Stellar Photospheres, Third Edition. Cambridge Univ. Press, Cambridge

\bibitem[\protect\citeauthoryear{Georgy et al.}{2013}]{geo2013} Georgy, C., Ekstr\"{o}m, S., Granada, A., Meynet, G., Mowlavi, N., Eggenberger, P., \& Maeder, A., 2013, A\&A, 553, A24

\bibitem[\protect\citeauthoryear{Granada et al.}{2013}]{gra2013} Granada, A., Ekstr\"{o}m, S., Georgy, C., Krti\v{c}ka, J., Owocki, S., Meynet, G., \& Maeder, A., 2013, A\&A, 553, A25


\bibitem[\protect\citeauthoryear{Harmanec}{2000}]{H00}Harmanec, P., 2000, in Smith, M.A., Henrichs, H.F., and Fabregat, J., eds, ASP Conf. Ser. Vol. 214, The Be Phenomena in Early-Type Stars, p. 13

\bibitem[\protect\citeauthoryear{Hubrig et al.}{2009}]{hub09} Hubrig, S., Briquet, M., De Cat, P., Sch\"{o}ller, M., Morel, T., \& Ilyin, I.\ 2009, Astr. Nachr., 330, 317


\bibitem[\protect\citeauthoryear{Hunter et al.}{2009}]{hun09} Hunter, I., et al., 2009, A\&A, 496, 841 

\bibitem[\protect\citeauthoryear{Hunter et al.}{2008a}]{hun08a} Hunter, I., et al., 2008a, A\&A, 479, 541 

\bibitem[\protect\citeauthoryear{Hunter et al.}{2008b}]{hun08b} Hunter, I., et al., 2008b, ApJ, 676, L29

\bibitem[\protect\citeauthoryear{Hunter et al.}{2007}]{hun07}  Hunter, I., et al., 2007, \aap, 466, 277


\bibitem[\protect\citeauthoryear{Keller}{2004}]{kel04} Keller, Stefan\ 2004, Publ.\ Astron.\ Soc.\ Australia, 21, 310

\bibitem[\protect\citeauthoryear{K\"{o}hler et al.}{2012}]{koh12} K\"{o}hler, K., Borzyszkowski, M., Brott, I., and Langer, N., 2012, A\&A, 544, A76


\bibitem[\protect\citeauthoryear{Kurucz}{1993}]{kur93} Kurucz, R. F.\ 1993, Kurucz CD-ROM 13, ATLAS9 Stellar Atmosphere Programs (Cambridge: SAO)

\bibitem[\protect\citeauthoryear{Kurucz}{1979}]{kur79} Kurucz, R. F.\ 1979, Model atmospheres for G, F, A, B, and O stars, ApJS, 40, 1

\bibitem[\protect\citeauthoryear{Langer \& Kudritzki}{2014}]{LK14}Langer, N., \& Kudritzki, R. P.\ 2014, A\&A 564, A52

\bibitem[\protect\citeauthoryear{Lanz \& Hubeny}{2007}]{Lanz07}  Lanz, T. \& Hubeny, I.\ 2007, ApJS, 83, 104.

\bibitem[\protect\citeauthoryear{Lefever et al.}{2010}]{lef10} Lefever, K., Puls, J., Morel, T., Aerts, C., Decin, L., \& Briquet, M.\ 2010, A\&A, 515, A74

\bibitem[\protect\citeauthoryear{Lennon et al.}{2005}]{len05} Lennon, D. J., Lee, J.-K., Dufton, P. L., and Ryans, R. S. I.\ 2005, A\&A, 438, 265


\bibitem[\protect\citeauthoryear{Levenhagen \& Leister}{2006}]{LL06} Levenhagen, R. S. \& Leister, N. V.\ 2006, MNRAS, 371, 252

\bibitem[\protect\citeauthoryear{Lodders}{2003}]{lod03} Lodders, K., 2003, ApJ, 591, 1220

\bibitem[\protect\citeauthoryear{Lyubimkov et al.}{2013}]{lyu12} Lyubimokv, Leonid S., \& 4 co-authors\ 2013, MNRAS, 428, 3497 


\bibitem[\protect\citeauthoryear{Maeder et al.}{2014}]{Mae14}Maeder, Andr\'{e}, et al., 2014, A\&A, 565, A39

\bibitem[\protect\citeauthoryear{Maeder et al.}{2009}]{mae09} Maeder, A., Meynet, G., Ekstr\"{o}m, S., \& Georgy, C.\ 2009, Communications in Asteroseismology, 158, 72

\bibitem[\protect\citeauthoryear{Maeder \& Meynet}{2012}]{MM12}Maeder, Andr\'{e}, \& Meynet, Georges \ 2012, Rev.\ Mod.\ Phys., 84, 25

\bibitem[\protect\citeauthoryear{Maeder}{2009}]{mae_b09} Maeder, A.\ 2009, Physics, Formation and Evolution of Rotating Stars. Astronomy and Astrophysics Library, Springer, Berlin


\bibitem[\protect\citeauthoryear{Martayan et al.}{2006}]{mar06} Martayan, C., Fr\'{e}mat, Y., Hubert, A.-M., Floquet, M., Zorec, J., and Neiner, C., 2006, A\&A, 452, 273

\bibitem[\protect\citeauthoryear{McGill}{2013}]{Meghan13}McGill, M. A., 2013, PhD Thesis, The University of Western Ontario

\bibitem[\protect\citeauthoryear{McGill et al.}{2013}]{MJS13}McGill, M. A., 2013, McGill, M.\ A., Sigut, T.\ A.\ A., \& Jones, C.\ E., 2011, ApJS, 204, 2

\bibitem[\protect\citeauthoryear{McGill et al.}{2011}]{MJS11}McGill, M. A., 2011, McGill, M.\ A., Sigut, T.\ A.\ A., \& Jones, C.\ E., 2011, ApJ, 743, 111

\bibitem[\protect\citeauthoryear{McSwain \& Gies}{2005}]{mcs05} McSwain, M. Virginia, \& Gies, Douglas R.\ 2005, ApJS, 161, 118

\bibitem[\protect\citeauthoryear{Meynet \& Maeder}{2000}]{mey00} Meynet, G., \& Maeder, Andr\'{e}, 2000, A\&A, 361, 101

\bibitem[\protect\citeauthoryear{Mihalas}{1978}]{mil78} Mihalas, Dimitri, 1978, Stellar Atmospheres, 2nd Edition. W.\ H.\ Freeman \& Company, New York


\bibitem[\protect\citeauthoryear{Moon \& Dworetsky}{1985}]{MD85} Moon, T. T. \& Dworetsky, M. M., 1985, MNRAS, 217, 305


\bibitem[\protect\citeauthoryear{Nieva \& Przybilla}{2014}]{NP14}Nieva, M.-F.\ \& Przybilla, Norbert\ 2014, A\&A, 566, A7 

\bibitem[\protect\citeauthoryear{Nieva \& Przybilla}{2012}]{nie12} Nieva, M.-F. \& Przybilla, N.\ 2012, A\&A, 539, A143

\bibitem[\protect\citeauthoryear{Nieva \& Sim\'{o}n-D\'{i}az}{2011}]{nie11} Nieva, M.-F. \& Sim\'{o}n-D\'{i}az, S.\ 2011, A\&A, 532, A2


\bibitem[\protect\citeauthoryear{Palacios}{2013}]{Pal13} Palacios, A., 2013, EAS Publications Series, 62, 227

\bibitem[\protect\citeauthoryear{Petit}{2011}]{pet11} Petit, V., Massa, L.D., Marcolino, W.L.F., Wade, G.A., \& Ignace, R., 2011, MNRAS, 412, 45

\bibitem[\protect\citeauthoryear{Porter}{1996}]{por96}Porter, John M., 1996, MNRAS, 250, L31

\bibitem[\protect\citeauthoryear{Przybilla et al.}{2010}]{prz2010}Przybilla, N., Firnstein, M., Nieva, M.F., \& Maeder, A., 2010, A\&A, 517, A38

\bibitem[\protect\citeauthoryear{Richards et al.}{2000}]{RICH00}Richards, M.\ T., Koubsk\'{y}, P., Vojt\v{e}ch, \v{S}., Peters, G., Ryuko, H., \v{S}koda, P., \& Masuda, S., 2000, ApJ 531, 1003

\bibitem[\protect\citeauthoryear{Rivinius et al.}{2013}]{Riv13}Rivinius, T., Carciofi, A. C., \& Martayan, C.\  2013, A\&ARv, 21, 69

\bibitem[\protect\citeauthoryear{Schmidt-Kaler}{1982}]{SK82}Schmidt-Kaler, T., 1982, in Landolt-B\"{o}rnstein. 2, Subvol.\ b (Berlin: Springer-Verlag)


\bibitem[\protect\citeauthoryear{Silja et al.\/}{2014}]{sil14} Silaj, J.; Jones, C. E.; Sigut, T. A. A.; Tycner, C., 2014, ApJ 795, 1

\bibitem[\protect\citeauthoryear{Silja et al.\/}{2010}]{sil10} Silaj, J.; Jones, C. E.; Tycner, C.; Sigut, T. A. A.; Smith, A. D., 2010, ApJS, 187, 228

\bibitem[\protect\citeauthoryear{Sigut \& Patel}{2013}]{sp13}Sigut, T. A. A., \& Patel, P.\ 2013, ApJ, 765, 41

\bibitem[\protect\citeauthoryear{Sigut}{2011}]{sig11}Sigut, T. A. A., 2011, in C. Neiner, G. Wade, G. Meynet, \& G. Peters, eds, IAU Symp. 272, Active OB Stars: Structure, Evolution, Mass Loss, and Critical Limits. Cambridge Univ. Press, p. 426

\bibitem[\protect\citeauthoryear{Sigut et al.}{2009}]{sig09}Sigut, T. A. A., McGill, M. A., \& Jones, C. E.\ 2009, ApJ, 699, 1973

\bibitem[\protect\citeauthoryear{Sigut \& Jones}{2007}]{sig07} Sigut, T. A. A., \& Jones, C. E.\ 2007, ApJ, 668, 481 

\bibitem[\protect\citeauthoryear{Sigut}{1996}]{sig96} Sigut, T. A. A.\ 1996, ApJ, 473, 452

\bibitem[\protect\citeauthoryear{Sim\'{o}n-D\'{i}az et al.}{2017}]{SD17} Sim\'{o}n-D\'{i}az, S., Godart, M., Castro, N., Herrero, A., Aerts, C., Puls, J., Telting, J., \& Grassitelli, L., 2017, A\&A 597, A22


\bibitem[\protect\citeauthoryear{Sterken \& Manfroid}{1996}]{SM96} Sterken, C., \& Manfroid, J., 1996, A\&A, 305, 481

\bibitem[\protect\citeauthoryear{Stoeckley}{1968}]{Sto68} Stoeckley, T.R., 1968, MNRAS, 140, 141

\bibitem[\protect\citeauthoryear{Struve}{1931}]{Str31} Struve, O.\ 1931, AJ, 73, 94

\bibitem[\protect\citeauthoryear{Takeda et al.}{2010}]{tak10} Takeda, Yoichi, Kambe, Eiji, Sadakane, Kozo \& Masuda, Seiji, 2010, PASJ , 62, 1239

\bibitem[\protect\citeauthoryear{Tarasov \& Malchenko}{2012}]{tar12} Tarasov, A. E., \& Malchenko, S. L.\ 2012, Astro.\ Letters, 38, 428

\bibitem[\protect\citeauthoryear{Thoul et al.}{2013}]{thou13} Thoul, A., \& 12 co-authors \ 2013, A\&A, 551, A12

\bibitem[\protect\citeauthoryear{Tody}{1993}]{tod93} Tody, Doug, 1993, in Hanisch, R.J., Brissenden, R.J.V., and Barnes, J., ASP Conference Series vol. 52, Astronomical Data Analysis Software and Systems II, p.\ 173

\bibitem[\protect\citeauthoryear{Townsend et al.}{2004}]{tow04} Townsend, R. H. D., Owocki, S. P., \& Howarth, I. D.\ 2004, MNRAS, 350, 189

\bibitem[\protect\citeauthoryear{Trundle et al.}{2007}]{tru07} Trundle, C., et al., 2007, \aap, 471, 625.  

\bibitem[\protect\citeauthoryear{Underhill et al.}{1979}]{und79} Underhill, A. B., Divan, L., \& Burnichon, M.-L.\ 1979, MNRAS, 189, 601

\bibitem[\protect\citeauthoryear{van Belle}{2012}]{bel12} van Belle, Gerard T.\ 2012, A\&AR, 20, 51

\bibitem[\protect\citeauthoryear{von Zeipel}{1924}]{von24} von Zeipel, H.\ 1924, MNRAS, 48, 665

\bibitem[\protect\citeauthoryear{Wade et al.}{2016}]{wade16} Wade G. A. et al.\ 2016, MNRAS 456, 2


\bibitem[\protect\citeauthoryear{Wenger et al.}{2000}]{W00} Wegner, M., et al., 2000, A\&AS, 143, 9

\bibitem[\protect\citeauthoryear{Wisniewski \& Bjorkman}{2006}]{wis06} Wisniewski, J. P., \& Bjorkman, K.S.\ 2006, ApJ, 652, 458

\bibitem[\protect\citeauthoryear{Yudin}{2001}]{yud01} Yudin, R. V.\ 2001, A\&A, 368, 912

\bibitem[\protect\citeauthoryear{Zorec et al.}{2009}]{Zor09} Zorec, J., Cidale, L., Arias, M.L., Fre\'{e}mat, Y., Muratore, M.F., Torres, A.F., and Martayan, C., 2009, A\&A, 501, 297

\end{thebibliography}


\clearpage




\bsp	
\label{lastpage}
\end{document}